\documentclass[a4paper,11pt]{article}
\usepackage{jheppub}
\usepackage{amsmath}
\usepackage{amssymb}
\usepackage{amsfonts}
\usepackage{amsthm}
\usepackage{lineno}
\usepackage{dsfont}
\usepackage{bm}
\usepackage{physics}
\usepackage{mathtools}
\usepackage{commath}
\usepackage{color}
\usepackage{xcolor}
\usepackage[normalem]{ulem}

\renewcommand{\Re}{\mathrm{Re}}
\renewcommand{\Im}{\mathrm{Im}}
\renewcommand{\bar}{\overline}

\newcommand{\ii}{\mathrm{i}}

\newcommand{\A}{\textsc{a}}
\newcommand{\B}{\textsc{b}}

\newcommand{\LL}{\mathcal{L}}
\newcommand{\OO}{\mathcal{O}}
\newcommand{\tv}{\tilde{v}}

\newcommand{\M}{\mathcal{M}}
\newcommand{\N}{\mathcal{N}}

\newcommand{\eg}{{e.g.,}\ }
\newcommand{\ie}{{i.e.,}\ }
\newcommand{\mt}[1]{\textrm{\tiny #1}}

\newcommand{\LA}{L_\mt{AdS}}

\title{Entanglement harvesting in conformal field theory}

\author[1,2,4]{Kelly Wurtz,}
\author[1,3,4]{Caroline Lima,}
\author[1,3]{Robert C. Myers,}
\author[1,2,3]{Eduardo Mart\'in-Mart\'inez}

\affiliation[1]{Perimeter Institute for Theoretical Physics, Waterloo, Ontario, N2L 2Y5, Canada}
\affiliation[2]{Department of Applied Mathematics, University of Waterloo,\\ Waterloo, Ontario, N2L 3G1, Canada}
\affiliation[3]{Department of Physics and Astronomy, University of Waterloo,\\ Waterloo, Ontario N2L 3G1, Canada}
\affiliation[4]{Institute for Quantum Computing, University of Waterloo, Waterloo, Ontario, N2L 3G1, Canada}

\emailAdd{kwurtz@uwaterloo.ca, clima@perimeterinstitute.ca, rmyers.perimeter@gmail.com,
emartinmartinez@uwaterloo.ca}

\abstract{We study entanglement harvesting in general $d$-dimensional conformal field theories using pointlike Unruh–DeWitt detectors coupled to scalar primary operators. This extends standard harvesting protocols beyond free fields to interacting conformal theories and arbitrary spatial dimensions. We find that increasing the operator scaling dimension suppresses both negativity and mutual information, reflecting the faster decay of correlations. For holographic CFTs, we show that bulk effective field theory enables a separation between field-harvested and communication-mediated entanglement. We also derive asymptotic, closed-form approximations that agree well with numerical results.}

\begin{document}
\maketitle
\flushbottom

\section{Introduction}
\label{sec:calculation}

A defining feature of quantum field theory (QFT) is the presence of entanglement between spacelike-separated regions. Many questions at the intersection of QFT, gravity, and quantum information concern how this entanglement is structured and how it can be accessed. In particular, entanglement plays a central role in holographic dualities, where it is tied to the emergence of spacetime geometry \cite{Maldacena_1999,Ryu_2006,HRTproposal,VanRaamsdonk}. Traditionally, entanglement in QFT is analyzed using regulator-dependent quantities \cite{Srednicki_1993}, geometric constructs \cite{Ryu_2006}, or algebraic frameworks \cite{fewster2019AQFT}. While these yield powerful formal insights, they often obscure how such correlations could be accessed operationally. This motivates an approach, common in relativistic quantum information, that models the explicit interaction between localized probes and quantum fields \cite{unruh1976notes,DeWitt:1980}.

Entanglement harvesting is an operational protocol in which initially uncorrelated localized quantum probes become entangled through their local interactions with a quantum field \cite{Reznik_2003, Pozas_Kerstjens_2015}. Crucially, when the detectors remain spacelike-separated throughout their interaction with the field, any entanglement resulting between them cannot have been created by causal signalling; it must reflect entanglement already present in the field. In this way, correlations between the detectors can serve as an operational probe into the entanglement structure of the field. In particular, this setup makes it possible to investigate how spacetime geometry, topology, causal structure, and detector motion influence how correlations can be extracted (see, among others, \cite{Steeg_2009,Salton_2015, Mart_n_Mart_nez_2016,caribe_2023}).

Studies on entanglement harvesting to date have coupled detectors to \emph{free} quantum fields, where the two-point functions are known in closed form and the final density matrix of the detector is relatively simple to compute. While free theories are the most analytically tractable, they reflect a limited range of the entanglement structure that quantum fields can exhibit. Interacting theories, by contrast, exhibit a far richer range of correlation structure and exhibit phenomena absent in free theories. However, applying the entanglement harvesting protocol in interacting settings presents new technical challenges, even when the relevant correlators can in principle be computed perturbatively.

One class of interacting theories where such calculations remain tractable is conformal field theories (CFTs). A defining feature of CFTs is the spectrum of local operators which, rather than being specified as composites of some underlying quantum fields, are instead characterized intrinsically by their transformation properties under conformal symmetry. The scalar primary operators play a fundamental role: acting on a primary with translation generators produces its descendants, and together these operators form an irreducible representation of the conformal group. This structure fixes the two-point function of primary operators to be a simple power law in the spacetime separation, with the exponent determined by the operators' scaling dimensions. As a result, CFTs provide analytic control that makes them a natural setting for harvesting calculations and a useful context in which to explore how harvested entanglement depends on operator dimension in an interacting CFT.

In this work, we study entanglement harvesting in CFTs by coupling a pair of local probes to scalar primary operators. We analyse the correlations generated between the detectors as a function of their spacetime configuration and the operators' scaling dimension, using negativity to quantify entanglement and mutual information to quantify total correlations.

We are able to take this analysis further for CFTs which possess a holographic dual, \ie a description of the CFT in terms of a gravitational theory in anti-de Sitter (AdS) spacetime. In the regime of large central charge and strong coupling, this dual takes the particularly simple form of Einstein gravity with a negative cosmological constant and weakly coupled to matter fields. The duality transforms certain difficult problems in strongly interacting QFTs into tractable calculations in semiclassical gravity. Leveraging this, we show that in holographic CFTs, one may separate detector correlations into contributions arising from harvested field entanglement and those due to causal communication --- a decomposition generally unavailable in interacting QFTs.

The remainder of this paper is organised as follows. In section \ref{sec:setup_density_matrix}, we present the framework of entanglement harvesting and outline the perturbative calculation of the final state of two detectors after each couples locally to a scalar conformal primary operator. Section \ref{sec:quantifying_entanglement} quantifies the resulting correlations using negativity and mutual information in the case that the detectors remain spacelike-separated. 
In section \ref{sec:separation}, we discuss how holographic CFTs are a special class of interacting QFTs for which we can separate the contribution from harvested entanglement from correlations due to communication. We conclude in section \ref{sec:discussion} with a discussion and a summary of our results. 

Various technical details and further remarks are collected in appendices. In appendix \ref{appx:CFTs}, we provide the interested reader with an introduction to some of the salient features of CFTs and the AdS/CFT correspondence. Appendix \ref{appx:hadamard_finite_part} presents a careful calculation of the diagonal elements of the density matrix in the case where twice the conformal dimension of the primary operators is an integer. Finally, appendix \ref{appx:leadingObehaviour} provides details on the analytical evaluation of the various components of the density matrix to leading order in certain parameter regimes.

\section{Entanglement harvesting with scalar primary operators in a CFT}
\label{sec:setup_density_matrix}

In this section, we present the framework for entanglement harvesting with Unruh-DeWitt detectors coupled to scalar primary operators in a CFT. In particular, we derive the reduced density matrix for two comoving inertial detectors after their interaction with the quantum field. Since most elements of the density matrix involve numerical integration, we also provide analytic expressions for the leading-order behaviour in various parameter regimes. The details of these expansions can be found in appendix \ref{appx:leadingObehaviour}.

\subsection{Density matrix derivation}

The standard setup in entanglement harvesting involves coupling local probes to local operators in a QFT, often the field amplitude or its conjugate momentum. In this paper, we will study entanglement harvesting by coupling a pair of  two-level quantum systems (the detectors) to a scalar primary operator of a CFT in $d$-dimensional Minkowski spacetime. (A brief introduction to CFTs is given in appendix \ref{appx:CFTs}.) 

We consider the two detectors, A and B, to be comoving with the inertial coordinate frame $(t,\bm x)$. In the interaction picture, the interaction Hamiltonian is given by
\begin{equation}
\label{eq:interaction_Hamiltonian}
    \hat{H}_\textrm{int}(t)\ =\ \sum_{i = \A, \B} \int_{\Sigma_t} \dd^{d-1} \bm{x} \, \lambda_i \, \chi_i(t) \, F_i(\bm{x}) \left[ \hat{\mu}_i(t) \otimes \hat{\mathcal{O}}_{\Delta}(t, \bm{x}) \right]\,,
\end{equation}
where $\Sigma_t$ denotes a constant time slice, $\lambda_i$ is the coupling strength (with dimensions of $[\text{energy}]^{1 - \Delta}$), 
$\chi_i(t)$ is a switching function controlling the time dependence of the interaction, and $F_i(\bm{x})$ is a spatial smearing function.\footnote{These local probes are often called generically Unruh-DeWitt detectors. For a general covariant description of how local detectors couple to field observables, see, e.g., \cite{Mart_n_Mart_nez_2018,Mart_n_Mart_nez_2020,Mart_n_Mart_nez_2021}} The operator $\hat{\mu}_i(t)$ is the monopole moment of detector $i$ with energy gap $\Omega$ (which we will take to be equal for all detectors), which evolves as 
\begin{equation}
\hat{\mu}_i(t)\ =\ \hat{\sigma}^+_i e^{\ii \Omega t} + \hat{\sigma}^-_i e^{-\ii \Omega t}\,,
\label{eq:monopole}
\end{equation}
where $\hat{\sigma}^+_i = \ketbra{e}{g}_i$ and $\hat{\sigma}^-_i = \ketbra{g}{e}_i$ are the raising and lowering operators for detector $i$ (here we denote the ground and excited states respectively by $\ket{g}_i$ and $\ket{e}_i$). The operator $\hat{\mathcal{O}}_{\Delta}(\mathsf{x})$ is a local scalar conformal primary in the CFT with scaling dimension $\Delta$. While in principle $\Delta \in \mathbb R^+$, unitarity requires
\begin{equation} \label{eq:unibomb0}
    \Delta\ge \tfrac12(d-2).
\end{equation}

At $t = -\infty$, we take the CFT to be in the Minkowski vacuum state and the detectors to be in their ground state:
\begin{equation}
    \hat{\rho}_0\ =\ \hat{\rho}_{0,\textsc{a}} \otimes \hat{\rho}_{0,\textsc{b}}\otimes \hat{\rho}_\textsc{cft}\ =\ |g\rangle \langle g |_\textsc{a} \otimes |g\rangle \langle g |_\textsc{b} \otimes |0\rangle \langle 0|_\textsc{cft}.
    \label{eq:initialstate}
\end{equation}
Through the interaction, the detectors and field evolve into an entangled state $\hat{\rho} = \hat{U}\hat{\rho}_0 \hat{U}^\dagger$, where $\hat{U}$ is the time evolution operator generated by the interaction Hamiltonian: 
\begin{equation}
    \hat{U}\ =\ \mathcal{T}\exp\left(-\mathrm{i} \int_{-\infty}^\infty \, \mathrm{d}t \, \hat{H}_\mathrm{int}(t)\right),
\end{equation}
where $\mathcal{T}$ is the time-ordering operator. By tracing over the field degrees of freedom, we obtain the final reduced density matrix of the two detectors:
\begin{equation}
    \hat{\rho}_\textsc{ab}\ =\ \operatorname{Tr}_\textsc{cft}[\hat{U}\hat{\rho}_0 \hat{U}^\dagger].
\end{equation}
We will evaluate $\hat{\rho}_\textsc{ab}$ perturbatively by way of a Dyson expansion of the time evolution operator:
\begin{equation}
    \hat{U}\ =\ \mathds{1} + \hat{U}^{(1)} + \hat{U}^{(2)} + \mathcal{O}(\lambda^3),
    \label{eq:Dyson}
\end{equation}
where
\begin{align}
    \hat{U}^{(1)}\ &=\ - \mathrm{i} \int_{-\infty}^{\infty} \mathrm{d}t \, \hat{H}_\mathrm{int}(t),\\
    \hat{U}^{(2)}\ &=\ - \int_{-\infty}^{\infty} \mathrm{d}t \int_{-\infty}^{t} \! \mathrm{d}t' \, \hat{H}_\mathrm{int}(t) \hat{H}_\mathrm{int}(t').
\end{align}
The final state of the detector-field system can then be written as 
\begin{subequations}
\label{eq:DysonExpansion}
\begin{align}
\hat{\rho}\ &=\ \hat{U}\,\hat{\rho}_0\,\hat{U}^\dagger \label{eq:DysonExpansion:a}\\
\ & =\ \hat{\rho}_0
  + \hat{U}^{(1)}\hat{\rho}_0 + \hat{\rho}_0 \hat{U}^{(1)\dagger}
  + \hat{U}^{(1)}\hat{\rho}_0\hat{U}^{(1)\dagger}
  + \hat{U}^{(2)}\hat{\rho}_0 + \hat{\rho}_0 \hat{U}^{(2)\dagger}
  + \mathcal{O}(\lambda^{3}). \label{eq:DysonExpansion:b}
\end{align}
\end{subequations}
Tracing out the field degrees of freedom, we obtain
\begin{equation}
    \hat{\rho}_{\A\B} = \hat{\rho}_{0,\A}\otimes \hat{\rho}_{0,\B}  + \hat{\rho}^{(2)}_{\A\B}+\mathcal{O}(\lambda^3).
\end{equation}
In the above, we have already used $\hat{\rho}^{(1)}_{\A\B} = 0$. This term vanishes because it is proportional to the one-point function and, in the case of CFTs, translation invariance forces this correlator to be constant, while scale invariance then requires that constant to vanish for any non-trivial operator~\cite{DiFrancesco:1997nk}. Next, the second-order contribution is
\begin{align}
    \hat{\rho}^{(2)}_{\A\B}\ &=\  
    \int_{-\infty}^{\infty} \dd t \int_{-\infty}^{\infty} \dd t' \int_{\Sigma_t}\!\! \dd^{d-1} \bm{x} \int_{\Sigma_{t'}}\!\! \dd^{d-1} \bm{x}'
    \sum_{i,j=\A,\B}
    \lambda_i \,  \lambda_j \, \chi_i (t) \, \chi_j (t') \, F_i(\bm{x}) F_j(\bm{x}') \nonumber\\
    &\hspace{5cm} \times \hat{\mu}_i(t) \left( \hat{\rho}_{0,\A}\otimes \hat{\rho}_{0,\B} \right) \hat{\mu}_j(t')\,\langle 0 | \hat{\mathcal{O}}_{\Delta}(t, \bm{x}) \hat{\mathcal{O}}_{\Delta}(t', \bm{x}')| 0 \rangle \nonumber \\
    & \qquad - \Bigg[ \int_{-\infty}^{\infty} \dd t \int_{-\infty}^{t} \dd t' \int_{\Sigma_t}\!\! \dd^{d-1} \bm{x} \int_{\Sigma_{t'}}\!\!\dd^{d-1} \bm{x}' \sum_{i,j=\A,\B}
    \lambda_i\, \lambda_j\, \chi_i (t)\, \chi_j (t') \,F_i(\bm{x}) F_j(\bm{x}') \nonumber\\
    &\hspace{3cm} \times  \hat{\mu}_i(t)\hat{\mu}_j(t') \hat{\rho}_{0,\A}\otimes \hat{\rho}_{0,\B}\, \langle 0 |  \hat{\mathcal{O}}_{\Delta}(t, \bm{x}) \hat{\mathcal{O}}_{\Delta}(t', \bm{x}') | 0 \rangle + \,\text{h.c.} \Bigg].
\end{align}
Denoting the vacuum two-point function $\langle 0 |  \hat{\mathcal{O}}_{\Delta}(t, \bm{x}) \hat{\mathcal{O}}_{\Delta}(t', \bm{x}') | 0 \rangle$ as $\langle  \hat{\mathcal{O}}_{\Delta}(\textsf{x}) \hat{\mathcal{O}}_{\Delta}(\textsf{x}') \rangle$, we can write this in matrix form (in the basis $\{\ket{gg}, \ket{ge}, \ket{eg}, \ket{ee}\}$) as
\begin{equation}
\hat{\rho}_{\A\B}\ =\ \left(
    \begin{matrix}
        1 - \mathcal{L}_{\A\A} - \mathcal{L}_{\B\B} & 0 & 0 & \mathcal{M}^*\\
        0 & \mathcal{L}_{\A\A} & \mathcal{L}_{\A\B} & 0 \\
        0 & \mathcal{L}^*_{\A\B} & \mathcal{L}_{\B\B} & 0 \\
        \mathcal{M} & 0 & 0 & 0
    \end{matrix} \right) + \mathcal{O}(\lambda^3),
    \label{eq:AB_density_matrix}
\end{equation}
where
\begin{multline}
    \label{eq:Lij_def}
    \mathcal{L}_{ij}\ =\ \lambda_i \lambda_j \int_{-\infty}^{\infty} \dd t \int_{-\infty}^{\infty} \dd t'\, \int_{\Sigma_t} \dd^{d-1} \bm{x} \int_{\Sigma_{t'}} \dd^{d-1} \bm{x}' \chi_i(t)\,\chi_j(t')\, e^{-\ii\Omega (t-t')}\, \\
    \times F_i\bigl(\bm{x}\bigr)\, F_j^*\bigl(\bm{x}'\bigr)\, \langle  \hat{\mathcal{O}}_\Delta (\textsf{x}) \hat{\mathcal{O}}_\Delta(\textsf{x}') \rangle,
\end{multline}
\begin{multline}
\mathcal{M}\ =\ -\lambda_\A\lambda_\B \int_{-\infty}^{\infty} \dd t \int_{-\infty}^{\infty} \dd t'\, \int_{\Sigma_t} \dd^{d-1} \bm{x} \int_{\Sigma_{t'}} \dd^{d-1} \bm{x}'\, \chi_\A(t)\,\chi_\B(t')\, e^{\ii\Omega (t+t')}\\
\times F_\A(\bm{x})\,F_\B(\bm{x}') \,
G_\textsc{f} (\textsf{x} ,\textsf{x}'),
\label{eq:M_def}
\end{multline}
and where $G_\textsc{f}$ is the time-ordered two-point function
\begin{equation}
    G_\textsc{f} (\textsf{x},\textsf{x}')\ =\ \theta(t - t') \langle \hat{\mathcal{O}}_{\Delta}(\textsf{x}) \hat{\mathcal{O}}_{\Delta}(\textsf{x}') \rangle + \theta(t' - t) \langle \hat{\mathcal{O}}_\Delta(\textsf{x}') \hat{\mathcal{O}}_\Delta(\textsf{x}) \rangle.
    \label{eq:Feyn}
\end{equation}
Here, $\theta(x)$ represents the Heaviside function. The diagonal terms $\mathcal{L}_{\A\A}$ and $\mathcal{L}_{\B\B}$ are the excitation probabilities of the individual detectors. $\mathcal{L}_\textsc{ab}$ and $\mathcal{M}$ are correlation terms; $\mathcal{L}_\textsc{ab}$ is used in the calculation of mutual information between the detectors, and $\mathcal{M}$ is involved in the calculation of negativity between the detectors. 

The two-point function of $\hat{\mathcal{O}}_{\Delta}$ can be written as \cite{simmonsduffin2019TASIlorentziancft}
\begin{equation}
\label{eq:CFT_2PF}
\langle \hat{\mathcal{O}}_{\Delta}(t, \bm{x}) \hat{\mathcal{O}}_{\Delta}(t', \bm{x}') \rangle \ =\ \lim_{\epsilon \to 0^+}
\frac{1}{[-(t-t' - \ii \epsilon)^2 + (\bm{x}-\bm{x}')^2]^\Delta},
\end{equation}
where the limit is to be understood in a distributional sense. Its structure is completely dictated by conformal invariance, up to normalisation. Since the two-point function is zero for different scalar primary operators, we have assumed the detectors couple to the same primary operator. Note that the case $\Delta = \tfrac12(d-2)$ reproduces the Wightman function of a free massless scalar field in $d$-dimensional Minkowski space, up to normalisation.\footnote{That is, the two-point function for the amplitude of a free massless scalar field in $d$-dimensional Minkowski space is equivalent to \eqref{eq:CFT_2PF} normalized by $\Gamma\!\left(\frac{d-2}{2}\right)/(4\pi^{d/2})$. \label{footnote:normalisation}} 

\begin{figure}[htbp]
    \centering
    \includegraphics[width=1\textwidth]{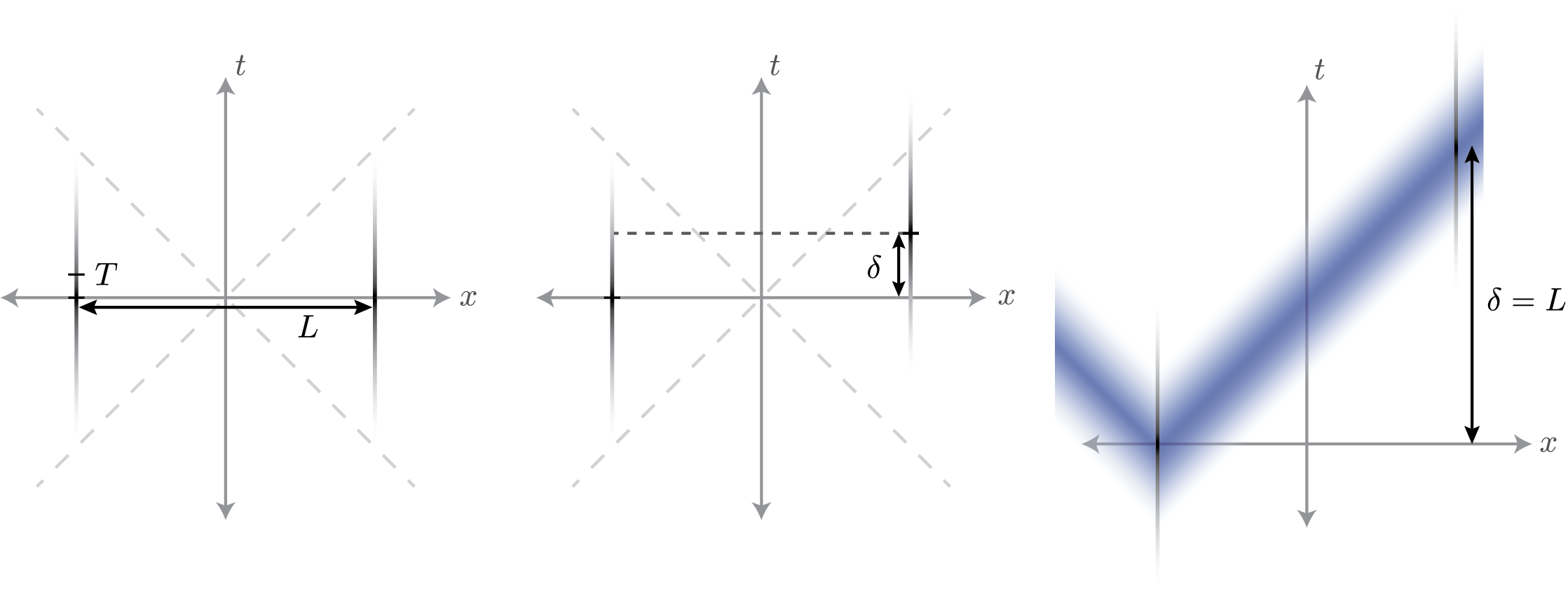}
    \caption{We consider two pointlike detectors with Gaussian switching functions with width $T$. The detectors are spatially separated by a distance $L$, and the temporal centres of the Gaussian switching functions are separated by $\delta$. At some separation $L\sim 10T$, the communication contribution from the Gaussian tails is negligible and the detectors can be taken to be effectively spacelike-separated. When $\delta = L$, the centre of one detector lies along the lightcone of the centre of the other.}
    \label{fig:setup_diagram}
\end{figure}

In this work, we consider two identical pointlike detectors, distinguished only by their locations in spacetime. A diagram of the setup is shown in figure \ref{fig:setup_diagram}. The detectors' smearing functions are taken to be spatial delta functions separated by a distance ${L}$ along the $x$-axis:
\begin{equation}
\label{eq:smear}
    F_\textsc{a}(\bm x) = \delta^{d-1}(\bm x), \qquad \qquad F_\textsc{b}(\bm x) = \delta^{d-1}(\bm{x} - \bm{L}),
\end{equation}
where $\bm{L} = (L,0, \ldots,0)$. The temporal switching functions are Gaussians of width $T$, with centres separated by $\delta$:
\begin{align}
    \chi_\textsc{a}(t) &= e^{-t^2 / T^2}, \qquad \qquad
    \chi_\textsc{b}(t) = e^{-(t - \delta)^2 / T^2}.
    \label{eq:switch0}
\end{align}
With identical detectors, we also have $\lambda_\textsc{a} = \lambda_\textsc{b} \equiv \lambda$. Generally in the following, we will use the \emph{dimensionless coupling}
\begin{equation}
    \bar{\lambda}\ \equiv \ \frac{\lambda}{T^{\Delta -1}}.
\end{equation}

In this paper, we will fix $T \Omega$ and look at how harvested correlations depend on $\Delta$, $L/T$, and $\delta/T$. To determine an appropriate value of $T \Omega$, we use guidance from entanglement-harvesting studies in free scalar field theories. Namely, with a smooth enough switching function, for a fixed $L/T$ (taking $\delta/T = 0$), it is always possible to find a value of $\Omega$ large enough so that entanglement can be harvested~\cite{Pozas_Kerstjens_2015,Maeso_2022}. This implies that there exists an optimal value of $T \Omega$ that maximises the amount of harvested entanglement. The optimisation is somewhat involved (for details in $3+1$ dimensions, see~\cite{Maeso_2022}), but roughly speaking, taking $T\Omega \lesssim L/T$ ensures that we are close to this optimal value. Although the analysis in~\cite{Maeso_2022} is of free scalar fields and thus does not obviously apply to the present work, we will use the regime $T\Omega \lesssim L/T$ as a reference point for our study. In our numerical calculations, we look at the range $0<L/T<20$, so throughout the paper we take $T\Omega  = 10$.

To compute the density matrix elements, we begin by substituting the smearing functions \eqref{eq:smear} into the components of the density matrix given in eqs.~\eqref{eq:Lij_def} and \eqref{eq:M_def}, obtaining
\begin{align}
    \mathcal{L}_{\A\A} =\mathcal{L}_{\B\B} &= \lambda^2 \int_{-\infty}^{\infty} \dd t \int_{-\infty}^{\infty} \dd t'\,  \chi_{\A}(t)\,\chi_{\A}(t')\, e^{-\ii\Omega (t-t')}\, 
    \langle  \hat{\mathcal{O}}_\Delta (t,\bm 0) \hat{\mathcal{O}}_\Delta(t',\bm 0) \rangle ,
    \label{eq:diag}\\
    \mathcal{L}_{\A\B}  &= \lambda^2 \int_{-\infty}^{\infty} \dd t \int_{-\infty}^{\infty} \dd t'\,  \chi_{\A}(t)\,\chi_{\B}(t')\, e^{-\ii\Omega (t-t')}\, 
    \langle  \hat{\mathcal{O}}_\Delta (t,\bm 0) \hat{\mathcal{O}}_\Delta(t',\bm L) \rangle ,
    \label{eq:offdiagL}\\
    \mathcal{M}&= -\lambda^2 \int_{-\infty}^{\infty} \dd t \int_{-\infty}^{\infty} \dd t'\,  \chi_\A(t)\,\chi_\B(t')\, e^{\ii\Omega (t+t')}\, G_\textsc{f} (t,\bm 0 ;t',\bm L). \label{eq:offdiagM}
\end{align}
Due to time-translation invariance of the two-point function, $\mathcal{L}_{\A\A} =\mathcal{L}_{\B\B}$. We observe that these expressions are completely independent of the spacetime dimension $d$, which follows from the fact that the detectors are pointlike and the CFT is scale-invariant. Hence the spacetime dimension enters only through the lower bound on $\Delta$ enforced by unitarity \eqref{eq:unibomb0}.

Using the two-point function \eqref{eq:CFT_2PF} and the switching functions \eqref{eq:switch0}, the self-\\excitation term \eqref{eq:diag} becomes 
\begin{align}
    \LL_{\A\A} 
    &= \bar{\lambda}^2 T^{2\Delta - 2} \int_{-\infty}^{\infty} \dd t \int_{-\infty}^{\infty} \dd t'\,  e^{-\frac{t^2 + t'^2}{T^2}-\ii\Omega (t-t')}\, \langle  \hat{\mathcal{O}}_{\Delta}(t, 0) \hat{\mathcal{O}}_{\Delta}(t', 0) \rangle
    \nonumber\\
    &= \tfrac12\, \bar{\lambda}^2 T^{2\Delta - 2} \int_{-\infty}^{\infty} \dd u \, e^{-\frac{u^2}{2T^2}} \int_{-\infty}^{\infty} \dd v \, e^{-\ii\Omega v} e^{-\frac{v^2}{2T^2}} \bigg( \lim_{\epsilon \to 0^+} \frac{1}{[-(v - \ii \epsilon)^2]^\Delta} \bigg)
    \nonumber\\
    &= \bar{\lambda}^2 T^{2\Delta - 1} \, \sqrt{\frac{\pi}{2}} \int_{-\infty}^{\infty} \dd v \, e^{-\ii\Omega v} e^{-\frac{v^2}{2T^2}}  \frac{1}{(|v|^2 e^{\ii \, \pi\,\text{sgn} v})^{\Delta}},
    \label{eq:diag22}
\end{align}
where in the second line we have introduced the coordinates $u = t + t'$ and $v = t - t'$, and all the integrals must be understood in the distributional sense. Treating these integrals properly can be somewhat nuanced, above all for integer and half-integer values of $\Delta$. The distributional treatment of these integrals is summarized in appendix~\ref{appx:hadamard_finite_part}. Despite these subtleties, we can find a closed-form expression that is valid for any $\Delta \in \mathbb{R}_+$:
\begin{align}
    \frac{\LL_{\A\A}}{\bar{\lambda}^2} &= \frac{\pi^{3/2}}{2^{\Delta}} \left[\,\frac{
    {}_1F_1\!\left(\tfrac{1}{2} - \Delta, \tfrac{1}{2}, -\tfrac{1}{2} T^2 \Omega^2\right)
    }{\Gamma\!\left(\tfrac{1}{2}+\Delta\right)}
     -  \frac{\sqrt{2} \, T \Omega}{\Gamma(\Delta)}\,{}_1F_1\!\left(1 - \Delta, \tfrac{3}{2}, -\tfrac{1}{2} T^2 \Omega^2\right)\right]\,.\label{eq:Lii}
\end{align}
Note that $\LL_{\A\A}$ does not depend on $L$ or $\delta$, as it is a property of the detectors' local interactions. 

\begin{figure}
    \centering
    \includegraphics{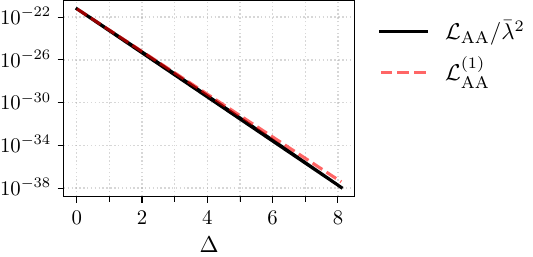}
    \caption{Here we plot the self-excitation probability (local noise term) $\mathcal{L}_{\A\A}$ given in eq.~\eqref{eq:Lii} as a function of the scaling dimension $\Delta$, using $T\Omega = 10$. The red dashed line shows the leading-order term in $1/(T \Omega)$, given in eq.~\eqref{eq:largeOmega0}. Since the subleading terms in this expansion scale with $\Delta$, we expect larger deviations of the first-order fit for larger $\Delta$.}
    \label{fig:Lii_Delta_analyticfit}
\end{figure}

While eq.~\eqref{eq:Lii} is an exact result for $\mathcal L_{\A\A}$ (to $\mathcal O(\bar{\lambda}^2)$), it is difficult to interpret in this form. The dominant exponential behaviour arises from the Gaussian switching functions and is controlled by the dimensionless combination $T\Omega$. In the regime $T\Omega \gg 1$, this structure leads to a controlled expansion of $\mathcal L_{\A\A}$ in inverse powers of $T\Omega$. To first order, $\mathcal L_{\A\A}/\bar{\lambda}^2$ takes the form (see the discussion leading to eq.~\eqref{eq:largeOmega}):
\begin{equation}
    \frac{\LL_{\A\A}}{\bar{\lambda}^2} \ =\ \LL_{\A\A}^{(1)} \bigg[\,1 + \mathcal O\bigg(\frac{1}{T^2\Omega^2}\bigg)\bigg],\qquad \LL_{\A\A}^{(1)} = \frac{\pi}{(T\Omega)^{2\Delta}}\, e^{-\tfrac{1}{2} T^{2}\Omega^{2}}.
    \label{eq:largeOmega0}
\end{equation}
In figure \ref{fig:Lii_Delta_analyticfit}, we plot $\mathcal L_{\A\A}$ as given in eq.~\eqref{eq:Lii} as a function of $\Delta$, along with the leading-order term given in eq.~\eqref{eq:largeOmega0}. We see that the leading-order expression provides a good approximation over a wide range of the scaling dimension. However, the fit begins to deviate for larger values, indicating that the subleading terms grow with $\Delta$, as shown explicitly in eq.~\eqref{eq:largeOmega}.
\vspace{5mm}

\begin{figure}
    \centering
    \includegraphics{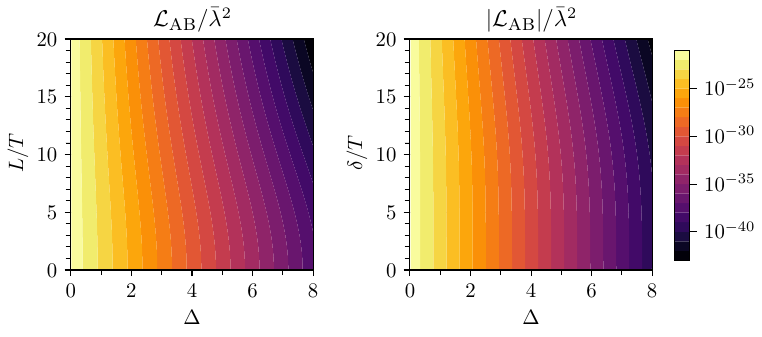}
    \caption{The above plots show $|\mathcal L_\textsc{ab}| / \bar{\lambda}^2$, used in the calculation of the mutual information, obtained via numerical analysis of eq.~\eqref{eq:Lij}. On the left side, $L/T$ and $\Delta$ are varied, and we have set $\delta/T = 0$. In this case, $\LL_{\A\B}$ is real. On the right side, $\delta/T$ and $\Delta$ are varied, and we have set $L/T = 10$. For $\delta/T \neq 0$, $\LL_{\A\B}$ is generally complex. For both plots, we use $\Omega T = 10$. The colour band widths are approximately equal for a fixed $L/T$ or $\delta/T$, indicating that $|\mathcal{L}_\textsc{ab}|$ decays roughly exponentially with the operator dimension $\Delta$ for our choices of parameters. In figure \ref{fig:Lab_L_and_delta_slices} of appendix \ref{appx:leadingObehaviour} we show cross-sections of these plots to compare the numerical results to the asymptotic behaviour given in eq.~\eqref{eq:Lij_leading_order}.}
    \label{fig:Lab_L_and_delta}
\end{figure}

We now consider the off-diagonal term $\mathcal{L}_\textsc{ab}$. Using the same change of variables as above, we can express eq.~\eqref{eq:offdiagL} as
\begin{align}
    \frac{\mathcal{L}_\textsc{ab}}{\bar{\lambda}^2}\ &=\  \lim_{\epsilon\to 0^+} \, \tfrac{1}{2} \, T^{2\Delta - 2} \, \int_{-\infty}^{\infty} \dd u \,e^{-\frac{(u-\delta)^2}{2T^2}}  \int_{-\infty}^{\infty} \dd v \, e^{-\frac{(v+\delta)^2}{2T^2} - \ii\Omega v}  \frac{1}{(-(v - \ii \epsilon)^2 + L^2)^\Delta}
    \nonumber\\
     &=\ \lim_{\epsilon\to 0^+} \, T^{2\Delta - 1} \sqrt{\frac{\pi}{2}} \int_{-\infty}^{\infty} \dd v \, e^{-\frac{(v+\delta)^2}{2T^2} - \ii\Omega v}  \frac{1}{(L^2 - v^2 + \ii \epsilon  v)^\Delta}.
    \label{eq:Lij}
\end{align}
We note that $\mathcal L_\textsc{ab}$ is real for $\delta = 0$ but otherwise generally complex. We cannot evaluate \eqref{eq:Lij} in closed form --- instead, we evaluate it numerically, the results of which are shown in figure \ref{fig:Lab_L_and_delta} as a function of $\Delta$, $L/T$, and $\delta/T$. The results show roughly exponential decay with $\Delta$ and comparatively very slow decay of $|\mathcal L_\textsc{ab}|$ as a function of $L/T$ and $\delta/T$.

While we are limited to numerical analysis to calculate $\mathcal L_\textsc{ab}$ in generality, we can again gain an analytic understanding by analysing the leading-order behaviour in certain regimes. The dominant exponential behaviour depends on
\begin{equation}
    D\ \equiv\ \frac{L^{2} - \delta^{2}}{T^2} + T^{2}\Omega^2 - 2\ii\Omega\delta.
    \label{eq:DDD}
\end{equation}
In the regime $|D|\gg 1$, the expression admits a controlled expansion in inverse powers of $D$. The details are given in the discussion leading to eq.~\eqref{eq:LAB_asymptotic}, where we find:
\begin{equation}
    \frac{\mathcal{L}_{\textsc{ab}}}{\bar{\lambda}^2}\ =\ \mathcal{L}^{(1)}_{\textsc{ab}}\left[1+\mathcal{O}\!\left(\tfrac{1}{D}\right)\right], \qquad {\rm with} \ \ \ \mathcal{L}^{(1)}_{\textsc{ab}}
    \ =\ \frac{\pi}{D^\Delta}\, e^{-\tfrac12 T^2 \Omega^2 + \ii\Omega\delta},
    \label{eq:Lij_leading_order}
\end{equation}
which holds for
\begin{equation}
 |D|^2\gg 4\left(\tfrac{\delta^2}{T^2} +  T^{2}\Omega^2\right)\,, \quad  |D|^2\gg 1.
\end{equation}
Both of these inequalities hold automatically here since we fix $\Omega T=10$ in all of our computations. In figure \ref{fig:Lab_L_and_delta_slices} of appendix \ref{appx:leadingObehaviour}, we compare the numerical results with $\mathcal{L}^{(1)}_{\textsc{ab}}$, showing that this approximation aligns very well for a wide range of $\Delta$, $L/T$, and $\delta/T$. Again, the approximations begin to deviate slightly for large values of $\Delta$, which is expected since the subleading corrections grow with $\Delta$, as shown in eq.~\eqref{eq:LAB_asymptotic}. 

\smallskip

\begin{figure}
    \centering
    \includegraphics{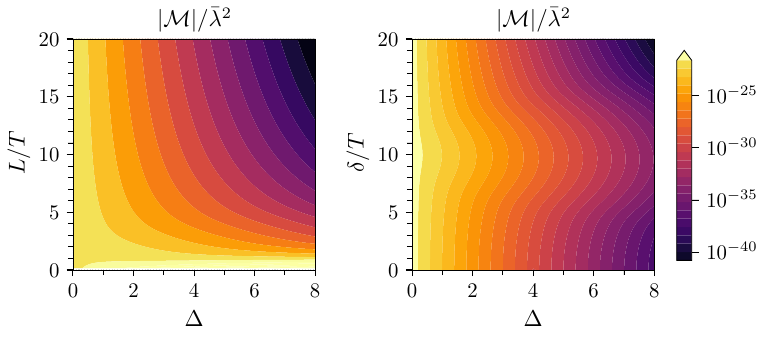}
    \caption{The above plots show $|\mathcal M| / \bar{\lambda}^2$, used in the calculation of negativity, obtained via numerical analysis of eq.~\eqref{eq:M_numerical}. On the left side, $L/T$ and $\Delta$ are varied, and we have set $\delta/T = 0$. On the right side, $\delta/T$ and $\Delta$ are varied, and we have set $L/T = 10$. For both plots, we use $\Omega T = 10$. In the right plot, for a fixed $\delta/T$, the colour band widths are approximately equal, indicating that $|\mathcal M|$ decays roughly exponentially with $\Delta$ in this regime. The same can be said in the left plot for a fixed $L/T \gtrsim 1$. For $L/T \lesssim 1$, we see $|\mathcal M|$ increases with $\Delta$. In figure \ref{fig:M_L_and_delta_slices} of appendix \ref{appx:leadingObehaviour} we show cross-sections of these plots to compare our numerical results to the asymptotic behaviour given in eq.~\eqref{eq:M_leading_order}.}
    \label{fig:M_L_and_delta}
\end{figure}

Finally, we look at $\mathcal{M}$, given by eq.~\eqref{eq:offdiagM}:
\begin{align}
    \frac{\mathcal{M}}{\bar{\lambda}^2}\ &= \ \lim_{\epsilon \to 0^+}\, -\tfrac{1}{2} T^{2\Delta - 2} \int_{-\infty}^{\infty} \dd u \, \,e^{-\frac{(u-\delta)^2}{2T^2} + \ii\Omega u}  \int_{-\infty}^{\infty} \dd v \, e^{-\frac{(v+\delta)^2}{2T^2}}  \frac{1}{(L^2 - v^2 + \ii\epsilon |v|)^\Delta}
    \nonumber\\
    &= \ \lim_{\epsilon \to 0^+}\, - T^{2\Delta - 1} \sqrt{\frac{\pi}2}\, e^{-\frac12\,T^2\Omega^2 + \ii \Omega\delta} \int_{-\infty}^{\infty} \dd v \, e^{-\frac{(v+\delta)^2}{2T^2}}  \frac{1}{(L^2 - v^2 + \ii\epsilon |v|)^\Delta}.
    \label{eq:M_numerical}
\end{align}
Again, we cannot evaluate $\mathcal{M}$ in closed form, so we evaluate it numerically. The results are shown in figure \ref{fig:M_L_and_delta} as a function of $\Delta$, $L/T$, and $\delta/T$. In the right plot, where $L=10 T$, we see that for fixed $\Delta$, $|\M|$ has a clear maximum around $\delta=10 T$, i.e.,~when the centres of the detectors are lightlike-separated. This is expected: as we will see in section \ref{sec:quantifying_entanglement}, $\mathcal{M}$ dictates the amount of entanglement between the probes. The case of $\Delta = 1$ corresponds to two detectors coupling to the amplitude of a free massless scalar field in $3+1$ dimensions, in which entanglement harvesting is known to be maximised on the lightcone~\cite{Pozas_Kerstjens_2015}. As noted after eq.~\eqref{eq:CFT_2PF}, integer and half-integer $\Delta$ correspond to amplitude coupling to a massless scalar field in $d=2+2\Delta$ dimensions, so we may expect this feature to apply for all values of $\Delta$. 

The leading-order behaviour of $\mathcal M$ can be evaluated in a similar manner to the previous density matrix elements (see the discussion leading to eq.~\eqref{eq:holler3} for details) to give
\begin{equation}
    \frac{\mathcal{M}_{\tilde{D}{\scriptscriptstyle\gg} 1}}{\bar{\lambda}^2} \ =\ \mathcal{M}^{(1)}_{\tilde{D}{\scriptscriptstyle\gg} 1}\left[1+\mathcal{O}\!\left(\tfrac{1}{\tilde{D}}\right)\right], \quad \mathcal{M}^{(1)}_{\tilde{D}{\scriptscriptstyle\gg} 1} = -\frac{\pi }{{\tilde{D}}^\Delta}\, e^{-\tfrac12 T^2 \Omega^2 + \ii\Omega\delta},
    \label{eq:M_leading_order}
\end{equation}
where
\begin{equation}
    \tilde{D} = \frac{L^{2} - \delta^{2}}{T^2}, \qquad |\tilde{D}|\gg1, \qquad |\tilde{D}|\gg 2|\delta|/T.
\end{equation}
Intuitively, the first condition is satisfied when the two detectors are far from each other's lightcone, and the second condition demands that the duration of their interaction has to be small compared to their spacetime separation. In the case that the detectors are near light contact, specifically in the regime $\tilde C \equiv 2 |\delta|/T \gg1$ with $|\tilde D|\ll \tilde C$, we may obtain a different asymptotic expression. This analysis is longer and is relegated to appendix \ref{appx:leadingObehaviour} (see the discussion preceding eq.~\eqref{eq:M_near_lightcone} for details). In figure \ref{fig:M_L_and_delta_slices} of appendix \ref{appx:leadingObehaviour}, we show that the expressions in eqs.~\eqref{eq:M_leading_order} and \eqref{eq:M_near_lightcone} closely approximate the numerical results in their respective regimes.

In the next section, we use the density matrix elements calculated in this section to evaluate correlations between the detectors.

\section{Correlations acquired between detectors}
\label{sec:quantifying_entanglement}

Having computed $\hat{\rho}_{\A\B}$ to order $\bar{\lambda}^2$, we may now quantify correlations between the detectors in their final state. We will analyse both the entanglement as measured by the negativity, as well as the total correlations (quantum and classical) as measured by the mutual information.

\subsection{Entanglement}
\label{subsec:negativity}

To quantify the entanglement between the two detectors, we use negativity. It is defined as
\begin{equation}
    \label{eq:negativity_def}
    \mathcal{N}(\hat{\rho})\ =\ \frac{\|\hat{\rho}^{T_2}\|_1 - 1}{2},
\end{equation}
where \(\hat{\rho}^{T_2}\) denotes the partial transpose of $\hat{\rho}$ with respect to a partition of the total Hilbert space, and \(\|\cdot\|_1\) is the trace norm. This measure is based on the Peres-Horodecki criterion, also called the positive partial transpose (PPT) criterion, which states that if a state $\hat{\rho}$ is separable, its partial transpose $\hat{\rho}^{T_2}$ must have no negative eigenvalues. In the specific case of $2 \times 2$ systems (such as our two detectors), this criterion is both necessary and sufficient: the presence of any negative eigenvalue in $\hat{\rho}^{T_2}$ is a definitive signature of entanglement, and, conversely, a state with a positive partial transpose is guaranteed to be separable. Up to order $\bar{\lambda}{}^2$, the negativity between detectors A and B can be expressed in terms of the excitation probability $\mathcal{L}_\textsc{aa}$ and the coherence $\mathcal{M}$: using the density matrix \eqref{eq:AB_density_matrix} in eq.~\eqref{eq:negativity_def}, for identical detectors we get
\begin{equation}
    \mathcal{N}(\hat{\rho}_{\A\B})\ =\ \max \left( 0, |\mathcal{M}| - \mathcal{L}_{\A\A} \right) + \mathcal{O}(\bar{\lambda}{}^4).
    \label{eq:negativity2}
\end{equation}
Thus, to have entanglement at leading order, the correlations captured in the term $|\mathcal{M}|$ must dominate the detector excitation probability $\mathcal{L}_{ii}$, which acts as a source of local noise. In what follows, we will write $\mathcal{N}$ in place of $\mathcal{N}(\hat{\rho}_{\A\B})$, and $g_+$ in place of $\max[0,g]$.

\begin{figure}
    \centering
    \includegraphics{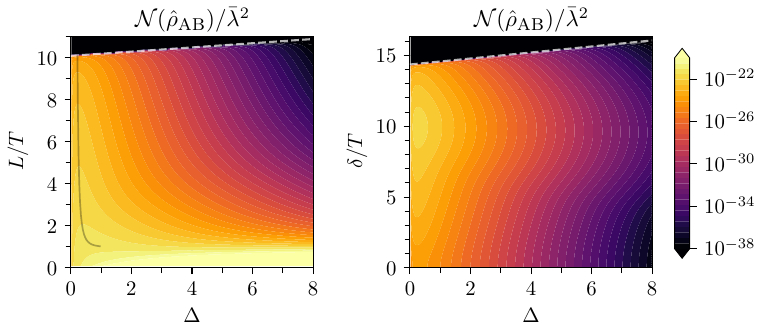}
    \caption{These plots show the negativity between the two detectors as defined in eq.~\eqref{eq:negativity2}. On the left side, $L/T$ and $\Delta$ are varied, and we have set $\delta/T = 0$. The solid black line marks the maximum $\mathcal N$ for a given $L/T$ as given by the approximation in eq.~\eqref{eq:max_Delta}. In the right panel, we set $L/T = 10$ and vary $\delta/T$. Here we see the negativity clearly maximised when the detectors are in light contact. In both plots, we have set $\Omega T = 10$. The gray dashed lines in each figure show an approximation of the boundary where the negativity vanishes, obtained in closed form in eqs.~\eqref{eq:nomore} and \eqref{eq:nomore2} for the left and right figures respectively. In figure \ref{fig:Negativity_L_and_delta_slices} of appendix \ref{appx:leadingObehaviour} we show cross-sections of these plots to compare our numerical results to the asymptotic behaviour given in eq.~\eqref{eq:hullabaloo}, as well as higher-order and near-lightcone approximations.}
    \label{fig:negativity}
\end{figure}

In the left panel of figure \ref{fig:negativity}, we show $\mathcal{N}$ as a function of $\Delta$ and $L/T$. For a fixed $\Delta$, as expected, we see that the negativity always decreases with $L/T$ and does so more rapidly for larger $\Delta$. For a fixed $L/T$, we can see that for most of the parameter regime, $\mathcal{N}$ decreases with $\Delta$. For $L/T \gtrsim 1$, the negativity initially rises and hits a maximum at some small $\Delta$, while for $L/T \lesssim 1$, the negativity strictly increases with $\Delta$ and grows without bound. This is expected: as the separation grows, the correlators decay as the inverse of a polynomial whose order increases with $\Delta$. Hence, as the detectors get closer and their spacetime distance falls below the duration of their interaction, the correlators grow as the inverse of a polynomial.

In the right panel, we set $L/T = 10 $ and plot the negativity as a function of $\Delta$ and $\delta/T$. Again, for a fixed $\Delta$, the negativity decreases with $\delta/T$ and does so more rapidly for larger $\Delta$. As foreshadowed by the behaviour of $\mathcal M$ with $\delta/T$ in figure \ref{fig:M_L_and_delta}, the negativity is indeed maximised when the centres of the detectors' switching functions are in light contact. However, this does not mean that the majority of this entanglement is genuinely harvested from the field.

When two detectors are in causal contact, the entanglement that they acquire can come from two very different mechanisms. The detectors can acquire entanglement by harvesting the entanglement already present in the state of the field or by exchanging quantum information through the field. In our case, since the detectors have Gaussian switching functions, technically they are always in causal contact. However, the amount of information they can exchange is strongly suppressed as the centres of the switching functions become more and more spacelike separated. This has been studied in previous literature \cite{Pozas_Kerstjens_2016,Tjoa_2021,Maeso_2022}, and it has been shown that one can consider the detectors to be effectively spacelike separated when the separation between the maxima of the interaction is at least approximately five times the switching timescale.\footnote{The exact scale at which one can consider Gaussian-smeared detectors in time or space to be effectively spacelike separated depends weakly on the energy gap $\Omega$. For more details, see~\cite{Maeso_2022}.} We do not assume this criterion a priori: in section \ref{sec:separation}, we will review a technique to separate these contributions and explore how much of this negativity can be truly attributed to pre-existing entanglement in the field~\cite{Tjoa_2021,caribe_2023,TeixidoBonfill2024}. 

We now turn to an asymptotic approximation of the negativity, obtained by subtracting $\LL_{\A\A}^{(1)}$ in eq.~\eqref{eq:largeOmega0} from $\mathcal M^{(1)}_{\tilde{D} {\scriptscriptstyle\gg} 1}$ in the far-from-lightcone regime, given in eq.~\eqref{eq:M_leading_order}. This yields
\begin{equation} \label{eq:hullabaloo}
    \frac{\N}{\bar{\lambda}^2}\ \approx\ \frac{\N^{(1)}_{\tilde{D} {\scriptscriptstyle\gg} 1}}{\bar{\lambda}^2} \ =\ \pi\,e^{-\tfrac12 T^2 \Omega^2} \left(\frac{T^{2\Delta} }{{|L^2-\delta^2|}^{\Delta}}- \frac{1}{(T\Omega)^{2\Delta}}\right)_+,
\end{equation}
which holds in the regimes that $\LL_{\A\A}^{(1)}$ and $\mathcal M^{(1)}_{\tilde{D} {\scriptscriptstyle\gg} 1}$ are valid: when $T \Omega \gg 1$, $|\tilde D| \gg 1$, and $|\tilde D| \gg 2|\delta|/T$. The subleading corrections to \eqref{eq:hullabaloo} grow with $\Delta$, as shown in eq.~\eqref{eq:N_NLO}. In figure \ref{fig:Negativity_L_and_delta_slices} of appendix \ref{appx:leadingObehaviour}, we compare our numerical results to the approximation provided by eq.~\eqref{eq:hullabaloo}, the next-to-leading-order expression in the $\tilde D \gg 1$ expansion given in eq.~\eqref{eq:N_NLO}, and the near-lightcone approximation in eq.~\eqref{eq:N_lightcone_approx}.

Despite the fact that the approximation $\N^{(1)}_{\tilde{D} {\scriptscriptstyle\gg} 1}$ in eq.~\eqref{eq:hullabaloo} was derived in an asymptotic far-from-lightcone regime, it reflects many of the features present in figure \ref{fig:negativity}. For example, when $\delta = 0$, $\N^{(1)}_{\tilde{D} {\scriptscriptstyle\gg} 1}$ has a unique maximum for $ 1 < L/T < \Omega T$ at 
\begin{equation}
\Delta\ =\ \Delta_{\max}\ =\ \frac{1}{2}\, \frac{\ln\!\big(\ln(T\Omega) /\ln(L/T)\big)}{\ln\!\big(T\Omega/(L/T)\big)},
\label{eq:max_Delta}
\end{equation} 
which is plotted by a solid black line in the left panel of figure \ref{fig:negativity}. Further, in the regime $ 1 < L/T < \Omega T$, $\N^{(1)}_{\tilde{D} {\scriptscriptstyle\gg} 1}$ goes to zero as $\Delta \to \infty$. For $0 < L/T < 1$, $\N^{(1)}_{\tilde{D} {\scriptscriptstyle\gg} 1}$ is strictly increasing and grows without bound as $\Delta \to \infty$. For $L/T = 1$, eq.~\eqref{eq:hullabaloo} asymptotes to
\begin{equation}
    \lim_{\Delta \to \infty} \mathcal{N}^{(1)}_{\tilde{D} {\scriptscriptstyle\gg} 1}/\bar{\lambda}^2 \big|_{L/T=1}\ =\ \pi e^{-\tfrac12 (T \Omega)^2}.
\end{equation}
We emphasise that the above observations take $\N^{(1)}_{\tilde{D} {\scriptscriptstyle\gg} 1}$ outside its formal regime of validity, yet do seem to closely align with the numerical results.

Further, we can see that eq.~\eqref{eq:hullabaloo} is positive  when $L/T\le T\Omega$. While the left panel of figure \ref{fig:negativity} shows that this is approximately true, we can refine this estimate by retaining the first subleading terms, which are given in eqs.~\eqref{eq:largeOmega} and \eqref{eq:holler3}. Doing so yields
\begin{equation}
    \frac{L}{T} \ \le\ T\Omega\left[1+\frac{\Delta+1}{(T\Omega)^2} +\mathcal O\!\left(\frac{1}{(T\Omega)^4}\right)\right].
    \label{eq:nomore}
\end{equation}
The line at which this inequality is saturated is marked in the left plot with a dashed line atop the numerical results, and we see that it provides a close approximation for the boundary. 

Turning to the right panel of figure \ref{fig:negativity}, even though near the lightcone we are outside the formal regime of validity of eq.~\eqref{eq:hullabaloo}, this expression is still helpful in understanding the behaviour approaching the lightcone. Notably, $\N^{(1)}_{\tilde{D} {\scriptscriptstyle\gg} 1}$ strictly decreases with $|L^2 - \delta^2|$, in line with the numerical results. Requiring eq.~\eqref{eq:hullabaloo} to be positive yields the condition $\delta^2/T^2 \le L^2/T^2 + T^2\Omega^2$. In the case considered in the plot, this yields a boundary of $\delta/T \leq \sqrt{2} \,\Omega T \approx 14.1$, in qualitative agreement with the observed behaviour. As before, this estimate can be refined by retaining the first subleading terms, which are given in eqs.~\eqref{eq:largeOmega} and \eqref{eq:holler3}. Doing so leads to
\begin{equation}
    \frac{\delta}{T}\le \alpha\left[1+(\Delta+1)\left(\frac{1}{\alpha^2}
    +\frac{1}{T^2\Omega^2}\right)+\mathcal O\left(\frac{1}{\alpha^4} , \frac{1}{(T\Omega)^4}\right)\right],
    \qquad 
    \alpha^2=\frac{L^2}{T^2}+T^2\Omega^2\,.
    \label{eq:nomore2}
\end{equation}
The line at which this inequality is saturated is marked in the right plot with a dashed line atop the numerical results. Again, this provides a close approximation for the numerical boundary.

\subsubsection*{Gap optimisation}

The dominant term in eq.~\eqref{eq:nomore} says that, for $\delta = 0$, producing entanglement between the detectors requires approximately $L/T \leq T \Omega$. At first glance, this seems to contradict the inequality $T\Omega \lesssim L/T$ mentioned in section \ref{sec:setup_density_matrix}. However, the latter was derived by tuning the gap to maximise the harvested entanglement at a fixed $L/T$ \cite{Maeso_2022}. This is an entirely different analysis: it determines, for a specified $L/T$, what value of $T\Omega$ will maximise the negativity, rather than just looking for a bound where $\mathcal N = 0$. The result for the optimal gap, that $T\Omega \lesssim L/T$, can be understood heuristically (for $\delta=0$) as tuning the detector's characteristic wavelength $\Omega^{-1}$ to match the separation $L$ between the detectors. Ref.~\cite{Maeso_2022} examined this question for a free scalar field in four dimensions by solving $\partial_\Omega \N^+=0$ for $T \Omega$. If we apply a similar analysis to $\N_{\tilde D \scriptscriptstyle \gg 1}$ (to the order written in eq.~\eqref{eq:N_NLO}), we find
\begin{equation} \label{eq:optimal_gap}
    T\Omega_{\rm opt} = \frac{L}{T} -\Delta\frac{T}{L} +\mathcal O\!\left(\frac{T^{3}}{L^{3}}\right).
\end{equation}
Following the discussion above, the analysis in 
\cite{Maeso_2022} should match $\Delta=1$, \ie $T\Omega_{\rm opt} \approx \frac{L}{T} -\frac{T}{L}$. However, their result, which corresponds to a numerical fit, was $T\Omega_{\rm opt} \approx \frac{L}{T} +a_1\frac{T}{L}$, where $a_1\simeq -1.39218$. The slight mismatch  can likely be explained by the absence of higher order terms in our expansion. 

Rearranging eq.~\eqref{eq:optimal_gap}, we find that for a fixed $T\Omega$ (as is the case throughout this paper), the separation $L/T$ at which that gap is optimal can be written as
\begin{equation}
    L/T= T\Omega_{\rm opt}\left[1+\frac{\Delta}{(T\Omega_{\rm opt})^2} + \mathcal O \left(\frac{1}{(T \Omega_\text{opt})^4}\right) \right].
\end{equation}
Note that this distance always lies in the allowed regime to get quantum entanglement given in eq.~\eqref{eq:nomore}. This is consistent with the fact that for detectors linearly coupled to the amplitude of free massless scalar fields with Gaussian switching functions, for any separation between the detectors, one can find a value of the gap for which there is always some degree of entanglement harvesting (see, e.g., ~\cite{Pozas_Kerstjens_2015,Maeso_2022}).

\subsection{Mutual information}
\label{subsec:mutual_info}

In this section, we will quantify the total correlations using mutual information. One can argue that mutual information provides a specific bound on the correlators of operators localised in two separate subregions~\cite{Wolf:2007tdq}. Further, mutual information has been used to quantify correlations between disconnected regions for holographic CFTs, \eg~\cite{Headrick:2010zt,VanRaamsdonk,Allais:2011ys,Morrison:2012iz}.  In holography, mutual information is widely used since it is easily computable using the Ryu-Takayanagi (RT) formula ~\cite{Ryu:2006bv,Rangamani:2016dms}. In contrast, negativity lacks a first-principles RT prescription and is only conjecturally tied to the entanglement wedge cross section~\cite{Kudler-Flam:2018qjo,Kusuki:2019zsp}. Mutual information captures both classical and quantum correlations and, as such, it is not a measure of entanglement for mixed states (including our pair of detectors). Nonetheless, it is a useful complementary diagnostic, allowing us to compare total correlations to the purely quantum correlations captured by negativity. Harvesting of mutual information has been studied before for the case of free scalar fields in, \eg \cite{Pozas_Kerstjens_2015}.

The mutual information between subsystems A and B is given by
\begin{equation}
\label{eq:mutual_info_def}
    I(\hat{\rho}_\textsc{ab})\ =\ S(\hat{\rho}_\textsc{a}) + S(\hat{\rho}_\textsc{b}) - S(\hat{\rho}_\textsc{ab}),
\end{equation}
where $S(\hat{\rho})\ =\ -\operatorname{Tr} (\hat{\rho} \log \hat{\rho})$ is the von Neumann entropy and 
\begin{equation}
    \hat{\rho}_{\A}\ =\ \hat{\rho}_\B\ =\ \begin{pmatrix}
        1 - \mathcal{L}_{\A\A} & 0\\
        0 & \mathcal{L}_{\A\A}
    \end{pmatrix} .
\end{equation}
Intuitively, this can be interpreted as the maximum possible reduction in uncertainty about subsystem A that could be achieved by any measurement on subsystem B. Henceforth we will write $I(\hat{\rho}_\textsc{ab})$ simply as $I$.

\begin{figure}
    \centering
    \includegraphics{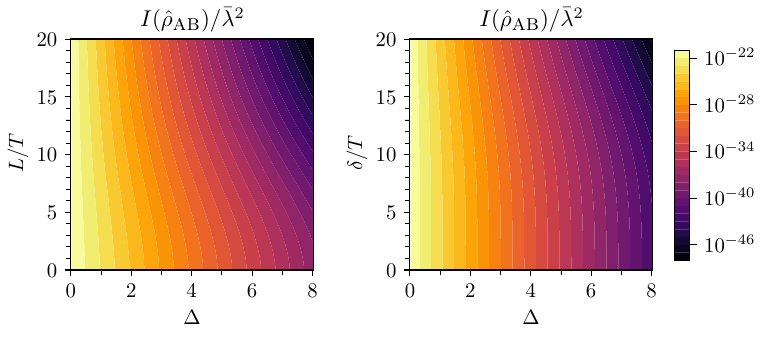}
    \caption{Here, we plot the mutual information between the two detectors as defined in eq.~\eqref{eq:mutual_info}.  On the left side, $L/T$ and $\Delta$ are varied, and we have set $\delta/T = 0$. On the right side, $\delta/T$ and $\Delta$ are varied, and we have set $L/T = 10$. For both plots, we use $\Omega T = 10$. We see that the mutual information follows qualitatively the same behaviour as $|\mathcal L_\textsc{ab}|$ in figure \ref{fig:Lab_L_and_delta}, as expected from the expression \eqref{eq:MutualInfo_LAB_LAA}. In figure \ref{fig:MutualInfo_L_and_delta_slices} of appendix \ref{appx:leadingObehaviour} we show cross-sections of these plots to compare our numerical results to the asymptotic behaviour given in eqs.~\eqref{eq:hullabaloo2} and \eqref{eq:MI_bigL}.}
    \label{fig:mutual_info}
\end{figure}

With our density matrix \eqref{eq:AB_density_matrix} (and for identical detectors), eq. \eqref{eq:mutual_info_def} becomes
\begin{align} 
    I &= (\mathcal{L}_\textsc{aa} + |\mathcal{L}_\textsc{ab} |) \log(\mathcal{L}_\textsc{aa} + |\mathcal{L}_\textsc{ab}|) \nonumber\\
    &\qquad\qquad +(\mathcal{L}_\textsc{aa} -|\mathcal{L}_\textsc{ab}|) \log(\mathcal{L}_\textsc{aa}-|\mathcal{L}_\textsc{ab}|) -2 \mathcal{L}_\textsc{aa} \log\mathcal{L}_\textsc{aa}+ \mathcal{O}(\lambda^4)
    \label{eq:mutual_info}\\
    &=\mathcal{L}_\textsc{aa}\left[ \left(1 + \frac{|\mathcal{L}_\textsc{ab}|}{\mathcal{L}_\textsc{aa}} \right) \log\! \left(1+ \frac{ |\mathcal{L}_\textsc{ab}| }{\mathcal{L}_\textsc{aa}} \right) + \left(1-\frac{|\mathcal{L}_\textsc{ab}|}{\mathcal{L}_\textsc{aa}} \right) \log \!\left( 1-\frac{|\mathcal{L}_\textsc{ab}|}{\mathcal{L}_\textsc{aa}} \right)\right] + \mathcal{O}(\lambda^4). \nonumber
\end{align}
In figure \ref{fig:mutual_info}, we show $I /\bar{\lambda}^2$ in our setup as a function of $\Delta$, $L/T$, $\delta/T$, with $T\Omega=10$. Comparing with figure \ref{fig:negativity}, we see that the detectors acquire mutual information in the entire parameter range considered---notably, outside the range in which they harvest entanglement. In these regions, the mutual information therefore reflects correlations that are not associated with entanglement between the detectors. It is possible that the detectors simply lose their ability to harvest entanglement in this region, even though there is still strong entanglement present in the field. It is also possible that in this regime, field correlations decay too quickly for the detectors to operationally access them. This distinction is particularly interesting in light of recent works showing that entanglement between disconnected regions of spacetime may decay exponentially with their separation even though the correlators decay polynomially~\cite{Klco_2021,klco2022entanglementstructuresquantumfield}.

The leading-order behaviour of the mutual information may be obtained by substituting in $\LL_{\A\A}^{(1)}$ and $\LL_{\A\B}^{(1)}$ from eqs.~\eqref{eq:largeOmega0} and \eqref{eq:Lij_leading_order}, respectively. We obtain
\begin{align}
     \frac{I^{(1)}}{\bar{\lambda}^2} \ &= \ \frac{\pi}{(T\Omega)^{2\Delta}}\,e^{-\tfrac12 T^2 \Omega^2} \left[\left(1+\frac{(T\Omega)^{2\Delta}}{|D|^\Delta} \right) \log\!\left(1+\frac{(T\Omega)^{2\Delta}}{|D|^\Delta} \right)
     \right. \label{eq:hullabaloo2}\\
     &\qquad\qquad\qquad\qquad\qquad\qquad\left.+\left(1-\frac{(T\Omega)^{2\Delta}}{|D|^\Delta} \right) \log\!\left(1-\frac{(T\Omega)^{2\Delta}}{|D|^\Delta} \right)
     \right],
     \nonumber
\end{align}
where $D$ is given in eq.~\eqref{eq:DDD}, and where the approximation is valid in the regimes that $\LL_{\A\A}^{(1)}$ and $\LL_{\A\B}^{(1)}$ are both valid. While this is not a particularly intuitive expression, a much simpler form can be obtained by working in the limit $L/T \gg \Omega T$. In this case (see the discussion leading to eq.~\eqref{eq:Iapprox2} for details), the mutual information reduces to
\begin{equation}
    \frac{I^{(1)}_{L/T {\scriptscriptstyle \gg} \Omega T}}{\bar{\lambda}^2} \ =\ \pi\,e^{-\tfrac12 T^2\Omega^2}\, \frac{(T\Omega)^{2\Delta}}{|D|^{2\Delta}}\left[ 1 + \mathcal{O}\left(\frac{(T\Omega)^{4\Delta}}{|D|^{4\Delta}} \right)\right].
    \label{eq:MI_bigL}
\end{equation}
In fact, this can we written as
\begin{equation} \label{eq:MutualInfo_LAB_LAA}
    \frac{I^{(1)}_{L/T {\scriptscriptstyle \gg} \Omega T}}{\bar{\lambda}^2} \ =\ \frac{|\mathcal L_\textsc{ab}^{1)}|^2}{\mathcal L_\textsc{aa}^{(1)}}\left[ 1 + \mathcal{O}\left(\frac{(T\Omega)^{4\Delta}}{|D|^{4\Delta}} \right)\right].
\end{equation}
The mutual information thus exhibits very similar qualitative behaviour to $|\mathcal L_\textsc{ab}|$ in the regime of validity of these approximations. For fixed $L/T$ and $\delta/T$, both quantities decay exponentially with increasing $\Delta$, with leading behaviour $|\mathcal L_\textsc{ab}|\propto |D|^{-\Delta}$ and $I(\hat{\rho}_{\A\B})\propto |D|^{-2\Delta}$. Consequently, at fixed $\Delta$, $I$ decreases more rapidly than $|\mathcal L_\textsc{ab}|$ as $|L|$ or $|\delta|$ increases.

Figure \ref{fig:MutualInfo_L_and_delta_slices} in appendix \ref{appx:leadingObehaviour} compares the results of numerical integration to the fits provided by eqs.~\eqref{eq:hullabaloo2} and \eqref{eq:MI_bigL}. We find that eq.~\eqref{eq:MI_bigL} fits the data well, even in some places outside of the regime $L/T \gg \Omega T$. As expected, eq.~\eqref{eq:hullabaloo2} provides a closer fit in regions of small $L/T$.

\section{Separating harvesting and communication in holographic CFTs}
\label{sec:separation}

As shown in the previous section (\eg figure \ref{fig:negativity}), the negativity between the detectors in their final state increases the closer the centres of the switching functions are to lightlike-separated. This suggests that much of the entanglement between the detectors is being generated by the exchange of quantum information mediated by excitations of the QFT. 

Indeed, the negativity between the detectors can be sourced by two mechanisms. As noted in section \ref{subsec:negativity}, it may arise due to the detectors acquiring pre-existing entanglement from the vacuum, which is considered genuine entanglement harvesting. Secondly, negativity may arise due to correlations acquired through field-mediated communication. Ideally, we wish to separate these two contributions, as only the former is what gives insight into the entanglement structure of the underlying quantum field theory itself. For free field theories, as proposed in~\cite{Tjoa_2021} (see also \cite{caribe_2023,TeixidoBonfill2024}), it is possible to estimate the magnitude of each contribution to the total negativity by separating the two-point function into commutator and anti-commutator terms. While the arguments justifying this separation were originally made for free theories, in this section we will discuss under what conditions the same arguments hold for a conformal field theory. Additionally, we will show the decomposition of our results of section \ref{subsec:negativity} into field-mediated communication and field-sourced entanglement.

\subsection{Separating communication from genuine harvesting in free fields}
\label{subsec:free_field_separation}

Before addressing the case of general CFTs, in this subsection, we will summarise the arguments made in \cite{Tjoa_2021,caribe_2023,TeixidoBonfill2024} for making a decomposition between harvested entanglement and communication-mediated entanglement in a free field theory. We consider the two-point function of a scalar operator $\hat{\mathcal{O}}$. The two-point function of \(\hat{\mathcal{O}}\) at spacetime points \(\mathsf{x}\) and \(\mathsf{x}'\) can be split into the vacuum expectation values of the anti-commutator and the commutator:
\begin{equation}
    \langle \hat{\mathcal{O}} (\mathsf{x}) \hat{\mathcal{O}} (\mathsf{x}') \rangle = \frac{1}{2} \langle \{ \hat{\mathcal{O}} (\mathsf{x}) , \hat{\mathcal{O}} (\mathsf{x}') \} \rangle + \frac{1}{2} \langle [ \hat{\mathcal{O}} (\mathsf{x}) , \hat{\mathcal{O}} (\mathsf{x}') ] \rangle.
    \label{gamma8}
\end{equation}
Assuming the operator \(\hat{\mathcal O}\) is self-adjoint, it follows that the real and imaginary parts correspond to the symmetric (anti-commutator) and anti-symmetric (commutator) contributions
\begin{equation}
\Re \langle \hat{\mathcal O}(\mathsf{x}) \hat{\mathcal O}(\mathsf{x}') \rangle
= \tfrac12 \,\langle \{\hat{\mathcal O}(\mathsf{x}),\hat{\mathcal O}(\mathsf{x}')\}\rangle,
\qquad
\Im \langle \hat{\mathcal O}(\mathsf{x}) \hat{\mathcal O}(\mathsf{x}') \rangle
= \tfrac{1}{2\ii}\,\langle [\hat{\mathcal O}(\mathsf{x}),\hat{\mathcal O}(\mathsf{x}')]\rangle .
\label{eq:reim}
\end{equation}
In what follows we will argue why, in free fields, these terms encode harvested entanglement and communication-mediated entanglement, respectively.

We will begin by looking at the support of each of these contributions. 
For any relativistic QFT, the commutator has support only when $\mathsf{x}$ and $\mathsf{x}'$  are timelike or null separated. Thus, when $\mathsf{x}$ and $\mathsf{x}'$  are spacelike separated, the two-point function reduces to its symmetric part. Since correlations between detectors that probe the field in spacelike-separated points cannot arise through communication, any entanglement contributions arising from such separations must be attributed to genuine entanglement harvesting. Hence, in this regime, we may identify the symmetric part of the two-point function as encoding entanglement harvested from the field.

Whether this identification can be extended to timelike-separated points is not a priori clear, as in the timelike separation regime both terms in eq.~\eqref{gamma8} are generally non-zero. In this case, we will justify our identifications by analysing the state-dependence of each of the contributions. Specifically, if the commutator of $\hat{\mathcal{O}}(\mathsf{x})$ at two different points is a multiple of the identity, the anti-symmetric contribution will be state independent. In free field theories, this condition is satisfied by local canonical field operators. Because of this state independence, one may conceivably replace the QFT vacuum with a state that possesses less correlations and this term will not change. Hence we cannot interpret any contribution to entanglement coming from the anti-symmetric part of the two-point function as representing correlations extracted from the field theory. Rather, the entanglement of the detectors resulting from the anti-symmetric part must come from the exchange of quantum information between the detectors mediated by excitations of the underlying QFT. Indeed, one can show that the leading-order contribution to any information exchange between the detectors depends only on the commutator \cite{Martin-Martinez:2015psa}. In contrast, the contribution from the symmetric part is generally state dependent. One can therefore argue that the symmetric part term is related to genuine entanglement harvesting, in agreement with the preceding discussion.

Accordingly, to quantify how much of the entanglement between the detectors is acquired through communication rather than harvesting, we can define the following quantities:
\begin{equation}
    \mathcal{N}^{\pm} = \max\left( 0, |\mathcal{M}^{\pm}| - \mathcal{L}_{\A\A} \right) + \mathcal{O}(\lambda^4),
    \label{taxihome}
\end{equation}
where $\mathcal M^\pm$ are found by using only the symmetric or the anti-symmetric part of the two-point function in eq.~\eqref{eq:M_numerical}. The ratio of $\mathcal{N}^-/\mathcal{N}$ then estimates how much of the entanglement is acquired through communication in the sense that when $\mathcal{N}^-/\mathcal{N}\approx 1$ the entanglement between the detectors is mostly coming from the exchange of information via the field. Notice that this separation is more subtle than a mere decomposition into two additive components, that is, $\mathcal{N}\neq \mathcal{N}^+ + \mathcal{N}^-$ (see \cite{Zambianco_2024} for details on this point). Since $\mathcal L_\textsc{aa}$ involves only coincident points and hence acts as a local noise term, it remains the same. A more in-depth discussion of this separation and how it is used to classify entanglement can be found in~\cite{Tjoa_2021,caribe_2023,TeixidoBonfill2024}.

In general, both the commutator and anticommutator of the scalar operator appearing in eq.~\eqref{gamma8} are state dependent. This is true even in a free theory, for example for composite operators such as $[\phi(\mathsf{x})]^n$. Consequently, the detector entanglement cannot be unambiguously attributed to either causal communication or entanglement harvesting with the same arguments. In the present case, where detectors are coupled to a generally interacting CFT through scalar conformal primary operators, no separation between these two physical mechanisms is available a priori. Nevertheless, as we will show next, holographic CFTs exhibit an additional structural simplification: the commutator of scalar primary operators can also be shown to be a multiple of the identity via holography, allowing for the same arguments to justify the decomposition between communication-induced and harvested correlations.

\subsection{Separating communication from genuine harvesting in holographic CFTs}

In the present work, we consider detectors that are coupled to scalar conformal primary operators in a CFT. At first sight, this would appear to place us squarely in the generic situation described above (apart from the cases of free conformally coupled scalar field theories). However, as we now explain, an analogous simplification occurs for a broad class of interacting CFTs that admit a holographic description.

We consider CFTs with large central charge and strong coupling, which are dual to semiclassical gravitational theories in asymptotically AdS spacetimes via the AdS/CFT correspondence (see appendix~\ref{appx:CFTs}). In this duality, scalar primary operators $\mathcal O(t,\bm{x})$ in the boundary CFT are mapped to scalar fields $\Phi(t,\bm{x},z)$ propagating in the $(d+1)$-dimensional bulk. The scaling dimension $\Delta$ of $\mathcal O$ is related to the mass $m$ of the dual bulk scalar by
\begin{equation}
\label{eq:Delta_d_m_relation}
    \Delta = \frac{d}{2} + \sqrt{\frac{d^2}{4} + m^2 L_\text{AdS}^2},
\end{equation}
where $L_\text{AdS}$ is the AdS curvature radius.

In AdS/CFT, real-time boundary correlators are easily obtained via the \emph{extrapolate dictionary} \cite{Banks:1998dd,Balasubramanian:1998de,Harlow:2011ke}. Using Poincar\'e coordinates \eqref{eq:AdSmetric}, one leverages the identification of bulk and boundary operators:  $\mathcal O(\mathsf{x})\sim \lim_{z\to0}z^{-\Delta}\,\Phi(\mathsf{x},z)$ in the absence of any sources. Hence, one begins by computing bulk correlators of the dual scalar field in a specified bulk quantum state and then takes the appropriate boundary limit. In particular, one quantises the bulk scalar field $\phi$ and evaluates the bulk Wightman function
$\langle \Phi(t,\bm{x},z)\Phi(t',\bm{x}',z') \rangle$. The corresponding boundary two-point function of  $\mathcal O$ is then obtained as
\begin{equation}
\label{eq:extrapolate}
    \langle \mathcal O(t,\bm{x})\mathcal O(t',\bm{x}')\rangle
    =
    2^{-\Delta}\mathcal N^{-1}_\Delta
    \lim_{z,z'\to 0}
    z^{-\Delta} z'^{-\Delta}
    \,
    \langle \Phi(t,\bm{x},z)\Phi(t',\bm{x}',z') \rangle ,
\end{equation}
where $\mathcal N_\Delta$ is a normalisation constant fixed by holographic renormalisation.

At leading order in the limit of large central charge,\footnote{This phrasing extends to the discussion of the 't~Hooft limit for $\mathcal N=4$ SYM in appendix \ref{appx:CFTs} to a general context where $c\sim L_\mt{AdS}^{d-1}/G_\mt{N}$ is large. The latter central charge controls a variety of expressions in the boundary CFT, including the two- and three-point correlator of the stress tensor, \eg \cite{Buchel:2009sk}, and universal contributions in the entanglement entropy, \eg \cite{Rangamani:2016dms}.}  the bulk theory is weakly coupled and the dynamics of $\Phi$ are governed by a free scalar field equation in AdS. Further, in our construction, when the bulk Wightman function is evaluated in the Poincar\'e AdS vacuum \eqref{eq:AdSmetric} and near the asymptotic boundary,  it takes the form \cite{DHoker:1998ecp}
\begin{equation}
    \langle\, \Phi(t,\bm{x},z)\, \Phi(t',\bm{x}',z')
    \,\rangle
    \simeq \lim_{\epsilon \to 0^+}{\mathcal N_\Delta}
    \left(\frac{2 z z'}{z^2+z'^2 - (t-t'-\mathrm{i}\epsilon)^2 + |\bm{x}-\bm{x}'|^2}\right)^\Delta\,.
    \label{eq:bulkW}
\end{equation}
Substituting this expression into eq.~\eqref{eq:extrapolate}, one immediately finds
\begin{equation}
\label{eq:holoWightman}
    \langle \mathcal O(t,\bm{x}) \mathcal O(t',\bm{x}') \rangle
    =
    \frac{1}{\bigl[-(t - t' - \mathrm{i}\epsilon)^2 + |\bm{x} - \bm{x}'|^2\bigr]^\Delta},
\end{equation}
which reproduces our expression for the CFT two-point function given in eq.~\eqref{eq:CFT_2PF} and used in our evaluation of the elements of the density matrix \eqref{eq:AB_density_matrix}.

In this holographic setting, where the boundary two-point function is related to a free two-point function in the bulk, we can employ the methods discussed above to disentangle the harvesting and communication contributions to the entanglement of the detectors. However, it is important to stress here that the extrapolate dictionary \cite{Banks:1998dd,Balasubramanian:1998de,Harlow:2011ke}  is a statement about the identifying boundary and bulk operators and it does not assume that the bulk scalar must be a free field. In general, when bulk interactions are included, the bulk commutator $[\Phi(X),\Phi(X')]$ becomes operator valued, and hence its expectation value becomes state dependent. 

However, holographic CFTs are typically considered in the limit of large central charge (\ie the 't~Hooft limit, when the boundary CFT is a gauge theory) so that the bulk description reduces to weakly coupled semiclassical gravity. At large $c$, correlation functions of single-trace operators exhibit \emph{generalised free field} behaviour \cite{Witten:1998qj,Heemskerk:2009pn,Fitzpatrick:2012yx}, factorizing into sums of pairwise contractions with connected contributions suppressed by powers of $1/c$,
\begin{equation}
    \langle \mathcal O_1 \cdots \mathcal O_n \rangle
    =
    \text{(all pairwise contractions)}
    + \mathcal O(1/c).
\end{equation}
Similarly, connected correlators, \eg $\langle \mathcal O_1 \mathcal O_2 \mathcal O_3 \mathcal O_4 \rangle_{\mathrm{conn}}$, are suppressed by $1/c$. As a result, to leading order in $1/c$, the operator algebra of $\mathcal O$ is effectively Gaussian, and the commutator $[\mathcal O(x), \mathcal O(x')]$ is a $c$-number distribution fixed entirely by the two-point function. Equivalently, for a broad class of states $|\psi\rangle$ that do not carry $\mathcal O(c)$ excitations,
\begin{equation}
    \langle \psi | [\mathcal O(x),\mathcal O(x')] | \psi \rangle
    =
    \langle 0 | [\mathcal O(x),\mathcal O(x')] | 0 \rangle
    + \mathcal O(1/c).
\end{equation}

We therefore conclude that, for holographic CFTs at leading order in the large-$c$ expansion, the commutator contribution in eq.~\eqref{gamma8} is state independent and can be cleanly identified with communication between the detectors, while the anti-commutator contribution encodes correlations associated with genuine entanglement harvesting. Subleading $1/c$ corrections may blur this separation, but are suppressed in the limit of interest.

\subsection{Results of the separation for holographic CFTs}

We now look at the decomposition of our negativity results in section \ref{subsec:negativity}. As argued above, the decomposition holds for CFTs when the CFT admits a holographic description.

First, we will look at the decomposition of $\mathcal M$ into $\mathcal M^\pm$ as defined after eq.~\eqref{taxihome}. $\mathcal M^+$ is obtained using the symmetric (real) part of the Feynman propagator in the integral, and, as argued in the prior part of this section, encodes genuine entanglement harvesting. $\mathcal M^-$ is obtained using the antisymmetric (imaginary) part of the Feynman propagator in the integral, and, as we have argued, encodes correlations arising from field-mediated communication between the detectors. 

\begin{figure}
    \centering
    \includegraphics{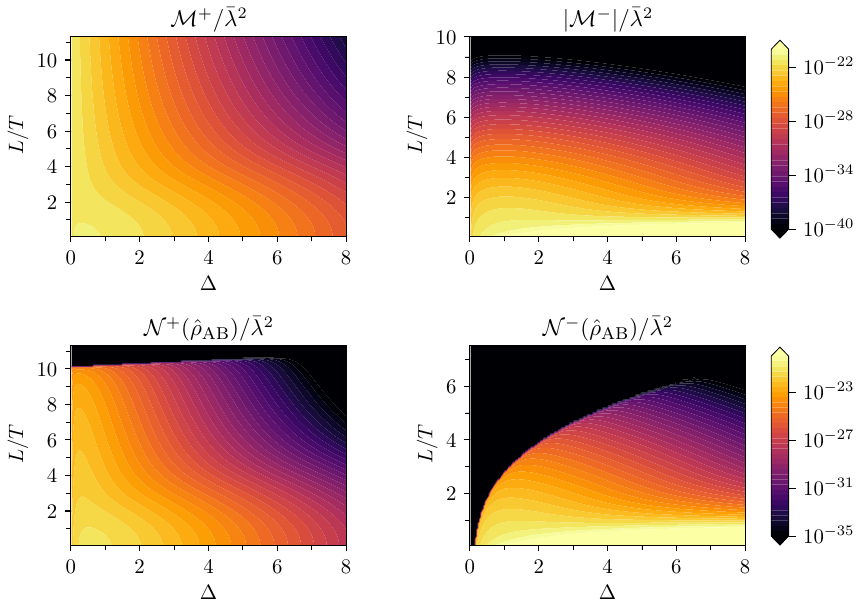}
    \caption{Here, we show $|\mathcal M^\pm|$ and $\mathcal N^\pm$ as a function of $\Delta$ and $L/T$. In all of these plots, $\delta=0$ and $T \Omega =10$, where the former makes $\mathcal M^+$ real. We see, as expected, that $|\mathcal M^-|$ decays much more rapidly than $\mathcal M^+$ as a function of $L/T$. Figures \ref{fig:M+_L_delta_Delta_slices}--\ref{fig:N-_L_delta_Delta_slices} in appendix \ref{appx:leadingObehaviour} show cross-sections of these plots to compare our numerical results to the asymptotic behaviour given in eqs.~\eqref{eq:M+_SL_leading0} and~\eqref{eq:N+_SL_0}.}
    \label{fig:M_plus_minus_and_N_plus_minus_Delta-L_heatmaps}
\end{figure}

The top row of figure \ref{fig:M_plus_minus_and_N_plus_minus_Delta-L_heatmaps} shows how $\mathcal M^\pm$ scale with $L/T$ and $\Delta$ (with $\delta/T = 0$ and $\Omega T = 10$). As $L/T$ increases, we move further and further into the regime of spacelike separation of the detectors. As expected, we see the communication term $|\mathcal M^-|$ decays far more rapidly with $L/T$ than the harvesting term $\mathcal{M}^+$ (which is purely real when $\delta = 0$, as can be seen from eq.~\eqref{eq:M+_SL_leading0}).  When the detectors become effectively spacelike separated, we see that the plots of $\mathcal M^+$ and $\mathcal N^+$ follow qualitatively similar behaviour to $\mathcal M$ and $\mathcal N$ in the left panels of figures \ref{fig:M_L_and_delta} and \ref{fig:negativity}, which results from the fact that $\mathcal M^-$ (and hence $\mathcal N^-$) is much smaller in magnitude at this point ($\M^-$ in this regime comes entirely from the tails of the switching functions that extend to timelike separations). Of course, this effect becomes more dramatic as $L/T$ becomes small, as is evident in the far right panel. Hence the majority of entanglement picked up in this regime is from genuine entanglement harvesting. On the other hand, when the detectors are close together ($L/T \lesssim 1$), $\mathcal M^-$ and $\mathcal N^-$ closely approximate $\mathcal M$ and $\mathcal N$, thus the majority of entanglement arises from communication. 

In appendix \ref{appx:leadingObehaviour}, we analyse the leading-order behaviour of these terms in various asymptotic regimes. Specifically, in eq.~\eqref{ohare22} we show the dominant behaviour of the harvesting-related term $\mathcal M^+$ in the far-spacelike regime (with $L^2 -\delta^2 \gg T^2$) is
\begin{equation}
    \frac{\mathcal{M}^+}{\bar{\lambda}^2} \ \approx\ \frac{\mathcal{M}^+_\textsc{sl}}{\bar{\lambda}^2}  \  = \ -\pi\,e^{-\frac12 T^2\Omega^2+\ii\Omega\delta}\; \frac{T^{2\Delta}}{(L^2-\delta^2)^\Delta} \ = \ \mathcal{M}^{(1)}_{\tilde{D}\scriptscriptstyle\gg 1},
    \label{eq:M+_SL_leading0}
\end{equation}
i.e.~it reduces to the leading-order expression for $\mathcal{M}_{\tilde{D}\scriptscriptstyle\gg 1}$ which we found in eq.~\eqref{eq:M_leading_order}. We compare eq.~\eqref{eq:M+_SL_leading0} with our numerical results for $\mathcal M^+$ in figure \ref{fig:M+_L_delta_Delta_slices} of appendix \ref{appx:leadingObehaviour}. Meanwhile, the behaviour of $\mathcal M^-$ in this regime is negligible compared to $\mathcal M^+$ since the anti-symmetric part of the two-point function vanishes exactly in spacelike separation. (For an analytic study of the spacelike behaviour of $\mathcal M^-$, see the discussion preceding eq.~\eqref{eq:M-_SL_leading} as well as figure \ref{fig:M-_L_delta_Delta_slices}.)

The bottom row of the same figure shows the negativities $\mathcal N^\pm$ obtained with these terms. The quantity $\mathcal{N}^+$ can be roughly interpreted as a contribution to the entanglement acquired by the detectors from genuine entanglement harvesting, and and $\mathcal{N}^-$ as a contribution to the detectors' entanglement acquired through communication via the field. Using the leading-order terms $\mathcal L_\textsc{aa}^{(1)}$ from \eqref{eq:largeOmega0} and $\mathcal{M}^+_\textsc{sl}$ from \eqref{eq:M+_SL_leading0}, the negativity becomes
\begin{equation}
    \frac{\mathcal N^{+}}{\bar{\lambda}^2}\approx \frac{\mathcal N^{+}_\textsc{sl}}{\bar{\lambda}^2}\ =\ 
    \pi\,e^{-\frac12 T^2\Omega^2} \Bigg[
    \frac{T^{2\Delta}}{(L^2-\delta^2)^\Delta}-\frac{1}{(T\Omega)^{2\Delta}} \Bigg]_+.
    \label{eq:N+_SL_0}
\end{equation}
If we set $\delta/T = 0$, as is the case in figure \ref{fig:M_plus_minus_and_N_plus_minus_Delta-L_heatmaps}, we arrive at the same boundary for $\mathcal N^{+\,(1)}_\textsc{sl} = 0$ as in eq.~\eqref{eq:nomore} for $\N^{(1)}_{\tilde{D} {\scriptscriptstyle\gg} 1}$. Higher-order corrections to $\mathcal N_\textsc{sl}$ are given in eq.~\eqref{Nplus55}, and we compare with our numerical analysis in figure \ref{fig:N+_L_delta_Delta_slices}. As for the communication contribution, in the far spacelike regime, 
\begin{align}
    \frac{\mathcal{N^-_\textsc{sl}}}{\bar{\lambda}^2}\ &\simeq\ 0,
    \label{eq:N-_SL_0}
\end{align}
hence $\N\approx\N^+ $ in the far-spacelike regime.

\begin{figure}
    \centering
    \includegraphics{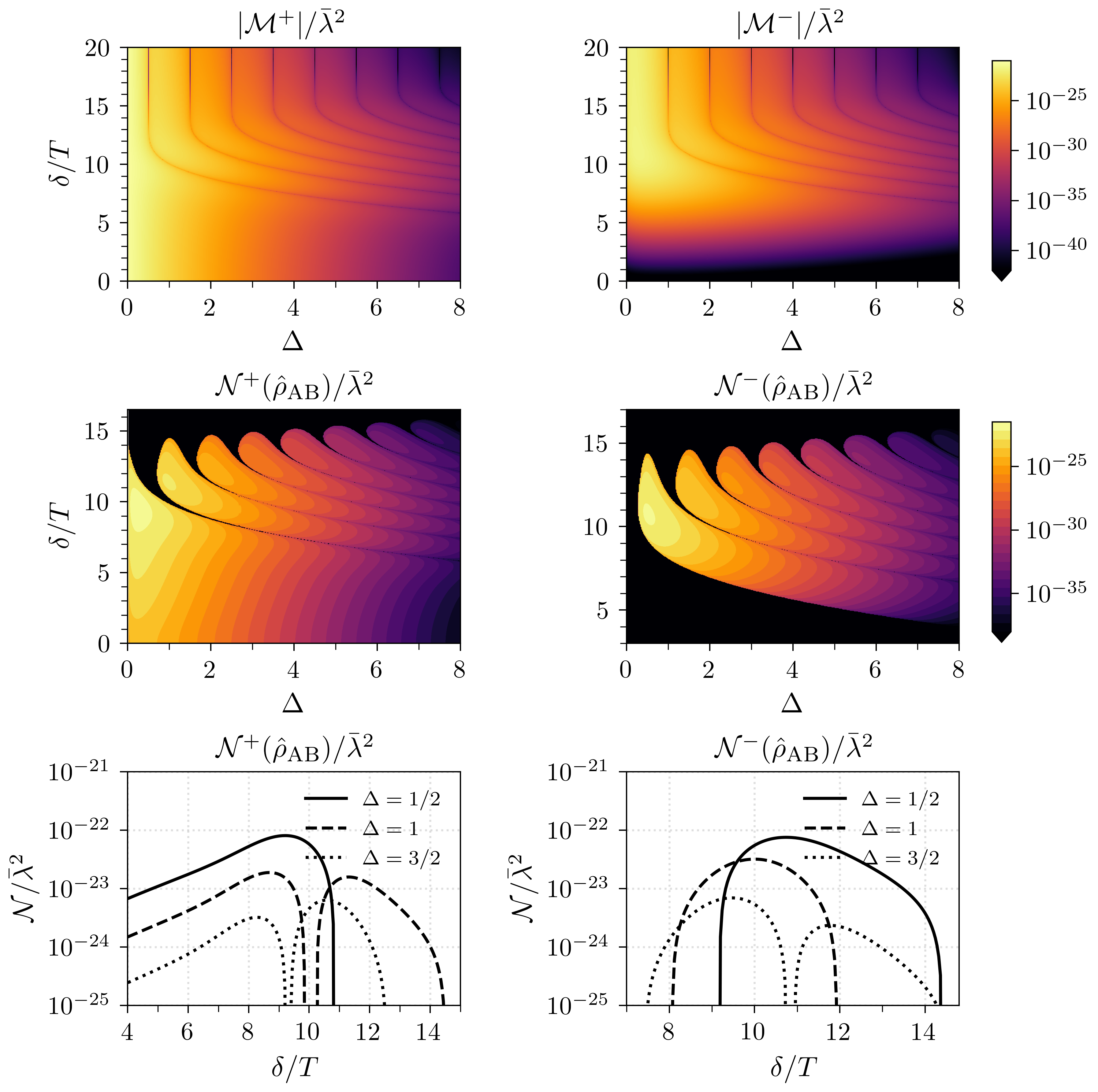}
    \caption{The above plots show $|\mathcal M^\pm|$ and $\mathcal N^\pm$ as a function of $\Delta$ and $\delta/T$. In all, we have set $L/T = T \Omega = 10$. When one detector sits in the lightcone of the other, we see that these quantities oscillate. We also see more bumps arising in the cross-sections in the bottom row as $\Delta$ increases. The values of $\Delta$ plotted in the cross-sections reproduce results in \cite{Tjoa_2021} for the case of a amplitude coupling to a free massless scalar field in flat space. Figures \ref{fig:M+_L_delta_Delta_slices}--\ref{fig:N-_L_delta_Delta_slices} in appendix \ref{appx:leadingObehaviour} show cross-sections of these plots to compare our numerical results to the asymptotic behaviour given in eqs.~\eqref{eq:M+_SL_leading0}--\eqref{eq:N-_TL_0}.}
    \label{fig:M_plus_minus_and_N_plus_minus_Delta-delta_heatmaps}
\end{figure}

The top row of figure \ref{fig:M_plus_minus_and_N_plus_minus_Delta-delta_heatmaps} shows how $\mathcal M^\pm$ scale with $\Delta$ and $\delta/T$ (with $L/T = 10$ and $\Omega T = 10$). The plots reveal that as the causal contact between the detectors increases, specifically as the detectors enter the regime where they are timelike connected, $\mathcal M^+$ and $\mathcal M^-$ oscillate. In eqs.~\eqref{Mplus99} and \eqref{Mminus99} respectively, we show that asymptotic behaviour of $\mathcal M^\pm$ in this regime, the leading-order terms of which are given by
\begin{align}
    \frac{\mathcal{M}^+_\textsc{tl}}{\bar{\lambda}^2}\ \simeq \ -\,\pi\,\cos(\pi\Delta)\,e^{-\frac12 T^2\Omega^2+\ii\Omega\delta} \,
    \frac{T^{2\Delta}}{(\delta^2-L^2)^{\Delta}}\ &=\ \cos(\pi \Delta) \mathcal{M}^{(1)}_{\tilde{D} {\scriptscriptstyle\gg} 1},
    \label{eq:M+_TL_leading0}\\
    \frac{\mathcal{M}^-_\textsc{tl}}{\bar{\lambda}^2}\ \simeq \ \ii\, \pi\,\sin(\pi\Delta)\,e^{-\frac12 T^2\Omega^2+\ii\Omega\delta} \,\frac{T^{2\Delta}}{(\delta^2-L^2)^{\Delta}} \ &=\ -\ii \sin(\pi \Delta) \mathcal{M}^{(1)}_{\tilde{D} {\scriptscriptstyle\gg} 1}. \label{eq:M-_TL_leading0}
\end{align}
These are compared to our numerical results in figures \ref{fig:M+_L_delta_Delta_slices} and \ref{fig:M-_L_delta_Delta_slices} of appendix \ref{appx:leadingObehaviour}. The source of this behaviour can most easily be seen from the phase behaviour of the two-point function given in eq.~\eqref{eq:CFT_2PF}. When $(t',\bm x')$ is spacelike-separated from $(t,\bm x)$, the phase of the denominator is zero. When $(t',\bm x')$ is in the causal future of $(t,\bm x)$, on the other hand, the denominator picks up a phase of $-\pi \Delta$. In the regimes where the above approximations are valid, this causes the relative contributions of $\mathcal M^+$ and $\mathcal M^-$ to oscillate with $\Delta$, leading to alternating dominance of communication-type and harvesting-type contributions. In particular, $\mathcal M$ is communication-dominated for half-integer $\Delta$ and harvesting-dominated for integer $\Delta$. The alternation of the cancellation of the leading order behaviour of $\mathcal{M}^\pm$ for integer and half-integer values of $\Delta$ (the ones that are analogous to detectors coupling to the amplitude of a massless scalar field in $2\Delta+1$ spatial dimensions) is consistent with the results in \cite{Tjoa_2021} where it was shown that generically when one of the quantities vanish the other one peaks.

In the middle and bottom rows of plots in figure \ref{fig:M_plus_minus_and_N_plus_minus_Delta-delta_heatmaps}, we show $\mathcal N^\pm$ as a function of $\Delta$ and $\delta/T$, with cross-sections of the plots at certain values of $\Delta$. Again, we note that these generalise the results in \cite{Tjoa_2021} for the cases of $2 \Delta + 1$ spatial dimensions. In eqs.~\eqref{Nplus44} and \eqref{Nminus44} respectively, we give that asymptotic behaviour of $\mathcal \N^\pm$ in the far-timelike regime, the leading-order terms of which are given by
\begin{align} 
    \frac{\mathcal{N}^{+}_\textsc{tl}}{\bar{\lambda}^2}\ &\simeq\  \pi\,e^{-\frac12 T^2\Omega^2} \Bigg[ 
    \frac{|\cos(\pi\Delta)|\,T^{2\Delta}}{(\delta^2-L^2)^{\Delta}} -\frac{1}{(T\Omega)^{2\Delta}} \Bigg]_+,\label{eq:N+_TL_0} \\
    \frac{\mathcal{N}^-_\textsc{tl}}{\bar{\lambda}^2}\ &\simeq\ \pi\,e^{-\frac12 T^2\Omega^2} \Bigg[ 
    \frac{|\sin(\pi\Delta)|\,T^{2\Delta}}{(\delta^2-L^2)^{\Delta}} -\frac{1}{(T\Omega)^{2\Delta}} \Bigg]_+\,.\label{eq:N-_TL_0}
\end{align}
We compare these (to the order given in eqs.~\eqref{Nplus44} and \eqref{Nminus44}) to our numerical results in figures \ref{fig:N+_L_delta_Delta_slices} and \ref{fig:N-_L_delta_Delta_slices}. The oscillatory terms in these expressions already show that the boundary at which the negativity vanishes oscillates, which can be seen in figure \ref{fig:M_plus_minus_and_N_plus_minus_Delta-delta_heatmaps}. We derive approximations for these boundaries in eqs.~\eqref{eq:N+_bdy_2ndO} and \eqref{eq:N-_bdy_2ndO}, which are shown in the respective figures of appendix \ref{appx:leadingObehaviour}.

\begin{figure}
    \centering
    \includegraphics{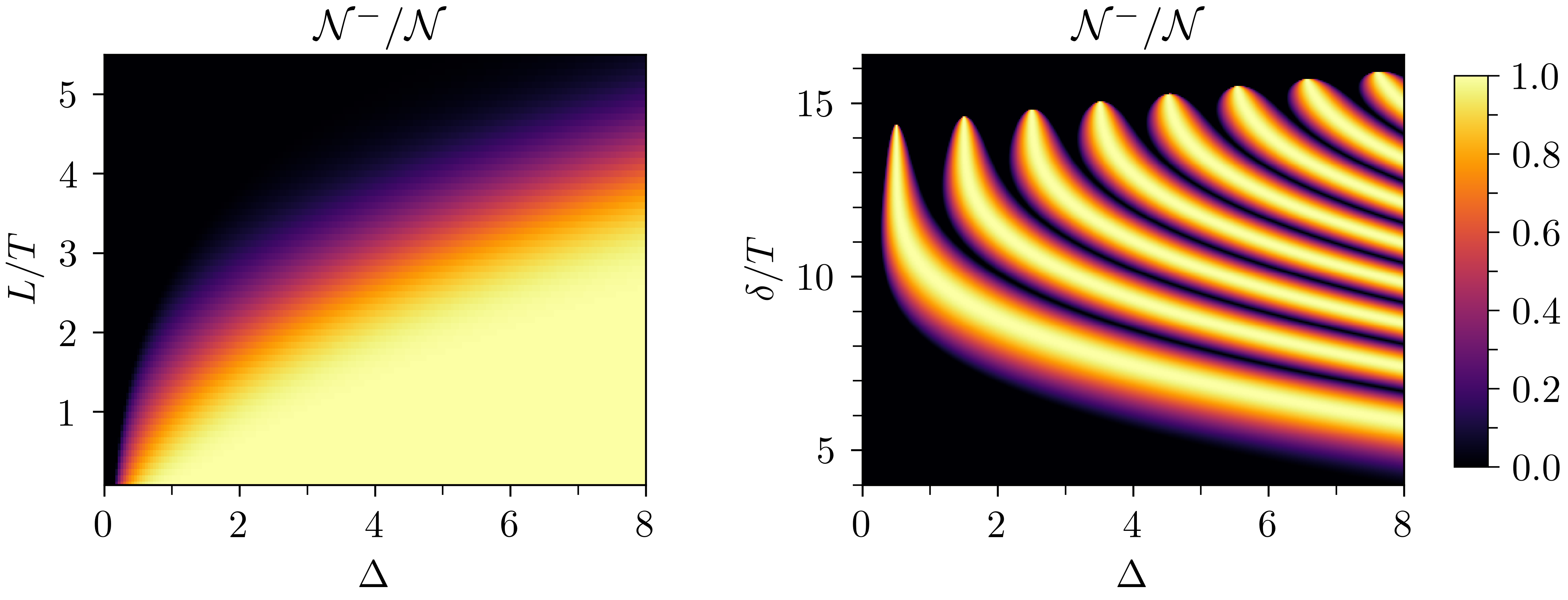}
    \caption{Here, we plot the ratio of the communication-sourced negativity $\mathcal N^-$ to the total negativity $\mathcal N$.  On the left side, $L/T$ and $\Delta$ are varied, and we have set $\delta/T = 0$. On the right side, $\delta/T$ and $\Delta$ are varied, and we have set $L/T = 10$. For both plots, we use $\Omega T = 10$. As mentioned in section \ref{subsec:free_field_separation}, this constitutes a measure of the proportion of harvested entanglement sourced by communication, in the sense that when $\mathcal N^-/\mathcal N = 1$, the entanglement is entirely coming from communication.}
    \label{fig:Nminus_over_N_ratio}
\end{figure}

Finally, in figure \ref{fig:Nminus_over_N_ratio}, we show the ratio $\mathcal N^-/\mathcal{N}$. Recall that at the points where \mbox{$\mathcal N^-/\mathcal{N}\simeq 1$}, the entanglement acquired between the detectors is not harvested but rather created through exchange of quantum information via the field, as discussed in subsection~\ref{subsec:free_field_separation}. Hence, in the regimes that our detectors become entangled, we take $\mathcal N^-/\mathcal N$ as a proxy for how much of that entanglement is sourced by communication.

\section{Discussion}
\label{sec:discussion}

In this work, we established a  framework to study entanglement harvesting in 
conformal field theories using pointlike Unruh–DeWitt detectors coupled to scalar primary operators. Conformal symmetry provides analytic control in this setting, and so we were able to extend the standard investigations of harvesting beyond free field theories to interacting CFTs. We also addressed the distinction between genuinely harvested entanglement and correlations arising from field-mediated communication, demonstrating that this separation can be made precise for CFTs with a holographic dual. Further, we derived asymptotic closed-form expressions in relevant regimes and found excellent agreement with numerical results. These approximations---and the new insights we derived from them---also apply to studies of entanglement harvesting from free massless scalar fields, since those are but a particular case of a CFT.

The behaviour of the detector's entanglement and classical correlations with the conformal dimension $\Delta$ matches closely that of the the two-point function. Namely, as the conformal dimension $\Delta$ increases, the two-point function \eqref{eq:CFT_2PF} decays more sharply with the separation, and as a consequence both the negativity and the mutual information follow the same behaviour, as shown in figures \ref{fig:negativity} and \ref{fig:mutual_info}. This may also be interpreted holographically: in the AdS/CFT dictionary, larger $\Delta$ corresponds to heavier bulk fields, whose correlations are shorter-ranged; our results are consistent with this bulk intuition. 

We emphasize that our analysis is restricted to the leading order in the perturbative Dyson expansion \eqref{eq:Dyson} and therefore depends only on the two-point correlation function \eqref{eq:CFT_2PF}. Consequently, our results are insensitive to effects arising from higher-point correlators, which enter at higher orders in the expansion in the coupling $\lambda$. This limitation is particularly relevant in holographic CFTs, where the large-$N$ structure suppresses connected higher-point functions, and phenomena such as the transition of mutual information between connected and disconnected phases \eg~\cite{Headrick:2010zt} would be expected to be associated with subtle changes in very high-order correlations.

Our work highlights several broader lessons. First, conformal field theories (CFTs) provide a natural arena for studying operational quantum-information protocols, such as entanglement harvesting, in genuinely interacting quantum field theories. The conformal symmetry of these theories strongly constrains the correlators of local primary operators, allowing analytic control to be maintained throughout the calculations. This stems from the fact that the two-point function takes a universal conformal form determined solely by the scaling dimension, as given in eq.~\eqref{eq:CFT_2PF}. In this setting, we have presented, to our knowledge, the first study of entanglement harvesting in interacting QFTs. Of course, this framework also opens the door to applying a wide range of other RQI protocols to interacting field theories. 

Second, for CFTs admitting a holographic dual, AdS/CFT maps calculations in an interacting boundary theory to a qualitatively new arena in the bulk, where novel geometric and field-theoretic structures can be exploited to gain new insights into quantum-information protocols in the boundary CFT. In particular, AdS/CFT allows aspects of the strongly coupled boundary CFT to be translated into (nearly) free-field dynamics in the bulk. This perspective allowed us to make a clean separation between correlations generated by direct communication and by genuine entanglement harvesting in section \ref{sec:separation}.

Taking our results together, we have shown that CFTs, and especially holographic CFTs, provide a controlled laboratory where detector-based protocols can meaningfully interrogate entanglement structure in interacting QFTs.
Our investigations can naturally be extended in a number of directions: 
(i) One may exploit conformal mappings to relate entanglement harvesting in different conformally related backgrounds: the Wightman function \eqref{eq:CFT_2PF} transforms covariantly, but one much properly take into account the scales involved in the detectors. Investigations in this direction were already pursued in \cite{Hotta:2020pmq}.
(ii) Beyond scalar primaries, the detectors can be coupled to conserved currents or to the stress tensor, whose two-functions are also fixed by conformal symmetry, enabling systematic comparisons with results obtained for free fields.
(iii) One can probe excited CFT states prepared by a Euclidean path integral with local operator insertions. These introduce higher-point functions and provide a route establishing to interesting diagnostics of state dependent correlations.
(iv) Finally, it is natural to ask whether suitably engineered couplings and protocols can give new insights into holography. For example, one might be able detect phase transitions in holographic entanglement entropy (\ie between connected versus disconnected phases).

\acknowledgments

We thank Mar\'ia Rosa Preciado-Rivas for useful discussions.
Research at Perimeter Institute is supported in part by the Government of Canada through the Department of Innovation, Science and Economic Development Canada and by the Province of Ontario through the Ministry of Colleges and Universities. 
EMM acknowledges support through the Discovery Grant Program of the Natural Sciences and Engineering Research Council of Canada (NSERC).
CL acknowledges the support from the Natural Sciences and Engineering Research Council of Canada (NSERC) through a Vanier Canada Graduate Scholarship [Funding Reference Number: CGV--192752].
RCM is also supported in part by an NSERC Discovery Grant, and by funding from the BMO Financial Group.

\appendix

\section{Conformal field theory primer}
\label{appx:CFTs}

Here we give an introduction to several basic aspects of conformal field theories (CFTs) that may be helpful to unfamiliar readers. The interested reader can find more comprehensive reviews in \cite{DiFrancesco:1997nk,Blumenhagen:2009zz}.

Typically, a quantum field theory in flat spacetime is invariant under the Poincar\'e group, which is comprised of coordinate transformations leaving the Minkowski metric invariant (\ie Lorentz boosts, rotations and translations). A conformal field theory enjoys a larger symmetry group: the conformal group, which consists of coordinate transformations that preserve the metric up to a local scale factor,
\begin{equation}
    g_{\mu\nu}(\mathsf{x})\ \mapsto\ \tilde g_{\mu\nu}(\mathsf{x})=e^{2\sigma(\mathsf{x})} g_{\mu\nu}(\mathsf{x})\, .
\end{equation}
In $d\ge3$,\footnote{CFTs in $d=2$ are special because the conformal group becomes infinite-dimensional, leading to far stronger constraints than in higher dimensions, \eg see \cite{Ginsparg:1988ui,Blumenhagen:2009zz}.} as well as the usual Poincar\'e transformations, this includes dilatations and special conformal transformations, respectively,
\begin{equation}
\mathsf{x}^\mu \mapsto \lambda \, \mathsf{x}^\mu
    \qquad{\rm and}\qquad
    \mathsf{x}^\mu \mapsto \frac{\mathsf{x}^{\mu} - b^{\mu}\, \mathsf{x}^{2}}{1 - 2\, b\!\cdot\!\mathsf{x} + b^{2}\, \mathsf{x}^{2}}\,.
    \label{eq:extrans}
\end{equation}
In group-theoretic terms, the Poincar\'e group is the semi-direct product of spacetime translations with the Lorentz group,
\begin{equation}
    \mathrm{ISO}(1,d-1)=\mathbb{R}^{1,d-1}\rtimes SO(1,d-1)\,.
\end{equation}
While in a CFT, this symmetry is enlarged to the conformal group, which is
locally isomorphic to the Lorentz group with two additional dimensions (one timelike and
one spacelike),
\begin{equation}
    \mathrm{Conf}(1,d-1)\simeq SO(2,d)\,.
\end{equation}

A CFT is organised into families of local operators, each consisting of a {primary} operator together with its {descendants}. A primary operator is an eigenoperator under dilatations, with eigenvalue known as its conformal dimension~$\Delta$. Under $\mathsf{x}^\mu\mapsto \lambda\,\mathsf{x}^\mu$, a \emph{scalar} primary transforms as
\begin{equation}
\mathcal{O}_{\Delta}(\mathsf{x}) \;\mapsto\; 
\mathcal{O}_{\Delta}(\lambda\,\mathsf{x})=\lambda^{-\Delta}\,\mathcal{O}_{\Delta}(\mathsf{x})\, .
\end{equation}
Descendants are obtained by acting with derivatives on the primary operator, with each derivative increasing the conformal dimension by one. Thus, a $k$-th descendant of the primary $\mathcal{O}_\Delta$ takes the form $\partial_{\mu_1} \cdots\partial_{\mu_k}\mathcal{O}_\Delta$ and has dimension $\Delta+k$. Let us add that for scalar operators, unitarity of the underlying CFT imposes the bound, \eg see \cite{Rychkov:2016iqz,Simmons-Duffin:2016gjk},
\begin{equation}
    \Delta \ge \frac{d-2}{2}\,.
    \label{unibomb}
\end{equation}

Conformal symmetry places strong constraints on correlation functions, in particular, in the vacuum state. For example, one-point functions of any primary operators vanish unless the operator is the identity. The two-point function of scalar primaries is fixed (up to normalisation) to be\footnote{In the main text, eq. \eqref{eq:CFT_2PF} absorbs $C$ into the normalisation of the operators, as is the standard convention, and we have already imposed $\Delta_i = \Delta_j$.}
\begin{equation}
    \langle \mathcal{O}_{\Delta_i}(x)\, \mathcal{O}_{\Delta_j}(x') \rangle
    = \frac{C\,\delta_{\Delta_i\Delta_j}}{|x-x'|^{2\Delta_i}} .
    \label{eq:CFT_2PFfull}
\end{equation}
Similarly, three-point functions are fixed up to an overall constant, but
higher-point functions are less constrained by conformal symmetry, \eg they depend only on so-called conformal cross-ratios. 

A central organizing tool for CFTs is the operator product expansion (OPE). When two operators approach each other, their product expands into a sum over local operators:
\begin{equation}
    \mathcal{O}_i(x)\, \mathcal{O}_j(0)
    = \sum_k C_{ijk}\, |x|^{\Delta_k - \Delta_i - \Delta_j} \, \mathcal{O}_k(0)
    + \text{(descendant\ contributions)} ,
    \label{eq:OPE}
\end{equation}
where the coefficients $C_{ijk}$ are the structure constants of the theory. Any CFT can then be specified by the spectrum of conformal primaries and their OPE data (\ie the set of dimensions $\Delta_i$ and coefficients $C_{ijk}$).

While the data described above provide a non-perturbative characterisation of conformal field theories, in practice many well-known CFTs are specified in the more conventional way, namely by writing down an action whose dynamics realise a conformal fixed point. A simple and familiar example is the free conformally coupled scalar field, \eg \cite{Osborn:1993cr,Birrell:1982ix}.  On a fixed background geometry, its action is
\begin{equation}
    S = \frac{1}{2}\int \dd^d x\, \sqrt{-g}\,
    \Big[ (\nabla\phi)^2 + \xi\, R\, \phi^2 \Big],
    \qquad
    \xi = \frac{d-2}{4(d-1)}\,,
\end{equation}
where $R$ is the Ricci scalar associated with the background metric. In flat space ($R=0$), the theory reduces to that of a free massless scalar. The field $\phi$ itself defines a scalar primary operator with conformal dimension\footnote{Hence, it lies precisely at the unitarity bound \eqref{unibomb}.}
\begin{equation}
    \Delta_\phi = \frac{d-2}{2}\,.
\end{equation}

While such free theories provide useful intuition, most CFTs of physical interest are interacting or even strongly interacting theories. A broad and important class consists of superconformal gauge theories, \eg see \cite{Minwalla:1997ka,Terning:2006bq}. The canonical example is $\mathcal{N}=4$ supersymmetric Yang-Mills (SYM) theory in four dimensions, \eg \cite{DHoker:2002nbb,Aharony:1999ti}, which is an interacting, finite, and exactly conformal theory for all values of the gauge coupling.

In standard notation, the field content consists of a gauge field $A_\mu$, six real scalars $\phi^I$ ($I=1,\ldots,6$)  and four Weyl fermions $\psi^a$ ($a=1,\ldots,4$). All of the fields transform in the adjoint representation of the gauge group, and the scalars and fermions transform in the vector and spinor representations of a (global) $SO(6)$ R-symmetry group. With gauge group $SU(N)$ and coupling $g_\mt{YM}$, the Lagrangian may be written as
\begin{equation}
    \mathcal{L} = \operatorname{Tr} \left[ -\frac{1}{4} F_{\mu\nu}F^{\mu\nu} - \frac{1}{2} D_\mu \phi^I D^\mu \phi^I + \frac{g_\mt{YM}^2}{4} [\phi^I,\phi^J]^2 + \ii\, \bar\psi\, \bar\sigma^\mu D_\mu \psi
    + g_\mt{YM}\, \bar\psi\, \Gamma^I [\phi^I,\psi] \right]\, .
\label{eq:N4SYM}
\end{equation}
Again, this theory is exactly conformal, \ie the beta function vanishes identically, for all values of $N$ and $g_\mt{YM}$, \eg see \cite{Marcus:1984ei,Howe:1983sr}.

As a simple example, a large class of scalar primary operators in $\mathcal{N}=4$ SYM is given by single-trace operators built from the six scalars, \eg \cite{Heslop:2003xu},
\begin{equation}
    \mathcal{O}^{(k)}(t,x)
    = t_{I_1\cdots I_k}\,\operatorname{Tr}\!\left[\phi^{I_1}(x)\cdots \phi^{I_k}(x)\right]\, ,
    \label{operator4}
\end{equation}
which have classical scaling dimension $\Delta_{\rm cl} = k$ (\ie when $g_\mt{YM}=0$). Here, $t_{I_1\cdots I_k}$ is a complex constant tensor. Quantum corrections generally shift the conformal dimension, $\Delta = k + \gamma$, where $\gamma$ is the anomalous dimension which depends on the value of coupling $g_\mt{YM}$. A special class of these operators \eqref{operator4} in $\mathcal{N}=4$ SYM are those where the polarisation is symmetric and traceless on any pair indices. For these so-called $1/2$-BPS operators, $\gamma=0$ for any value of $g_\mt{YM}$ because supersymmetry forbids them from acquiring an anomalous dimensions.

The examples above illustrate both ends of the spectrum of possibilities: massless free fields furnish simple realisations of CFTs, whereas superconformal gauge theories give rise to rich families of interacting CFTs with extensive operator spectrum. The general CFT framework outlined in this appendix encompasses all such cases.

\subsection*{Holographic CFTs}

A remarkable development in the study of strongly coupled quantum field theories was the discovery of the
AdS/CFT correspondence, originally proposed in \cite{Maldacena:1997re,Witten:1998qj,Gubser:1998bc}.  
In its most familiar form, this duality asserts an exact equivalence between certain 
conformal field theories in $d$ dimensions and string theory or quantum gravity defined on 
$(d{+}1)$-dimensional anti-de Sitter (AdS) spacetime. The correspondence is holographic: the CFT lives in one dimension less on the conformal boundary of the AdS
spacetime.

The AdS/CFT correspondence is a broad and far-reaching subject and comprehensive reviews can be found in \cite{Aharony:1999ti,Ammon:2015wua}. In this brief introduction, we focus on the canonical example of the correspondence: the duality between $\mathcal{N}=4$ supersymmetric Yang-Mills theory \eqref{eq:N4SYM} in four dimensions with gauge group $SU(N)$ and type~IIB string theory on $\mathrm{AdS}_5\times S^5$. This duality arises from two equivalent descriptions of the low-energy physics for $N$ coincident D3-branes in type~IIB string theory \cite{Maldacena:1997re}.  The worldvolume theory on the D3-branes is $\mathcal{N}=4$ SYM, while the near-horizon geometry produced by the same D3-branes is $\mathrm{AdS}_5\times S^5$ supported by $N$ units of flux of the self-dual Ramond-Ramond five-form.  

The first indication of a duality between these two theories arises from their symmetry structures.  As a four-dimensional CFT, $\mathcal{N}=4$ SYM enjoys the conformal symmetry group $SO(2,4)$ and, as discussed above, possesses a global $SO(6)$ $R$-symmetry. In the dual $\mathrm{AdS}_5\times S^5$ description, these symmetries are realised geometrically: $SO(2,4)$ appears as the isometry group of $\mathrm{AdS}_5$ (see below), while $SO(6)$ is identified with the rotational symmetry of the $S^5$.

The $\mathrm{AdS}_5$ and $S^5$ geometries share a common radius of curvature, which we denote by $\LA$.  Another key element of the AdS/CFT dictionary relates this geometric scale to the parameters of the dual $\mathcal{N}=4$ SYM theory. In particular, one finds
\begin{equation}
    \lambda = g_{\mt{YM}}^2 N = \frac{\LA^4}{\ell_s^4},
    \qquad\qquad
    N = 2\pi^{5/2}\,\frac{\LA^4}{\ell_{\mt{Planck}}^4}\,,
    \label{eq:houseonfire}
\end{equation}
where $\ell_s$ is the string length (\ie the characteristic size of a string state), and the ten-dimensional Planck length is related to the Newton constant of type~IIB string theory by $\ell_{\mt{Planck}}^8 = 8\pi G_{10}$.  On the field-theory side, $\lambda$ is the ’t~Hooft coupling and $N$ is the rank of the $SU(N)$ gauge group.

Equation~\eqref{eq:houseonfire} naturally motivates consideration of the ’t~Hooft limit of the boundary theory \cite{tHooft:1973alw}.  Taking $N\to\infty$ ensures that the curvature scale of the bulk geometry is much larger than the Planck scale, thereby suppressing quantum gravity corrections.  Likewise, taking the ’t~Hooft coupling $\lambda$ to be large guarantees $\LA \gg \ell_s$, which suppresses stringy corrections.  In this combined limit then, type~IIB string theory on $\mathrm{AdS}_5\times S^5$ reduces to (semi)classical supergravity, and the corresponding bulk fields become weakly coupled, with connected
correlators suppressed by powers of $1/N$.

To proceed further, let us consider a general $(d+1)$-dimensional AdS geometry with curvature scale $\LA$.  In Poincar\'e coordinates, the metric may be written as
\begin{equation}
    \dd s^2
    = \frac{\LA^2}{z^2}\left(\dd z^2 + \eta_{\mu\nu}\,\dd \mathsf{x}^\mu \dd \mathsf{x}^\nu\right),
    \qquad z>0,
    \label{eq:AdSmetric}
\end{equation}
where $\eta_{\mu\nu}$ is the $d$-dimensional Minkowski metric and the asymptotic boundary is located at $z\to0$.  Eq.~\eqref{eq:AdSmetric} is manifestly invariant under the Poincar\'e transformations of the boundary spacetime (\ie Lorentz transformations and translations acting on the $\mathsf{x}^\mu$ coordinates). In addition, one may verify that the metric is invariant under the
following coordinate transformations:
\begin{equation}
    (z,\mathsf{x}^\mu) \mapsto (\lambda\, z,\lambda\, \mathsf{x}^\mu),
    \qquad
    (z,\mathsf{x}^\mu) \mapsto
    \left(
        \frac{z}{1 - 2\, b\!\cdot\!\mathsf{x} + b^{2}\bigl(\mathsf{x}^{2}+z^{2}\bigr)},
        \frac{\mathsf{x}^\mu - b^\mu \bigl(\mathsf{x}^{2}+z^{2}\bigr)}
             {1 - 2\, b\!\cdot\!\mathsf{x} + b^{2}\bigl(\mathsf{x}^{2}+z^{2}\bigr)}
    \right).
    \label{eq:bulkspec}
\end{equation}
The first transformation implements the dilatation symmetry of the boundary theory, while
the second realises the special conformal transformation, reducing to
eq.~\eqref{eq:extrans} in the limit $z\to0$.  This explicitly demonstrates that
AdS$_{d+1}$ is invariant under the same $SO(2,d)$ symmetry group as the dual CFT$_d$
defined on its boundary.

Bulk fields propagating in the AdS geometry \eqref{eq:AdSmetric} are dual to local operators
in the boundary CFT.  As a simple illustration, consider a scalar field
$\Phi(z,\mathsf{x})$ with mass $m$ satisfying the Klein-Gordon equation in the
AdS$_{d+1}$ bulk:
\begin{equation}
    0=\left(\Box - m^2\right)\Phi
    = \frac{z^{d+1}}{\LA^{d+1}}\,\partial_z\!\left(\frac{\LA^{d-1}}{z^{d-1}}\partial_z\Phi\right)
      + \frac{z^2}{\LA^{2}}\,\eta^{\mu\nu}\partial_\mu\partial_\nu\Phi
      - m^2\,\Phi\,.
\end{equation}
Near the boundary ($z\to0$), the dependence of the solutions on the boundary coordinates
$\mathsf{x}$ becomes subleading, and one may look for power-law behaviour of the form
$\Phi \sim z^\Delta$.  Substituting this ansatz into the equation above yields the
relation
\begin{equation}
    \Delta(\Delta - d) = m^2 \LA^{2}\,.
    \label{eq:mass-dimension}
\end{equation}
The root
\(
\Delta = \tfrac{d}{2} + \sqrt{\tfrac{d^{2}}{4} - m^{2}\LA^{2}}
\)
corresponds to the conformal dimension of the dual scalar primary operator
$\mathcal{O}(\mathsf{x})$.  The resulting near-boundary expansion of $\Phi$ takes the form
\begin{equation}
    \Phi(z,\mathsf{x})
    \sim z^{d-\Delta}\,\phi_0(\mathsf{x})
      + z^{\Delta}\,\phi_1(\mathsf{x})\,,
    \label{eq:asymp-modes}
\end{equation}
where the coefficient $\phi_0(\mathsf{x})$ of the so-called non-normalisable mode acts as a
source for $\mathcal{O}(\mathsf{x})$, while the coefficient $\phi_1(\mathsf{x})$ of the
normalisable mode encodes the expectation value
$\langle\mathcal{O}(\mathsf{x})\rangle$
\cite{Witten:1998qj,Gubser:1998bc,Skenderis:2002wp}.  One may further verify that under the
bulk transformation in eq.~\eqref{eq:bulkspec}, which implements a boundary dilatation,
$\phi_0(\mathsf{x})$ and $\phi_1(\mathsf{x})$ scale with dimensions $d-\Delta$ and
$\Delta$, respectively, as expected for a source and its associated one-point function.

Correlation functions of the boundary operator $\mathcal{O}$ may be obtained directly from
the bulk two-point function of the dual scalar field
via the \emph{extrapolate dictionary} \cite{Banks:1998dd,Balasubramanian:1998de,Harlow:2011ke}. Consider the bulk Wightman function of a free scalar of mass $m$ in
Poincar\'e coordinates \eqref{eq:AdSmetric}, which takes the standard form \cite{DHoker:1998ecp}
\begin{equation}
    \langle\, \Phi(t,\bm{x},z)\, \Phi(t',\bm{x}',z')
    \,\rangle \;=\;\lim_{\epsilon \to 0^+}
\mathcal N_{\Delta}\ 
\xi^{\Delta}\ 
{}_2F_1\!\left(
\Delta,\ \Delta-\frac d2+1;\ 2\Delta-d+1;\ \xi^{2}
\right),
\label{eq:wightman-exact}
\end{equation}
with $\xi$ defined (including the Wightman $i\epsilon$ prescription) by
\begin{equation}
\xi \;=\;
\frac{2\, z\, z'}{z^2+z'^2 - (t-t'-i\epsilon)^2 + |\vec x-\vec x'|^2}\, .
\label{eq:xixi}
\end{equation}
Further, $\mathcal N_\Delta$ is a normalisation constant\footnote{Using the conventions of \cite{DHoker:1998ecp}, $\mathcal N_\Delta
=
\tfrac{2^\Delta\,\Gamma(\Delta)\,\Gamma\left(\Delta-\tfrac{d-1}2\right)}
{(4\pi)^{(d+1)/2}\,\Gamma\left(2\Delta - d+1\right)}$.} and 
$\Delta$ fixed by eq.~\eqref{eq:mass-dimension}.  The boundary two-point function is obtained by taking both points to the boundary and
rescaling by the appropriate powers of $z$ and $z'$ to remove the scaling of $\phi_1(\mathsf{x})$ 
\begin{equation}
    \langle \mathcal{O}_{\Delta}(t, \bm{x}) \mathcal{O}_{\Delta}(t', \bm{x}') \rangle
     =
    2^{-\Delta}\mathcal N_\Delta^{-1} \lim_{z,z'\to0} z^{-\Delta} z'^{-\Delta}\,
    \langle\, \Phi(t,\bm{x},z)\, \Phi(t',\bm{x}',z')
    \,\rangle\,.
    \label{eq:2pt-limit}
\end{equation}
We observe that with $z,z'\to 0$, $\xi\to 0$ and hence $\langle \Phi(t,\bm{x},z)\, \Phi(t',\bm{x}',z')\rangle\sim \xi^\Delta\, (1 + \mathcal O(\xi^2))$.
Hence substituting eq.~\eqref{eq:wightman-exact},
one recovers the CFT result \eqref{eq:CFT_2PF}, 
\begin{equation}
\label{eq:CFT_2PF22}
\langle \mathcal{O}_{\Delta}(t, \bm{x}) \mathcal{O}_{\Delta}(t', \bm{x}') \rangle \ =\ \lim_{\epsilon \to 0^+}
\frac{1}{[-(t-t' + \ii \epsilon)^2 + (\bm{x}-\bm{x}')^2]^\Delta}\,.
\end{equation}
Thus in a holographic CFT, the boundary two-point
function \eqref{eq:CFT_2PF} can be understood as the boundary limit of the
bulk Wightman function of a free scalar field in AdS with mass and conformal dimension
related by eq.~\eqref{eq:mass-dimension}.

To close, let us return to the operators introduced in eq.~\eqref{operator4} for $\mathcal{N}=4$ SYM.  The six real scalars transform in the vector representation of the
$SO(6)$ $R$-symmetry.  Accordingly, these operators \eqref{operator4} may be organised into
irreducible $SO(6)$ representations arising in the tensor product of $k$ vectors, as
determined by the properties of the polarisation tensor $t_{I_1\cdots I_k}$.  In
particular, choosing $t_{I_1\cdots I_k}$ to be completely symmetric and traceless (for BPS
operators) projects onto the rank-$k$ symmetric (traceless)
representation of $SO(6)$.
The bulk description of these operators arises from the Kaluza--Klein (KK) reduction of
type~IIB supergravity on $S^5$, \eg \cite{Kim:1985ez,Gunaydin:1984fk,Lee:1998bxa}.  For
the traceless symmetric representations, the spherical harmonics on $S^5$ are labelled by a positive
integer $k$ (the principal angular-momentum quantum number).  This label fixes the KK mass
as $m^2 \LA^2 = k(k-4)$ which, through eq.~\eqref{eq:mass-dimension}, yields the expected
conformal dimension $\Delta = k$.

To close, let us return to the operators introduced in \eqref{operator4} for
$\mathcal{N}=4$ SYM.  The six real scalars transform in the vector representation of the
$SO(6)$ $R$-symmetry.  In general then, the operators \eqref{operator4} can be organised to transform in 
irreducible $SO(6)$ representations appearing in the tensor product of $k$ vectors, as
selected by the properties of the polarisation tensor $t_{I_1\cdots I_k}$.  In
particular, taking $t_{I_1\cdots I_k}$ to be completely symmetric (and, for BPS operators,
also traceless) projects onto the rank-$k$ symmetric (traceless) representation of
$SO(6)$. Their dual description in the bulk arises from the Kaluza--Klein reduction of type~IIB
supergravity on $S^5$, \eg \cite{Kim:1985ez,Gunaydin:1984fk,Lee:1998bxa}.  For the symmetric representations, the spherical harmonics on the $S^5$ are labelled by a
positive integer $k$ (the principal angular-momentum quantum number).  This label fixes
the KK mass via $m^2 L^2 = k(k-4)$, which is consistent with eq.~\eqref{eq:mass-dimension}
and yields the conformal dimension $\Delta = k$.

In summary, $\mathcal{N}=4$ SYM provides a concrete example of a strongly coupled CFT admitting a dual description in terms of weakly coupled (super)gravity theory on a higher-dimensional curved spacetime. The AdS geometry and its Kaluza--Klein structure encode the symmetries, spectrum and correlation functions of the boundary CFT, offering a useful geometric
perspective on conformal dynamics.

\section{\texorpdfstring{$\mathcal{L}_{\mathrm{AA}}$ for integer and half-integer values of $\Delta$}{LAA for integer and half-integer values of Delta}}
\label{appx:hadamard_finite_part}

As mentioned in the main text, the two-point correlator is not an ordinary, but rather a generalised function. Thus, when computing the reduced final density matrix $\hat{\rho}_{\A\B}$, one finds that all terms ($\mathcal{L}_{\A\A}$, $\mathcal{L}_{\A\B}$, and $\mathcal{M}$) contain integrals that should be interpreted in a distributional sense, not in the Lebesgue sense. In order to compute those integrals in position space, we must resort to distribution theory techniques. In this appendix, we review some important concepts of this theory and apply it to $\mathcal{L}_{\A\A}$ term, since for this term we can obtain a closed form final expression. In the discussion below, we closely follow Chapter 39 of reference~\cite{Barata:NotasFisMat}, complemented by Chapter 1 of reference~\cite{Gelfand:1964}.

To motivate the following, recall the expression for $\LL_{\A\A}$ given in eq.~\eqref{eq:diag22}:
\begin{align}\label{eq:repLAA}
\LL_{\A\A} 
&= \bar{\lambda}^2 T^{2\Delta - 1} \, \sqrt{\frac{\pi}{2}} \int_{-\infty}^{\infty} \dd v \, e^{-\ii\Omega v} e^{-\frac{v^2}{2T^2}} \bigg( \lim_{\epsilon \to 0^+} \frac{1}{[-(v - \ii \epsilon)^2]^\Delta} \bigg).
\end{align}
We will be interested in prescriptions for dealing with distributions in the form of the one between the parentheses. As shall be clear later in the text, extra care should be taken when dealing with cases where $\Delta \in \frac{1}{2}\mathbb{Z}_+$.

\subsection*{Principal value prescription}
Let $B(y, r)$ be the open ball of radius $r>0$ centred at $y \in \mathbb{R}^n$, so that the closed set $B(y, r)^c = \mathbb{R}^n \setminus B(y, r)$ is its complement. Consider a function $f(x)$ that is singular in $y$ but integrable in the sets $B(y, r)^c$ for all $r>0$. If the limit
\begin{align}
    \lim_{r\to 0} \int_{B(y, r)^c} \dd^n x \; f(x)
\end{align}
exists, it is defined to be the \textbf{Cauchy principal value} of the integral $\int_{\mathbb{R}^n} \dd^n x \; f(x)$. That is,
\begin{align}
    \text{PV} \left(\int_{\mathbb{R}^n} \dd^n x \; f(x)\right) \coloneqq \lim_{r\to 0} \int_{B(y, r)^c} \dd^n x \; f(x)
\end{align}
provided that the limit exists. In this paper, we are only interested in the one-dimensional integral. In this case,
\begin{align}
    \text{PV} \left(\int_{-\infty}^\infty \dd x \; f(x)\right) \coloneqq \lim_{r\to 0^+} \left(\int_{-\infty}^{y-r} \dd x \; f(x) + \int_{y+r}^{\infty} \dd x \; f(x)\right).
\end{align}
Based on the definition above, for a fixed $y \in \mathbb{R}$, we can define the \textbf{Cauchy principal value distribution} $\text{PV}\left(\frac{1}{x - y}\right)$ by
\begin{align}
    \left\langle \text{PV}\left(\frac{1}{x - y}\right), \varphi \right\rangle \coloneqq \text{PV} \left(\int_{-\infty}^\infty \dd x \; \frac{1}{x-y} \varphi(x)\right),
\end{align}
where $\varphi(x)$ is a test function. Notice the following property given a test function $\varphi(x)$:
\begin{align}
    \int_{-\infty}^{y-r} \dd x \; \frac{\varphi(x)}{x-y} + \int_{y+r}^{\infty} \dd x \; \frac{\varphi(x)}{x-y} &= \int_{-\infty}^{-r} \dd z \; \frac{\varphi(y+z)}{z} + \int_{r}^{\infty} \dd z \; \frac{\varphi(y+z)}{z} \nonumber\\
    &= \int_{r}^{\infty} \dd z \; \frac{\varphi(y+z) - \varphi(y-z)}{z},
\end{align}
so that
\begin{align} \label{eq:propertyPV}
    \left\langle \text{PV}\left(\frac{1}{x - y}\right), \varphi \right\rangle = \lim_{r\to 0^+}\int_{r}^{\infty} \dd z \; \frac{\varphi(y+z) - \varphi(y-z)}{z},
\end{align}
which is well defined because $\varphi(x)$ is differentiable.

Some comments are in place. First, the integral over the whole space need not converge. When it converges, it equals its principal value. Second, we can generalise the definition to the case in which we have multiple simple poles, so long as the same conditions are met for each pole. Finally, this prescription may not work for higher order poles. Indeed, in general, the principal value will not converge for higher order poles, thus requiring another prescription to handle those cases. To tackle those cases, we introduce the Hadamard finite part.

\subsection*{Hadamard finite part prescription}
Using the same definitions as before, suppose that $f: \mathbb{R}^n \mapsto \mathbb{C}$ is integrable in the sets $B(y, r)^c$ for all $r>0$, and further that for all $r>0$,
\begin{align}
    \int_{B(y, r)^c} \dd^n x \; f(x) = F(r) + D(r),
\end{align}
such that $\lim_{r\to 0} F(r)$ exists and is finite, whereas $D(r)$ diverges as $r\to 0$. We would like to claim that the finite part of the integral in the whole space is
\begin{align}\label{eq:Hadamardfinitepart}
    \text{FP}\left(\int_{\mathbb{R}^n} \dd^n x \; f(x)\right)=\lim_{r\to 0} F(r),
\end{align}
but that can only make sense if the above decomposition into $F(r)$ and $D(r)$ is unique. For some classes of divergent functions, this can be achieved. In particular, this is the case for $f(x)=\frac{g(x)}{(x-y)^m}$, where $g(x) \in C^\infty$ and $m \in \mathbb{N}$. Therefore, based on the one-dimensional version of eq. \eqref{eq:Hadamardfinitepart}, we introduce the \textbf{Hadamard finite part distribution} for a fixed $y$, $\text{FP}\left(\frac{1}{(x - y)^m}\right)$, by
\begin{align}
    \left\langle \text{FP}\left(\frac{1}{(x - y)^m}\right), \varphi \right\rangle \coloneqq \text{FP} \left(\int_{-\infty}^\infty \dd x \; \frac{1}{(x-y)^m} \varphi(x)\right),
\end{align}
where $\varphi(x)$ is a test function. Notice, in particular, that for $m=1$, the Hadamard finite part recovers the Cauchy principal value.

One can also show that 
\begin{align}\label{eq:FPandPVrelation}
    \left\langle \text{FP}\left(\frac{1}{(x - y)^m}\right), \varphi \right\rangle = \frac{1}{(m-1)!} \left\langle \text{PV}\left(\frac{1}{x - y}\right), \varphi^{(m-1)} \right\rangle.
\end{align}
This equation, for the case $y=0, m=2$, together with eq. \eqref{eq:propertyPV} (also for $y=0$) allow us to derive the well-known relation:
\begin{align}
    \left\langle \text{FP}\left(\frac{1}{x^2}\right), \varphi \right\rangle &= \left\langle \text{PV}\left(\frac{1}{x}\right), \varphi' \right\rangle \nonumber\\
    &=\lim_{r\to 0^+}\int_{r}^{\infty} \dd z \; \frac{\varphi'(z) - \varphi'(-z)}{z} \nonumber\\
    &=\lim_{r\to 0^+}\int_{r}^{\infty} \dd z \; \frac{1}{z} \; \frac{\dd}{\dd z}\left[\varphi(z) + \varphi(-z)\right] \nonumber\\
    &= \lim_{r\to 0^+} \left(\frac{\varphi(z)+\varphi(-z)}{z} \Bigg{|}_{r}^\infty + \int_{r}^{\infty} \dd z \; \frac{\varphi(z) + \varphi(-z)}{z^2} \right) \nonumber \\
    &= \lim_{r\to 0^+} \left(-\frac{\varphi(r)+\varphi(-r)}{r} + \int_{r}^{\infty} \dd z \; \frac{\varphi(z) + \varphi(-z)}{z^2} \right) \nonumber \\
    &= \lim_{r\to 0^+} \left(-\frac{\varphi(r)-\varphi(0)}{r}-\frac{\varphi(-r)-\varphi(0)}{r} + \int_{r}^{\infty} \dd z \; \frac{\varphi(z) + \varphi(-z) - 2\varphi(0)}{z^2} \right) \nonumber \\
    &= -\varphi'(0)+\varphi'(0) +\lim_{r\to 0^+}  \int_{r}^{\infty} \dd z \; \frac{\varphi(z) + \varphi(-z) - 2\varphi(0)}{z^2} \nonumber \\
    &=\lim_{r\to 0^+}  \int_{r}^{\infty} \dd z \; \frac{\varphi(z) + \varphi(-z) - 2\varphi(0)}{z^2}.
\end{align}
This can be generalised to higher orders as
\begin{align}
    \left\langle \text{FP}\left(\frac{1}{x^m}\right), \varphi \right\rangle = \lim_{r\to 0^+}  \int_{r}^{\infty} \dd z \; \frac{\varphi_{m-1}(z) + (-1)^m\varphi_{m-1}(-z)}{z^m},
\end{align}
where
\begin{align}
    \varphi_{q}(z)\coloneqq \varphi(z) - \sum_{p=0}^q \frac{\varphi^{(p)}(0)}{p!} z^p.
\end{align}
Notice that the second term in this expression is the Taylor polynomial of order $q$, centered in $z=0$. The Hadamard finite part prescription above can be extended to any complex exponent (see~\cite{Gelfand:1964} for details). For the cases of $\Delta \notin \frac{1}{2}\mathbb{Z}_+$, the Hadamard finite part prescription suffices to compute $\mathcal{L}_{\A\A}$. The reason is that the singularity becomes a branch point, making the function multi-valued and requiring the introduction of a branch cut. In this case, the distributional structure is often simpler to analyse because the singular behaviour is encoded in the discontinuity across the cut. However, in the cases of integer and half-integer $\Delta$, the distribution has isolated poles. To deal with their singular behaviour, we must also add an imaginary term that is proportional to Dirac delta distributions or derivatives thereof. Here, a general version of the Sokhotski–Plemelj theorem will prove handy. We review it below.

\subsection*{Sokhotski–Plemelj theorem}
To start, there is a well known identity, the Breit-Wigner formula, which states that\footnote{The proof relies on considering families of functions like the one in the left-hand side, and showing that for all of them the integral over the whole space is 1, and when we integrate in the whole space except for an open ball around the point $x_0$ the integral vanishes, which are precisely the properties of delta functions.}
\begin{align}
    \lim_{\epsilon\to 0} \frac{1}{\pi} \frac{\epsilon}{(x-x_0)^2 + \epsilon^2} = \delta(x-x_0).
\end{align}
Moreover, note that
\begin{align}
    \lim_{\epsilon \to 0}\int_{-\infty}^\infty \dd x\; \frac{x-x_0}{(x-x_0)^2+\epsilon^2} \; \varphi(x)&= \lim_{\epsilon \to 0} \int_{-\infty}^\infty \dd z\; \frac{z}{z^2+\epsilon^2} \; \varphi(z+x_0) \nonumber \\
    &= \lim_{\epsilon \to 0} \int_{-\infty}^\infty \dd z\; \frac{z^2}{z^2+\epsilon^2} \; \frac{\varphi(z+x_0)}{z} \nonumber \\
    &= \lim_{\epsilon \to 0} \int_0^\infty \dd z\; \frac{z^2}{z^2+\epsilon^2} \; \frac{\varphi(z+x_0)-\varphi(-z+x_0)}{z} \nonumber \\
    &= \int_0^\infty \dd z\; \left(\lim_{\epsilon \to 0}\frac{z^2}{z^2+\epsilon^2}\right) \; \frac{\varphi(z+x_0)-\varphi(-z+x_0)}{z} \nonumber \\
    &= \int_0^\infty \dd z \; \frac{\varphi(z+x_0)-\varphi(-z+x_0)}{z} \nonumber \\
    &=     \left\langle \text{PV}\left(\frac{1}{x - x_0}\right), \varphi \right\rangle.
\end{align}
Therefore, $\lim_{\epsilon\to 0} \frac{x-x_0}{(x-x_0)^2+\epsilon^2} = \text{PV}\left(\frac{1}{x - x_0}\right)$. With all of that, we can easily see the\\Sokhotski-Plemelj theorem:
\begin{align}
    \lim_{\epsilon \to 0}\frac{1}{(x-x_0) \pm \ii \epsilon} &=  \lim_{\epsilon \to 0} \frac{x-x_0}{(x-x_0)^2+\epsilon^2} \mp \ii  \lim_{\epsilon \to 0} \frac{\epsilon}{(x-x_0)^2+\epsilon^2} \nonumber \\
    &=  \text{PV}\left(\frac{1}{x - x_0}\right) \mp \ii \pi \delta(x-x_0).
\end{align}
Now, to obtain its generalisation, we take derivatives of the above distribution and use the identity
\begin{align}
    \frac{\dd^m}{\dd x^m} \text{PV}\left(\frac{1}{x-x_0}\right) &= (-1)^m m! \; \text{FP} \left(\frac{1}{(x-x_0)^{m+1}}\right),
\end{align}
yielding
\begin{align}\label{eq:S-Pgeneral}
     \lim_{\epsilon \to 0}\frac{1}{[(x-x_0) \pm \ii \epsilon]^m} &= \lim_{\epsilon \to 0}\frac{(-1)^{m-1}}{(m-1)!} \frac{\dd^{m-1}}{\dd x^{m-1}}\left[\frac{1}{(x-x_0) \pm \ii \epsilon}\right] \nonumber\\
    &= \frac{(-1)^{m-1}}{(m-1)!} \frac{\dd^{m-1}}{\dd x^{m-1}}\left[\lim_{\epsilon \to 0} \frac{1}{(x-x_0) \pm \ii \epsilon}\right] \nonumber\\
    &= \frac{(-1)^{m-1}}{(m-1)!} \frac{\dd^{m-1}}{\dd x^{m-1}}\left[\text{PV}\left(\frac{1}{x - x_0}\right) \mp \ii \pi \delta(x-x_0)\right] \nonumber\\
    &= \frac{(-1)^{m-1}}{(m-1)!}\left[(-1)^{m-1} (m-1)! \; \text{FP} \left(\frac{1}{(x-x_0)^{m}}\right) \mp \ii \pi \delta^{(m-1)}(x-x_0)\right] \nonumber\\
    &= \text{FP} \left(\frac{1}{(x-x_0)^{m}}\right) \mp \ii \pi \frac{(-1)^{m-1}}{(m-1)!} \delta^{(m-1)}(x-x_0).
\end{align}
This expression can be used to compute $\LL_\mt{AA}$ in the special cases when $\Delta \in \frac{1}{2}\mathbb{Z^+}$. Let us show some examples. First, for integers, we write $\Delta = n$, where n is a positive integer number, in which case the distribution in $\LL_\mt{AA}$ reduces to 
\begin{align}
    \lim_{\epsilon \to 0^+} \frac{1}{[-(v - \ii \epsilon)^2]^n}&= \lim_{\epsilon \to 0^+}\frac{(-1)^n}{(v - \ii \epsilon)^{2n}}=(-1)^n \ \text{FP} \left(\frac{1}{v^{2n}}\right) - \ii \pi \, \frac{(-1)^{n}}{(2n-1)!} \delta^{(2n-1)}(v).
\end{align}
Now, using equations \eqref{eq:propertyPV} and \eqref{eq:FPandPVrelation}, we can write
\begin{align}\label{eq:FPusingPVderiv}
    \left\langle \text{FP}\left(\frac{1}{v^{2\Delta}}\right), \varphi \right\rangle = \frac{1}{(2\Delta-1)!} \lim_{r\to 0^+}\int_{r}^{\infty} \dd z \; \frac{\varphi^{(2\Delta-1)}(z) - \varphi^{(2\Delta-1)}(-z)}{z}.
\end{align}
This equation holds both for integer and half-integer values $\Delta$.
In the case of positive integer $\Delta=n$, we are only interested in odd order derivatives of $\varphi$. So, by defining
\begin{align}\label{eq:faux}
    f(v)\coloneqq \varphi(v) + \varphi(-v),
\end{align}
we obtain
\begin{align}
    f^{(2n -1)}(v)\coloneqq \varphi^{(2n -1)}(v) - \varphi^{(2n -1)}(-v).
\end{align}
In the case of $\LL_{\A\A}$, the test function reads (from eq. \eqref{eq:diag22}, also reproduced in eq. \eqref{eq:repLAA})
\begin{align}
    \varphi(v)=e^{-\frac{v^2}{2 T^2}-\ii \Omega v},
\end{align}
yielding
\begin{align}
    f(v)=2 e^{-\frac{v^2}{2 T^2}} \cos (v \Omega ).
\end{align}

Therefore, we obtain: 
\begin{align}
    \frac{\LL_{\A\A}}{\bar{\lambda}^2} 
    &=  T^{2\Delta - 1} \, \sqrt{\frac{\pi}{2}} \frac{(-1)^\Delta}{(2\Delta-1)!} \left(\lim_{r\to 0^+}\int_{r}^{\infty} \dd z \; \frac{f^{(2\Delta-1)}(z)}{z}
    - \ii \pi \int_{-\infty}^{\infty} \delta^{(2\Delta-1)}(v) \varphi(v)\right) \nonumber\\
    &= T^{2\Delta - 1} \, \sqrt{\frac{\pi}{2}} \frac{(-1)^\Delta}{(2\Delta-1)!} \left(\lim_{r\to 0^+}\int_{r}^{\infty} \dd z \; \frac{f^{(2\Delta-1)}(z)}{z}
    + \ii \pi \varphi^{(2\Delta-1)}(0)\right).
\end{align}

Let us evaluate this expression for a few values of $\Delta$.
For $\Delta=1$,
\begin{align}
    f^{'}(v)&=-2 \Omega  e^{-\frac{v^2}{2 T^2}} \sin (v \Omega )-\frac{2 v e^{-\frac{v^2}{2 T^2}} \cos (v \Omega )}{T^2}\nonumber\\
    \varphi'(0)&=-\ii\Omega,
\end{align}
so that we obtain
\begin{align}
    \frac{\LL_{\A\A}^{\Delta=1}}{\bar{\lambda}^2} 
    &= T \, \sqrt{\frac{\pi}{2}} \lim_{r\to 0^+}\int_{r}^{\infty} \dd v \; \left( \frac1v  2 \Omega  e^{-\frac{v^2}{2 T^2}} \sin (v \Omega )+\frac{2 v e^{-\frac{v^2}{2 T^2}} \cos (v \Omega )}{T^2}
    - \pi \Omega \right)\nonumber\\
    &= T \, \sqrt{\frac{\pi}{2}} \left(\pi \Omega  \;\text{erf}\left(\frac{T \Omega }{\sqrt{2}}\right)+\frac{\sqrt{2 \pi } e^{-\frac{1}{2} T^2 \Omega ^2}}{T} - \pi \Omega\right)\\
    &= \pi e^{-\frac{1}{2} T^2 \Omega ^2} - \pi^{3/2} \frac{T\Omega}{\sqrt{2}} \,  \;\text{erfc}\left(\frac{T \Omega }{\sqrt{2}}\right).
\end{align}

For $\Delta=2$:
\begin{align}
    f^{(3)}(v)&=2 e^{-\frac{v^2}{2 T^2}} \Big(\frac{\Omega}{T^4}  \left(T^4 \Omega ^2+3(T^2-v^2)\right) \sin (v \Omega ),\\
    &\hspace{5cm}- \frac{v}{T^6} \left(v^2-3T^2 (T^2 \Omega ^2+1)\right) \cos (v \Omega )\Big)\\
    \varphi^{(3)}(0)&=\ii \Omega \left(\frac{3}{T^2}+\Omega^2\right),
\end{align}
so that we obtain
\begin{align}
    \frac{\LL_{\A\A}^{\Delta=2}}{\bar{\lambda}^2} 
    &= \frac{T^{3}}{6} \, \sqrt{\frac{\pi}{2}} \Bigg[- \pi \Omega \left(\frac{3}{T^2}+\Omega^2\right)\nonumber\\
    & \qquad + \lim_{r\to 0^+}\int_{r}^{\infty} \dd v \; \frac1v \bigg[ 2 e^{-\frac{v^2}{2 T^2}} \Big(\frac{\Omega}{T^4}  \left(T^4 \Omega ^2+3(T^2-v^2)\right) \sin (v \Omega ) \nonumber\\
    & \hspace{4.5cm} - \frac{v}{T^6} \left(v^2-3T^2 (T^2 \Omega ^2+1)\right) \cos (v \Omega )\Big) \bigg] \Bigg]\\
    &= \pi  e^{-\frac{1}{2} T^2 \Omega ^2}\; \frac{2+T^2 \Omega ^2}{6} - \pi ^{3/2} \frac{T \Omega}{\sqrt{2}}\; \text{erfc}\left(\frac{T \Omega}{\sqrt{2}}\right)\frac{3+T^2 \Omega ^2}{6}.
\end{align}

Similarly, we can compute $\LL_{\A\A}$ for other positive integer values of $\Delta=n$, and the results are of the form
\begin{equation}
\frac{\LL_{\A\A}}{\bar{\lambda}^2} 
=
\pi\ A_n(\tfrac12T^{2}\Omega^{2})\ e^{-\frac{1}{2}T^{2}\Omega^{2}}
-\pi ^{3/2} \frac{T \Omega}{\sqrt{2}} 
B_n(\tfrac12T^{2}\Omega^{2})\
{\rm erfc}\!\left(\frac{T\Omega}{\sqrt{2}}\right),
\label{eq:gamma4}
\end{equation}
where $A_n(x)$ and $B_n(x)$ are polynomials in $x$ of degree $n-1$. For the first few values of $n$, we have
\begin{align}
A_1(x) = 1,
&\qquad
B_1(x) = 1, \nonumber\\[5pt]
A_2(x) = \frac{x+1}{3},
&\qquad
B_2(x) = \frac{x}{3}+\frac{1}{2}, \label{eq:Alist}\\[5pt]
A_3(x) = \frac{x^{2}}{30}+\frac{3x}{20}+\frac{1}{15},
&\qquad
B_3(x) = \frac{x^{2}}{30}+\frac{x}{6}+\frac{1}{8}, 
\nonumber\\[5pt]
A_4(x) = \frac{x^{3}}{630}+\frac{x^{2}}{63}+\frac{29x}{840}+\frac{1}{105},
&\qquad
B_4(x) = \frac{x^{3}}{630}+\frac{x^{2}}{60}+\frac{x}{24}+\frac{1}{48}. \nonumber
\end{align}
\bigskip

Consider now half-integer values, and write $\Delta=n-\frac12$, where $n$ is a positive integer. In this case, we can write the distribution in $\LL_{\A\A}$ as
\begin{align}
    \lim_{\epsilon \to 0^+} \frac{1}{[-(v - \ii \epsilon)^2]^\frac{2n-1}{2}}&= \lim_{\epsilon \to 0^+}\frac{\ii \, (-1)^n}{(v - \ii \epsilon)^{2n - 1}}=\ii \,(-1)^n \ \text{FP} \left(\frac{1}{v^{2n-1}}\right) - \pi \, \frac{(-1)^{n}}{(2n-2)!} \delta^{(2n-2)}(v).
\end{align}

We can use this identity to compute closed-form expressions for half-integers, in a very similar way to the example shown for integers. Indeed, equation~\eqref{eq:FPusingPVderiv} applies here, and for half-integers we are only interested in even order derivatives of $\varphi(v)$. Therefore, by defining an auxiliary function similar to $f(v)$ in equation~\eqref{eq:faux}, but with a relative minus sign, we can take derivatives and deal with the Hadamard finite part. The contribution from the (derivative of the) $\delta(v)$ function is straightforward to compute. For half-integers, we can write the results in the form
\begin{equation}
\frac{\LL_{\A\A}}{\bar{\lambda}^2} 
=
- \pi\ \frac{T\Omega}{\sqrt{2}} \ \hat{A}_n(\tfrac12T^{2}\Omega^{2})\ e^{-\frac{1}{2}T^{2}\Omega^{2}}
+\pi^{3/2}\ 
\hat{B}_n(\tfrac12T^{2}\Omega^{2})\
{\rm erfc}\!\left(\frac{T\Omega}{\sqrt{2}}\right),
\end{equation}
where $\hat{A}_n(x)$ and $\hat{B}_n(x)$ are polynomials in $x$ of degree $n-2$ and $n-1$, respectively. For the first few values of $n$, we obtain
\begin{align}
\hat A_1(x) = 0,
&\qquad
\hat B_1(x) = \frac{1}{\sqrt{2}}, \nonumber\\[5pt]
\hat A_2(x) = \frac1{\sqrt{2}},
&\qquad
\hat B_2(x) = \frac1{2\sqrt{2}}\,(2x+1), 
\label{eq:Blist}\\[5pt]
\hat A_3(x) = \frac1{12\sqrt{2}}\,(2x+5),
&\qquad
\hat B_3(x) = \frac1{24\sqrt{2}}\,(4x^{2}+12x+3), \nonumber\\[5pt]
\hat A_4(x) = \frac1{360\sqrt{2}}\,(4x^{2}+28x+33),
&\qquad
\hat B_4(x) = \frac1{720\sqrt{2}}\,(8x^{3}+60x^{2}+90x+15).
\nonumber
\end{align} 

To conclude this appendix, recall that eq.~\eqref{eq:Lii} in the main text provides a general result for $\LL_{\A\A}$:
\begin{align}
    \frac{\LL_{\A\A}}{\bar{\lambda}^2} &= \frac{\pi^{3/2}}{2^{\Delta}} \left[\,\frac{
    {}_1F_1\!\left(\tfrac{1}{2} - \Delta, \tfrac{1}{2}, -\tfrac{1}{2} T^2 \Omega^2\right)
    }{\Gamma\!\left(\tfrac{1}{2}+\Delta\right)}
     -  \frac{\sqrt{2} \, T \Omega}{\Gamma(\Delta)}\,{}_1F_1\!\left(1 - \Delta, \tfrac{3}{2}, -\tfrac{1}{2} T^2 \Omega^2\right)\right]\,.\label{eq:Lii99}
\end{align}
To obtain this expression, we have already dealt with the subtleties that arise for the integer and half-integer cases of $\Delta$.  That is, this formula is valid for all $\Delta \in \mathbb{R}_+$. As expected, this result coincides with the ones shown above, which were obtained by using distribution theory.

\section{Derivation of asymptotic expressions}
\label{appx:leadingObehaviour}

In this appendix, we analyse asymptotic expansions of the density matrix elements derived in section \ref{sec:setup_density_matrix} to get approximations for certain parameter regimes, and compare to the results of our numerical analysis. There is a priori no reason one should be restricted to these parameter regimes in entanglement harvesting, however, the exact expressions for the density matrix elements are rather opaque, so the analysis in this section serves to provide intuition for their dominant behaviour in certain cases. Additionally, we use these approximations to derive approximations for the negativity and mutual information, as well as their communication/harvesting decompositions presented in section \ref{sec:separation}.

\subsection*{Density matrix elements}

We begin by analysing $\mathcal{L}_{\A\A}$ which is given in closed form in eq.~\eqref{eq:Lii}. We make use of the large-$x$ asymptotic behaviour of ${}_1F_1(a,b,-x)$ when $x>0$ (see section 13.7 of \cite{DLMF}):
\begin{align}
    {}_1F_1(a,b,-x) \ &=\ \frac{\Gamma(b)}{\Gamma(a)} e^{-x} x^{a-b}\left(1+\frac{(b-a)(1-a)}{x}+\OO(x^{-2})\right)
   \label{eq:hoot}\\
   &\qquad\qquad+ \frac{\Gamma(b)}{\Gamma(b-a)} x^{-a} \left(1-\frac{a(a-b+1)}{x}+\OO(x^{-2})\right).
    \nonumber
\end{align}
In our case, the power-law branch (beginning with $x^{-a}$) cancels term by term in the linear combination appearing in eq.~\eqref{eq:Lii}. Hence with $T\Omega \gg 1$ and the above expansion for the exponentially-suppressed branch, eq.~\eqref{eq:Lii} becomes
\begin{equation}
    \frac{\LL_{\A\A}}{\bar{\lambda}^2} \ =\ \LL_{\A\A}^{(1)} \bigg[\,1 - \frac{2\Delta(\Delta+\tfrac12)}{(T\Omega)^2}+
    \mathcal O\bigg(\frac{1}{(T\Omega)^4}\bigg)\bigg], \qquad \LL_{\A\A}^{(1)} = \frac{\pi}{(T\Omega)^{2\Delta}}\, e^{-\tfrac{1}{2} T^{2}\Omega^{2}}.
    \label{eq:largeOmega}
\end{equation}
Because we set $T \Omega = 10$ in the numerical results of this paper, we will take $T\Omega \gg 1$ to be satisfied throughout. Figure \ref{fig:Lii_Delta_analyticfit} plots $\LL_{\A\A}^{(1)}$ alongside the exact values from eq.~$\eqref{eq:Lii}$; we find that this provides a good fit over a wide range of the scaling dimension. However, the agreement begins to worsen for larger values of $\Delta$, as is expected from the expansion above, where the first subleading correction grows quadratically with $\Delta$.

\bigskip
Next, we consider the leading-order analysis of $\mathcal L_{\A\B}$. Starting from eq.~(\eqref{eq:Lij}), we complete the square in the exponential to identify the centre of the Gaussian envelope in the complex $v$-plane. This yields  
\begin{equation}
    \frac{\mathcal{L}_\textsc{ab}}{\bar{\lambda}^2}\ \simeq \ \lim_{\epsilon\to 0^+} T^{2\Delta -1} \sqrt{\frac{\pi}{2}} e^{-\frac12 T^2 \Omega^2 + \ii \Omega \delta} \int_{-\infty}^{\infty} \dd v \, e^{-\frac{(v + \delta + \ii T^2 \Omega)^2}{2T^2}}  \frac{1}{(L^2 - v^2 + \ii \epsilon  v)^\Delta}.
\end{equation}
Next, we shift the integration variable to the centre of the Gaussian, $\tv \equiv (v + \delta + \ii T^2 \Omega)/T$. This is a contour deformation in the complex plane that does not cross any poles for $L\gg T$, allowing the limit $\epsilon\to 0^+$ to be taken safely.
This yields
\begin{equation}
    \frac{\mathcal{L}_{\textsc{ab}}}{\bar{\lambda}^{2}}\ =\ \sqrt{\frac{\pi}{2}}\, e^{-\frac12 T^{2}\Omega^{2} + \ii\Omega\delta} \int_{-\infty}^{\infty} \dd \tv\; e^{-\frac12\tv^{2}} \frac{1}{(D + C \,\tv - \tv^{2})^{\Delta}},
    \label{eq:holler}
\end{equation}
where
\begin{equation}
    D\ \equiv\ \frac{L^{2} - \delta^{2}}{T^2} + T^{2}  \Omega^2 - 2\ii \Omega \delta, \qquad C\ \equiv\ 2\left(\frac\delta T + \ii\, T\Omega\right).
    \label{eq:D_def}
\end{equation}
We may rewrite the denominator as
\begin{equation}
    ( D + C \tv - \tv^{2})^{-\Delta}\ =\ D^{-\Delta}(1+\varepsilon)^{-\Delta}, \qquad \varepsilon\ \equiv\ \frac{C}{D}\,
    {\tv} - \frac{1}{ D}\,\tv^{2}.
    \label{eq:Lij_denom}
\end{equation}
Using the binomial expansion, we may approximate $(1+\varepsilon)^{-\Delta} \approx 1 - \Delta\varepsilon $ as long as $\varepsilon\ll 1$. Further, we may take $|\tv|\lesssim 1$ as the Gaussian factor $e^{-\tv^{2}/2}$ is concentrated in this region. Then $\varepsilon \ll 1$ is ensured by
\begin{equation}
    |D| \gg |C|, \qquad |D| \gg 1.
    \label{eq:Lij_approx_conditions}
\end{equation}
\begin{figure}
    \centering
    \includegraphics[width=\textwidth]{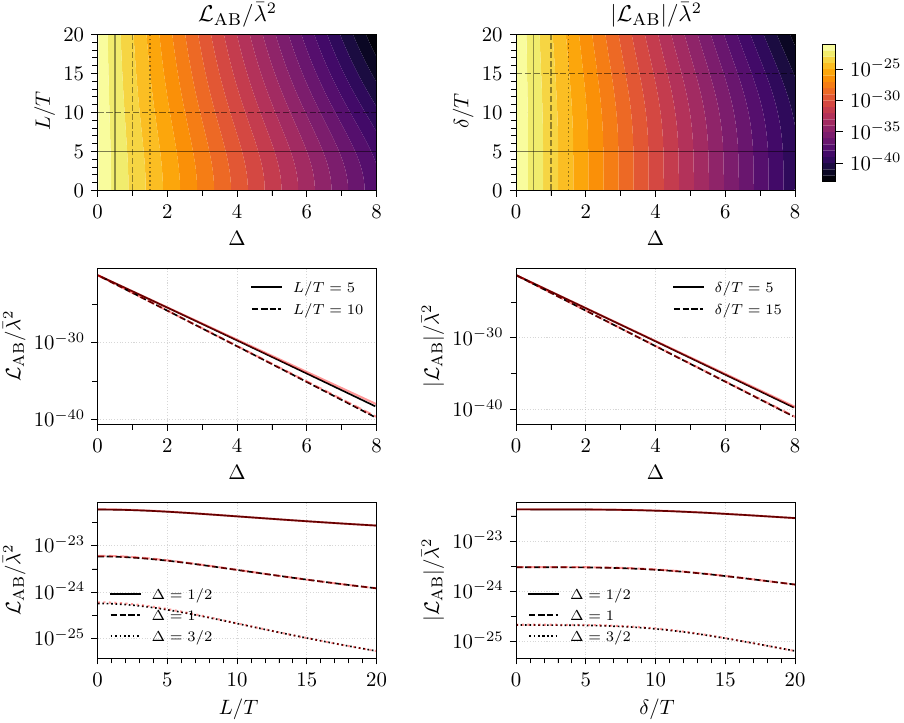}
    \caption{The plots above show $\mathcal L_\textsc{ab}/\bar{\lambda}^2$ as a function of $\Delta$, $L/T$, $\delta/T$, comparing the numerical analysis of eq.~\eqref{eq:Lij} with the approximation given by the leading-order term in $\tfrac1D$, given in eq.~\eqref{eq:LAB_leadingO} ($D$ is defined in eq.~\eqref{eq:D_def}). The top row of plots are the same as those shown in the main text in figure \ref{fig:Lab_L_and_delta}. Below them, we show cross-sections of the top plots at various fixed values. In the left column, we set $\delta/T = 0$ and vary $L/T$ and $\Delta$. In the right column, we set $L/T = 10$ and vary $\delta/T$ and $\Delta$. In all plots, we set $\Omega T = 10$. Subleading terms in the expansion grow with $\Delta$, explaining the larger deviations from the numerical results seen at higher values of $\Delta$.}
    \label{fig:Lab_L_and_delta_slices}
\end{figure}
When $\delta = 0$ for example (as on the left side of figures \ref{fig:Lab_L_and_delta} and \ref{fig:Lab_L_and_delta_slices}), these conditions can be simplified as $(L/T)^{2} + (T  \Omega)^2 \gg 1$ and $(L/T)^{2} + (T\Omega)^2 \gg 2T \Omega$. Assuming the conditions \eqref{eq:Lij_approx_conditions} hold, we expand \eqref{eq:Lij_denom} as
\begin{equation}
    (D + C \tv - \tv^{2})^{-\Delta}\ =\ \frac{1}{D^\Delta}\bigg[ 1 - \Delta\frac{C}{D}\,\tv + \Delta\left(\frac{D+\frac12(\Delta+1) C^2}{D^2}\right)\tv^2
    + \mathcal{O}\!\left(\frac{C}{D^2}\,\tv^{3},\,\frac{C^3}{D^3}\,\tv^{3}\right) \bigg].
\end{equation}
The linear term is odd in $\tv$ so its contribution to the integral \eqref{eq:holler} vanishes. Keeping the leading term and the $\tv^2$ terms the Gaussian integral yields
\begin{gather}
    \frac{\mathcal{L}_{\textsc{ab}}}{\bar{\lambda}^2} = \mathcal L_\textsc{ab}^{(1)} \left[1+\frac{\Delta}{D^2}\left(D+\tfrac12(\Delta+1) C^2\right)+\mathcal{O}\!\left(\frac{D^2,\,DC^2,\,C^4}{D^4}\right)\right],\label{eq:LAB_asymptotic}\\
    L_\textsc{ab}^{(1)} = \frac{\pi}{D^\Delta}\, e^{-\tfrac12 T^2\Omega^2 + \ii\Omega\delta}.
    \label{eq:LAB_leadingO}
\end{gather}
In figure \ref{fig:Lab_L_and_delta_slices}, we compare the leading-order exponential of $\mathcal{L}_{\textsc{ab}}/\bar{\lambda}^2$ with the results of numerical integration of eq.~\eqref{eq:Lij}. In the lower panels, we see that the leading-order term given in eq.~\eqref{eq:LAB_leadingO} provides a good approximation for a wide range of $L/T$, $\delta/T$ and $\Delta$. The approximation deviates more for larger values of $\Delta$, which is expected due to the fact that the subleading corrections grow with $\Delta$.

\bigskip

Finally, consider $\mathcal M$ in eq.~\eqref{eq:M_numerical}. To arrive at a similar large-$D$ expansion easily, we may observe that the integral in the final line is identical to that in eq.~\eqref{eq:Lij} if we set $\Omega=0$ in the latter. This allows us to evaluate the leading order behaviour for $\M$ using the results above for $\LL_{\A\B}$. This integral can be understood as the action of a distribution that contains singular values (the poles at $\tilde{v} = (\delta \pm L)/T$). However, as long as $\delta \pm L\gg T$, the contributions coming from the singular values are suppressed when acting on functions (like Gaussian) whose tails decay sufficiently fast. Therefore, it is a reasonable assumption---supported by comparison with the numerical results---that we may neglect the contributions from the singular parts. In far-from-lightcone regimes, we then have
\begin{gather}
    \frac{\mathcal{M}}{\bar\lambda^2}\approx\frac{\mathcal{M}_{\tilde{D}\scriptscriptstyle\gg 1} }{\bar{\lambda}^2}  = \mathcal{M}^{(1)}_{\tilde{D}\scriptscriptstyle\gg 1} \left[1+\frac{\Delta}{\tilde{D}^2}\left(\tilde{D}+2(\Delta+1) \frac{\delta^2}{T^2}\right)+\mathcal{O}\!\left(\frac{1}{\tilde{D}^2},\,\frac{1}{\tilde{D}^3}\frac{\delta^2}{T^2},\,\frac{1}{\tilde{D}^4}\frac{\delta^4}{T^4}\right)\right],
    \label{eq:holler3}\\
    \mathcal{M}^{(1)}_{\tilde{D}\scriptscriptstyle\gg 1} = -\frac{\pi}{\tilde{D}^\Delta}\, e^{-\tfrac12 T^2\Omega^2 + \ii\Omega\delta},\label{eq:M_D_first_order}
\end{gather}
where $\tilde{D} \equiv (L^{2} - \delta^{2})/{T^2}$. This equation is valid for
\begin{equation}
    |\tilde{D}| \gg 2|\delta|/T, \qquad |\tilde{D}| \gg 1.
    \label{eq:M_conditions}
\end{equation}
The conditions \eqref{eq:M_conditions} are certainly satisfied when either $|L|\gg|\delta|$ or $|\delta|\gg |L|$. However, the approximation above fails when $|L|\approx |\delta|$ as this makes $\tilde{D} \approx 0$. It is not surprising that special behaviour occurs around $|L|= |\delta|$ since at this point the centres of the switching functions are in light contact.

\begin{figure}
    \centering
    \includegraphics[width=\textwidth]{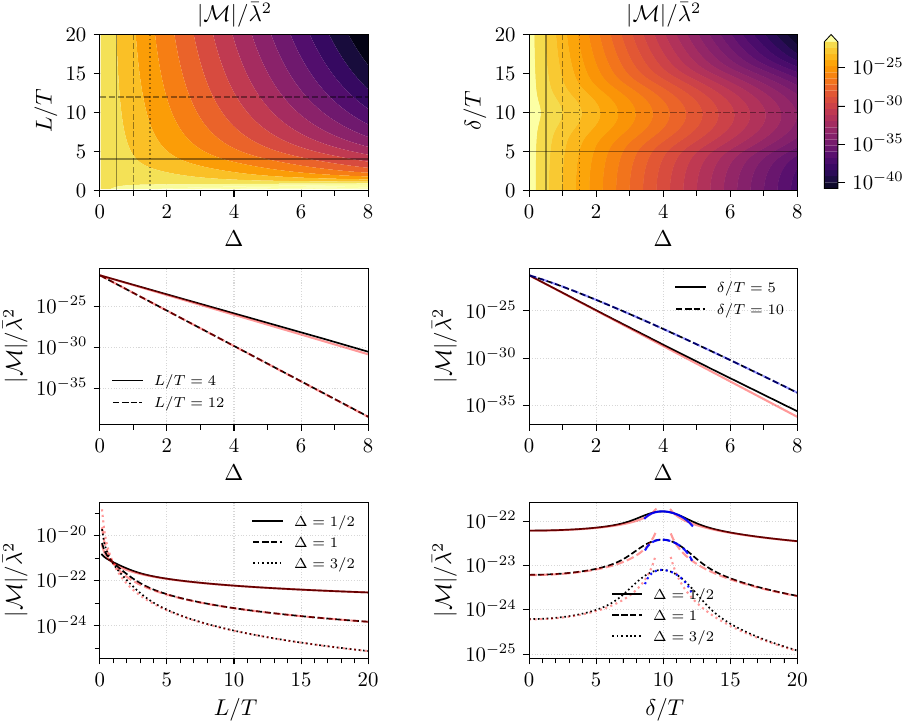}
    \caption{The plots above show $|\mathcal M|/\bar{\lambda}^2$ as a function of $\Delta$, $L/T$, $\delta/T$, comparing the numerical analysis of eq.~\eqref{eq:M_numerical} with closed-form approximations in certain asymptotic regimes. The top row of plots are identical to those shown in the main text in figure \ref{fig:M_L_and_delta}. Below them, in black, we show cross-sections of the top plots. In the left column, we set $\delta/T = 0$ and vary $L/T$ and $\Delta$. In the right column, we set $L/T = 10$ and vary $\delta/T$ and $\Delta$. In all plots, we set $\Omega T = 10$. In red, we plot the far-from-lightcone approximation given in eq.~\eqref{eq:M_D_first_order}, which holds for $|\tilde{D}| \gg 2|\delta|/T$ and $|\tilde{D}| \gg 1$, where $\tilde{D} \equiv (L^{2} - \delta^{2})/{T^2}$. In blue, we plot the near-lightcone approximation given in eq.~\eqref{eq:M_near_lightcone}, which holds for $\tilde{C} = 2 \delta/T \gg 1$.}
    \label{fig:M_L_and_delta_slices}
\end{figure}

To obtain a closed-form expression for the behaviour near the lightcone, can also consider the regime $\tilde C \equiv 2 |\delta|/T \gg1$ with $|\tilde D|\ll \tilde C$. We begin from the integral representation
\begin{equation}
    \frac{\mathcal{M}}{\bar\lambda^2}
    = -\sqrt{\frac{\pi}{2}}\,
    e^{-\frac12 T^2\Omega^2+\ii\Omega\delta}
    \int_{-\infty}^{\infty}\! \dd v\,e^{-v^2/2}\, \frac{1}{    (\tilde D+\tilde C\, v-v^2-\ii 0\,\tilde C)^\Delta}.
\end{equation}
In the regime $\tilde C\gg1$ with $|\tilde D|\ll \tilde C$, the term $\tilde C(v-\ii 0)$ dominates the denominator. Factoring this term out and expanding in powers of $1/\tilde C$ yields
\begin{align}
    (\tilde D+\tilde C &\, v-v^2-\ii 0\,\tilde C)^{-\Delta}
    =\nonumber \\
    &\tilde C^{-\Delta}(v-\ii 0)^{-\Delta}
    \Bigg[ 1 -\Delta\,\frac{\tilde D-v^2}{\tilde C\,(v-\ii 0)}
    +\frac{\Delta(\Delta+1)}{2}\, \frac{(\tilde D-v^2)^2}{\tilde C^2\,(v-\ii 0)^2} +\mathcal O(\tilde C^{-3})
    \Bigg].
\end{align}
The problem thus reduces to the evaluation of the tempered distributions $(v-\ii 0)^{-\Delta}$, $(v-\ii 0)^{-(\Delta+1)}$, and $(v-\ii 0)^{-(\Delta+2)}$ acting on a Gaussian test function. Evaluating them by splitting the integration domain into $v>0$ and $v<0$ provides closed-form expressions in terms of Gamma functions. To second order in $1/\tilde C$, we obtain
\begin{multline}
    \frac{\mathcal{M}}{\bar\lambda^2} \ \approx\ \frac{\mathcal{M}_{\tilde C \gg 1}^{(2)}}{\bar\lambda^2} \ =\ -\frac{\pi^{3/2}}{2^{\Delta/2} \, \Gamma\! \left(\frac{1+\Delta}{2} \right)}\; e^{-\frac12 T^2\Omega^2+ \ii \Omega\delta}\, e^{\frac{\ii\pi\Delta}{2}}\,
    \frac{1}{\tilde C^{\Delta}}
    \\
    \times \Bigg[ 1
    +\frac{\ii \,\Delta}{\sqrt{2}\,\tilde C}\,
    \frac{\Gamma\!\left(\frac{1+\Delta}{2}\right)}{\Gamma\!\left(1+\frac{\Delta}{2}\right)}\,
    (\tilde D+\Delta)
    +\frac{\Delta(\Delta+1)}{2\,\tilde C^2}
    \left( (1-\Delta)-2\tilde D-\frac{\tilde D^2}{1+\Delta} \right) +\mathcal O\left( \frac{1}{\tilde C^3} \right) \Bigg].
    \label{eq:M_C_expansion_full}
\end{multline}
Here the $O(\tilde C^{-1})$ correction is purely imaginary, corresponding to a phase shift of $\mathcal M$, while the $O(\tilde C^{-2})$ term provides the leading real correction to the magnitude. Thus, taking the absolute value, this becomes
\begin{multline}
    \frac{\left|\mathcal{M}_{\tilde C \gg 1}^{(2)}\right|}{\bar\lambda^2} \ =\ \frac{\pi^{3/2}}{2^{\Delta/2} \,\Gamma\!\left(\frac{1+\Delta}{2}\right)} e^{-\frac12 T^2\Omega^2}\, \tilde C^{-\Delta} \\
    \times \Bigg[ 1+\frac{1}{\tilde C^2}\Bigg(
    -\frac{\Delta}{2}\big(\tilde D^{2}+2(\Delta+1)\tilde D+(\Delta^{2}-1)\big) \\
    +\frac14\Bigg( \Delta\,
    \frac{\Gamma\!\left(\frac{1+\Delta}{2}\right)}{\Gamma\!\left(1+\frac{\Delta}{2}\right)}\,
    (\tilde D+\Delta) \Bigg)^{2}\; \Bigg)
    +\mathcal O\left( \frac{1}{\tilde C^3} \right)\Bigg].
    \label{eq:M_near_lightcone}
\end{multline}
where $\mathcal{M}_{\tilde C \gg 1}^{(2)}$ denotes the full approximation to second order (and not just the second order correction). This expression provides the appropriate asymptotic approximation for $|\mathcal M|$ in the near-lightcone regime $\tilde C\gg1$ with $|\tilde D|\ll\tilde C$. The identities
\begin{equation}
    \cos\!\left(\frac{\pi\Delta}{2}\right)
    \Gamma\!\left(\frac{1-\Delta}{2}\right)
    =\frac{\pi}{\Gamma\!\left(\frac{1+\Delta}{2}\right)},
    \qquad \sin\!\left(\frac{\pi\Delta}{2}\right)
    \Gamma\!\left(-\frac{\Delta}{2}\right) =
    -\frac{\pi}{\Gamma\!\left(1+\frac{\Delta}{2}\right)}
\end{equation}
have been used to simplify the Gamma-function coefficients.

In figure \ref{fig:M_L_and_delta_slices}, we compare the results of numerically integrating eq.~\eqref{eq:M_numerical} to the first-order terms in eqs.~\eqref{eq:M_D_first_order} (shown in red) and \eqref{eq:M_near_lightcone} (shown in blue), illustrating the accuracy of the approximations in their respective regimes of validity. 

\subsection*{Negativity}

\begin{figure}
    \centering
    \includegraphics[width=\textwidth]{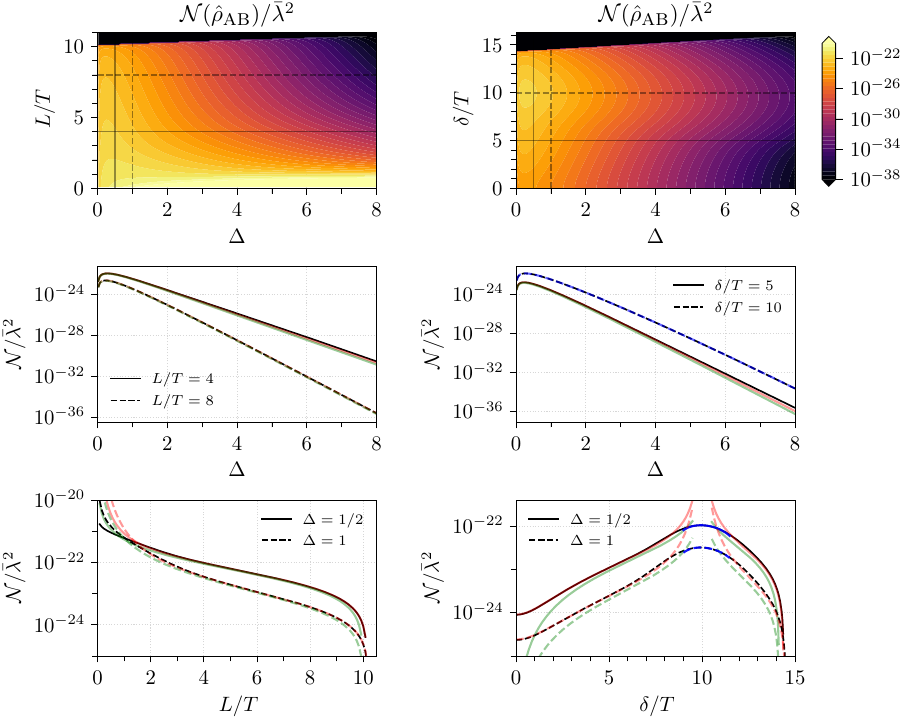}
    \caption{The plots above show $\mathcal N(\hat{\rho}_\textsc{ab})$ as a function of $\Delta$, $L/T$, $\delta/T$, comparing the numerical analysis of eq.~\eqref{eq:negativity_def} with the approximation given by eqs.\eqref{eq:N_LO}--\eqref{eq:N_lightcone_approx}. The plots in the top row are identical to those shown in the main text in figure \ref{fig:negativity}. Below them, in black, we show cross-sections of the top plots at various fixed values. In the left column, we set $\delta/T = 0$ and vary $L/T$ and $\Delta$. In the right column, we set $L/T = 10$ and vary $\delta/T$ and $\Delta$. In all plots, we set $\Omega T = 10$. In green, we plot the far-from-lightcone approximation to first order as given in eq.~\eqref{eq:N_LO}, and in red, the approximation to second order as given in eq.~\eqref{eq:N_NLO}. In blue, we plot the on-lightcone approximation given in eq.~\eqref{eq:N_lightcone_approx}.}
    \label{fig:Negativity_L_and_delta_slices}
\end{figure}

We now use the results obtained above to derive approximations for the negativity. We will first look at the far-from-lightcone regime defined in eq.~\eqref{eq:M_conditions}. Substituting $\LL_{\A\A}^{(1)}$ from eq.~\eqref{eq:largeOmega} and $\mathcal{M}^{(1)}_{\tilde{D} \scriptscriptstyle\gg 1}$ from eq.~\eqref{eq:M_D_first_order} into the negativity \eqref{taxihome} yields
\begin{equation} \label{eq:N_LO}
    \frac{\mathcal N^{(1)}_{\tilde{D}\scriptscriptstyle\gg 1}}{\bar{\lambda}^2} \ =\ \pi\,e^{-\tfrac12 (T\Omega)^2}\! \left[ \frac{T^{2\Delta}}{|L^2-\delta^2|^{\Delta}} -\frac{1}{(T\Omega)^{2\Delta}}
    \right]_+,
\end{equation}
where $g_+$ is shorthand for $\text{max}[g,0]$. We expect this to provide a good approximation in the regime where both $|D| \gg 1$ and $T \Omega \gg 1$. This first-order approximation is plotted in green in figure \ref{fig:Negativity_L_and_delta_slices} alongside the results of calculating $\mathcal N$ numerically. In the bottom left plot, since $\delta/T = 0$, we have $\tilde{D} = (L/T)^2$, and indeed we can see the fit improving for larger $L/T$ until we approach the regime $L \approx \Omega$, at which eq.~\eqref{eq:N_LO} goes to zero. 

We also plot eq.~\eqref{eq:N_LO} in green in the bottom right plot of figure \ref{fig:Negativity_L_and_delta_slices}. However, this is not quite in the regime for \eqref{eq:N_LO} to provide a good fit, at least at first order. If we include the next leading terms for $\mathcal L_\textsc{aa}$ and $\mathcal{M}_{\tilde{D}\scriptscriptstyle\gg 1}$ given terms in eq.~\eqref{eq:largeOmega} and eq.~\eqref{eq:holler3} respectively, the negativity becomes
\begin{multline} \label{eq:N_NLO}
    \frac{\mathcal N^{(2)}_{\tilde{D}\scriptscriptstyle\gg 1}}{\bar{\lambda}^2} \ =\ \pi\,e^{-\tfrac12 (T\Omega)^2}\!
    \Bigg[ \frac{T^{2\Delta}}{|L^2-\delta^2|^{\Delta}}
    \left(1+\frac{\Delta}{\tilde D} + \frac{2\Delta(\Delta+1)}{\tilde D^{2}}\left(\frac{\delta}{T}\right)^{\!2}\right) \\
    -\frac{1}{(T\Omega)^{2\Delta}}
    \left(1-\frac{2\Delta(\Delta+\tfrac12)}{(T\Omega)^2}\right)
    \Bigg]_+ ,
\end{multline}
where again $\mathcal N^{(2)}_{\tilde{D}\scriptscriptstyle\gg 1}$ denotes the full approximation to second order. This is plotted in red in the cross-sections of figure \ref{fig:Negativity_L_and_delta_slices}, and provides a significantly closer approximation to the negativity away from the lightcone.

To obtain an expression for the negativity near the lightcone, we instead use $|\mathcal{M}_{\tilde{C}\scriptscriptstyle\gg 1}^{(2)}|$ from eq.~\eqref{eq:M_near_lightcone}. This yields
\begin{multline} \label{eq:N_lightcone_approx}
    \frac{\mathcal N^{(2)}_{\tilde C\scriptscriptstyle\gg 1}}{\bar{\lambda}^2}
    =e^{-\tfrac12 (T\Omega)^2}
    \Bigg[\frac{\pi^{3/2}}{2^{\Delta/2}\,\Gamma\!\left(\frac{1+\Delta}{2}\right)}\,\tilde C^{-\Delta}\\
    \times\Bigg\{
    1+\frac{1}{\tilde C^2}\Bigg(
    -\frac{\Delta}{2}\big(\tilde D^{2}+2(\Delta+1)\tilde D+(\Delta^{2}-1)\big)
    +\frac14\left(
    \Delta\,
    \frac{\Gamma\!\left(\frac{1+\Delta}{2}\right)}{\Gamma\!\left(1+\frac{\Delta}{2}\right)}\,
    (\tilde D+\Delta)
    \right)^{\!2}
    \Bigg)
    \Bigg\}
    \\[4pt]
    -\frac{\pi}{(T\Omega)^{2\Delta}}
    \Bigg]_+ .
\end{multline}
which is valid when $\tilde{C} \gg 1$, $\tilde C \gg \tilde D$, and $T \Omega \gg 1$. In figure \ref{fig:Negativity_L_and_delta_slices} we show this in blue.

\subsection*{Mutual information}

In section \ref{subsec:mutual_info} we discussed the asymptotic analysis of mutual information, the results of which we repeat here for reference. 

\begin{figure}
    \centering
    \includegraphics[width=\textwidth]{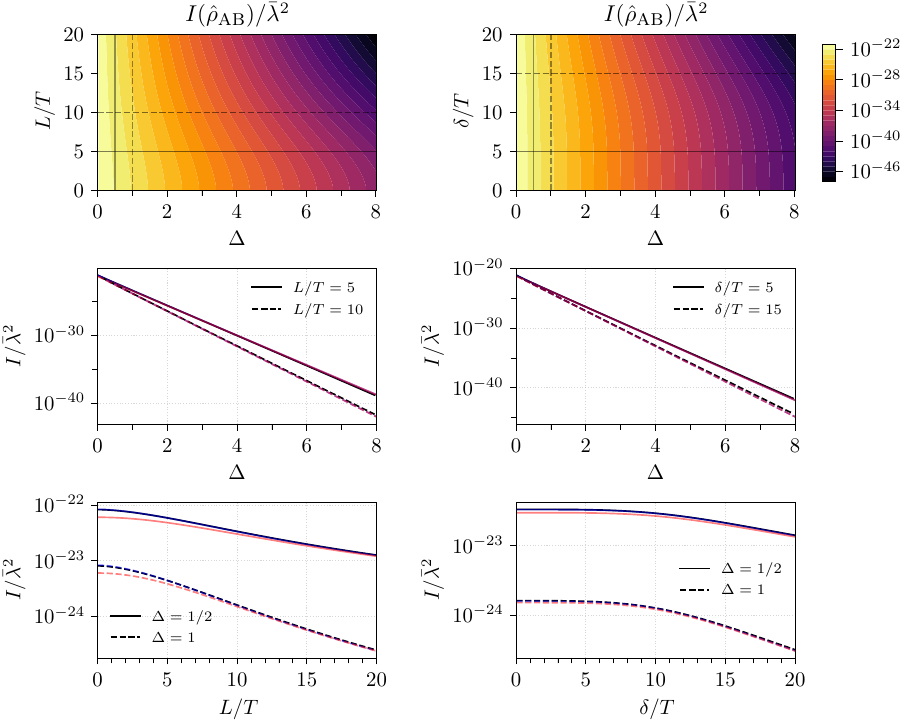}
    \caption{The plots above show mutual information $I(\hat{\rho}_\textsc{ab})$ as a function of $\Delta$, $L/T$, $\delta/T$, comparing the numerical analysis of eq.~\eqref{eq:negativity_def} with the approximation given by eqs.\eqref{eq:N_LO}--\eqref{eq:N_lightcone_approx}. The plots in the top row are identical to those shown in the main text in figure \ref{fig:MutualInfo_L_and_delta_slices}. Below them, in black, we show cross-sections of the top plots at various fixed values. In the left column, we set $\delta/T = 0$ and vary $L/T$ and $\Delta$. In the right column, we set $L/T = 10$ and vary $\delta/T$ and $\Delta$. In all plots, we set $\Omega T = 10$. In blue, we plot the approximation obtained with $\mathcal L_\textsc{aa}$ and $\mathcal L_\textsc{ab}^{(1)}$, given in eq.~\eqref{eq:Iapprox1}. In red, we plot the approximation obtained by further imposing $L/T \gg \Omega T$, given in eq.~\eqref{eq:Iapprox2}.}
    \label{fig:MutualInfo_L_and_delta_slices}
\end{figure}

Substituting $\LL_{\A\A}^{(1)}$ and $\LL_{\A\B}^{(1)}$ from eqs.~\eqref{eq:largeOmega} and \eqref{eq:LAB_leadingO} into the definition of the mutual information provided in eq.~\eqref{eq:mutual_info} yields
\begin{align}
     \frac{I(\hat{\rho}_{\A\B})^{(1)}}{\bar{\lambda}^2} \ &= \ \frac{\pi}{(T\Omega)^{2\Delta}}\,e^{-\tfrac12 T^2 \Omega^2} \left[\left(1+\frac{(T\Omega)^{2\Delta}}{|D|^\Delta} \right) \log\!\left(1+\frac{(T\Omega)^{2\Delta}}{|D|^\Delta} \right)
     \right. \label{eq:Iapprox1}\\
     &\qquad\qquad\qquad\qquad\qquad\qquad\left.+\left(1-\frac{(T\Omega)^{2\Delta}}{|D|^\Delta} \right) \log\!\left(1-\frac{(T\Omega)^{2\Delta}}{|D|^\Delta} \right)
     \right],
     \nonumber
\end{align}
which is valid in the regimes that $\LL_{\A\A}^{(1)}$ and $\LL_{\A\B}^{(1)}$ are both valid: $T \Omega \gg 1$ and $D \gg 1$ (where $D$ is given in \eqref{eq:D_def}). In the cross-sections of figure \ref{fig:mutual_info}, eq.~\eqref{eq:Iapprox1} is plotted in blue, and in most cases lies on top of the results obtained numerically. If we further impose $L/T \gg \Omega T$, this reduces to 
\begin{align}
    \frac{I(\hat{\rho}_{\A\B})^{(1)}_{L/T {\scriptscriptstyle \gg} \Omega T}}{\bar{\lambda}^2} \ &=\ \pi\,e^{-\tfrac12 T^2\Omega^2}\, \frac{(T\Omega)^{2\Delta}}{|D|^{2\Delta}}\left[ 1 + \mathcal{O}\left(\frac{1}{|D|}, \frac{1}{(T\Omega)^2} \right)\right],\label{eq:Iapprox2}\\
    & =\ \frac{|\mathcal L_\textsc{ab}^{(1)}|^2}{\mathcal L_\textsc{aa}^{(1)}}\left[ 1 + \mathcal{O}\left(\frac{1}{|D|}, \frac{1}{(T\Omega)^2} \right)\right].
\end{align}
This is plotted in red in the cross-sections of figure \ref{fig:mutual_info}, and still provides a good approximation in most of the parameter regime considered.

\subsection*{Decomposition of $\mathcal M$}

We will now consider the leading-order behaviour for  $\M^+$ and $\mathcal M^-$ defined in section \ref{sec:separation} by replacing  the two-point function in the integrand of $\M$ in eq.~\eqref{eq:M_numerical} by its real and imaginary parts respectively. These expressions may the be written as
\begin{gather}
\frac{\mathcal M^+}{\bar\lambda^2}
\ =\ -T^{2\Delta-1}\sqrt{\frac{\pi}{2}}\; e^{-\frac12T^2\Omega^2+\ii\Omega\delta}
\left[ \int_{-L}^{L}\!\dd v\;
\frac{e^{-\frac{(v+\delta)^2}{2T^2}}}{(L^2-v^2)^\Delta}
\;+\;
\cos(\pi\Delta)\!\!\int_{|v|>L}\!\!\!\dd v\;
\frac{e^{-\frac{(v+\delta)^2}{2T^2}}}{(v^2-L^2)^\Delta}
\right]\,,
\label{Mplus}\\ 
\frac{\mathcal M^-}{\bar\lambda^2} = T^{2\Delta-1}\sqrt{\frac{\pi}{2}}\; \sin(\pi\Delta)\;
e^{-\frac12T^2\Omega^2+\ii\Omega\delta}
\int_{|v|>L}\!\dd v\; \frac{e^{-\frac{(v+\delta)^2}{2T^2}}}{(v^2-L^2)^\Delta} \,.   
\label{Mminus}
\end{gather}

\begin{figure}[t]
    \centering
    \includegraphics[width=\textwidth]{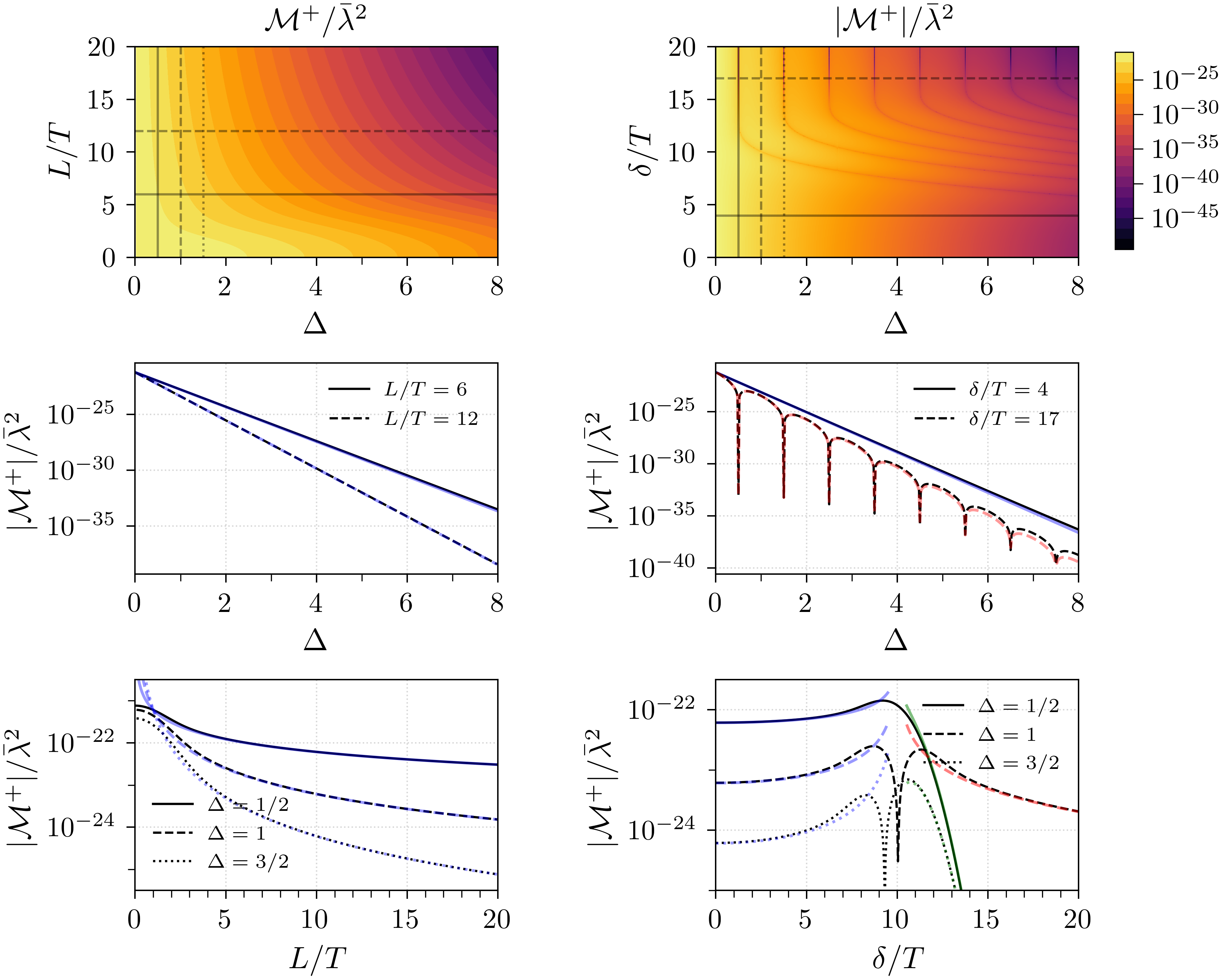}
    \caption{The plots above show $|\mathcal M^+|$ as a function of $\Delta$, $L/T$, $\delta/T$. The top row of plots appear in figures \ref{fig:M_plus_minus_and_N_plus_minus_Delta-L_heatmaps} and \ref{fig:M_plus_minus_and_N_plus_minus_Delta-delta_heatmaps} in the main text. Below them, in black, we show cross-sections of the top plots at various fixed values. In the left column, we set $\delta/T = 0$ (in which case $\mathcal M^+$ is real) and vary $L/T$ and $\Delta$. In the right column, we set $L/T = 10$ and vary $\delta/T$ and $\Delta$. In all plots, we set $\Omega T = 10$. In blue, we show $\mathcal{M}^{+ \, (1)}_\textsc{sl}$, the first-order approximation for $\mathcal M^+$ in the far-spacelike limit, as given in eq.~\eqref{eq:M_SL_leading}. In red, we show $\mathcal{M}^{+ \, (1)}_\textsc{tl}$, the first-order approximation for $\mathcal M^+$ in the far-timelike limit, as given in eq.~\eqref{eq:M+_TL_leading}. In green, we show the endpoint approximation \eqref{Mplus_endpoint}, which gives the leading behaviour of $\mathcal M^+$ in the near-lightcone timelike regime when the standard timelike contribution vanishes at half-integer values of $\Delta$.
    }
    \label{fig:M+_L_delta_Delta_slices}
\end{figure}

Let us focus on $\mathcal{M}^+$ and first consider the far-spacelike regime
$L^2-\delta^2\gg T^2$. In this regime, eq.~\eqref{Mplus} is dominated by the integral over $-L\le v\le L$, as the contribution from the timelike regions $|v|>L$ is exponentially suppressed. The Gaussian factor localizes the $v$-integral to a neighbourhood of $v=-\delta$ of width $\mathcal O(T)$, which lies well inside the spacelike region and sufficiently far from the endpoints $v=\pm L$. In this case we can approximate
\begin{gather}
\begin{multlined}
    \frac{\mathcal M^+}{\bar\lambda^2}\approx\frac{\mathcal{M}^+_\textsc{sl}}{\bar{\lambda}^2} = \mathcal{M}^{+ \, (1)}_\textsc{sl} \bigg[1+\Delta\,\frac{T^2}{L^2-\delta^2} +2\Delta(\Delta+1)\,\frac{\delta^2\,T^2}{(L^2-\delta^2)^2} \qquad\qquad\qquad\\
    +\mathcal O\!\left(\frac{T^4}{(L^2-\delta^2)^2},\, e^{-\delta^2/(2T^2)}\right)\bigg],
\end{multlined}
    \label{ohare22}\\
    \mathcal{M}^{+ \, (1)}_\textsc{sl} \ =\ -\pi\,e^{-\frac12 T^2\Omega^2+\ii\Omega\delta}\; \frac{T^{2\Delta}}{(L^2-\delta^2)^\Delta} \ = \ \mathcal{M}^{(1)}_{\tilde{D}\scriptscriptstyle\gg 1}.\qquad \label{eq:M_SL_leading}
\end{gather}
Note that $\mathcal{M}^{+ \, (1)}_\textsc{sl}$ matches the leading-order behaviour of $\mathcal{M}_{\tilde{D} \scriptscriptstyle\gg 1}$, given in eq.~\eqref{eq:M_D_first_order}. In fact, with a careful examination, we see that the full expansion in eq.~\eqref{ohare22} matches that for $\mathcal{M}$ in eq.~\eqref{eq:holler3}. This reflects the fact that in the far-spacelike regime, $\mathcal M^+ \approx \mathcal M$, since the antisymmetric part of the two-point function appearing in $\M^-$ is not supported in spacelike separation. As we will see below, $\M^-_\textsc{sl}$ is also strongly suppressed  in this regime. We plot $\mathcal{M}^{+ \, (1)}_\textsc{sl}$ in figure \ref{fig:M+_L_delta_Delta_slices} in blue alongside the numerical results in the spacelike regime. 

Next consider instead the far-timelike regime $\delta^2-L^2\gg T^2$, taking
$\delta>0$ for definiteness. In this case, the Gaussian factor in
eq.~\eqref{Mplus} is centred at $v=-\delta$, which lies deep inside the region $v\le -L$. Consequently, the contribution from $v\ge L$ is exponentially suppressed, and the leading behaviour comes from the integral over $v\le -L$:
\begin{equation}
    \frac{\mathcal M^+}{\bar\lambda^2} \approx \frac{ \mathcal{M}^+_\textsc{tl}}{\bar{\lambda}^2} = -\,T^{2\Delta-1} \sqrt{\frac{\pi}{2}}\,\cos(\pi\Delta)\; e^{-\frac12 T^2\Omega^2 + \ii\Omega\delta}\; \int_{-\infty}^{-L}\!\dd v\; \frac{e^{-\frac{(v+\delta)^2}{2T^2}}}{(v^{2}-L^2)^\Delta},
    \label{Mplus2}
\end{equation}
up to exponentially suppressed corrections. Since $\delta\gg L$, the Gaussian localises the remaining $v$-integral to a neighbourhood of $v=-\delta$ of width $\mathcal{O}(T)$, so the denominator may be expanded about $v=-\delta$. Further, extending the upper limit of integration from $-L$ to $+\infty$ introduces only exponentially small errors. One then finds
\begin{gather}
    \frac{\mathcal{M}^+_\textsc{tl}}{\bar{\lambda}^2}\ =\ \mathcal{M}^{+\,(1)}_\textsc{tl}\left[
    1-\Delta\,\frac{T^2}{\delta^2-L^2}
    +2\Delta(\Delta+1)\,\frac{\delta^2\,T^2}{(\delta^2-L^2)^2}
    +\mathcal O\!\left(\frac{T^4}{(\delta^2-L^2)^2}\right)
    \right],\label{Mplus99} \\
    \mathcal{M}^{+\,(1)}_\textsc{tl} \ =\
    -\,\pi\,\cos(\pi\Delta)\,e^{-\frac12 T^2\Omega^2+\ii\Omega\delta} \,
    \frac{T^{2\Delta}}{(\delta^2-L^2)^{\Delta}}\ =\ \cos(\pi \Delta) \mathcal{M}^{(1)}_{\tilde{D} {\scriptscriptstyle\gg} 1}.
    \label{eq:M+_TL_leading}
\end{gather}
Note this goes to zero when $\Delta$ is a half-integer. We discuss this case in more detail below. In figure \ref{fig:M+_L_delta_Delta_slices}, $\mathcal{M}^{+\,(1)}_\textsc{tl}$ is plotted in red alongside the numerical results for timelike parameter regimes. The factor $\cos(\pi\Delta)$ are evident, for example, in the $\delta = 17$ cross-section in the right panel of the middle row of figure \ref{fig:M+_L_delta_Delta_slices}.

When $\Delta$ is a half-integer,  \eqref{Mplus99} vanishes due to the factor $\cos(\pi\Delta)=0$. In this case, $\mathcal M^+$ is instead dominated by the contributions from the lightcone endpoints at $v=\pm L$. 
For $\delta>0$, the relevant contribution arises from the endpoint at $v=L$, which is probed by the Gaussian in the regime $\delta-L=\mathcal O(T)$. To capture this behaviour, we return to the integral \eqref{Mplus2} and consider the timelike regime $\delta>L$ with $\delta-L=\mathcal O(T)$, in which the Gaussian factor $e^{-(v-\delta)^2/(2T^2)}$ has appreciable support extending to the lower integration limit at $v=L$. In this regime, the singular behaviour of the denominator at the lightcone endpoint becomes relevant. Writing $v=L+s$ with $s\ge0$ and $s\ll L$, the denominator may be expanded as $v^2-L^2=2Ls+\mathcal O(s^2)$. Evaluating the resulting endpoint contribution yields the leading behaviour
\begin{equation}
    \frac{\mathcal M^{+\,(1)}_{\textsc{tl},\text{end}}}{\bar\lambda^2}\ =\
    -\,T^{2\Delta-1}\sqrt{\frac{\pi}{2}}\, e^{-\frac12 T^2\Omega^2+\ii\Omega\delta}\, \Gamma(1-\Delta)\, \frac{1}{(2L)^{\Delta}}\,
    \left(\frac{T^2}{\delta-L}\right)^{1-\Delta},
    \label{Mplus_endpoint}
\end{equation}
up to exponentially suppressed corrections. At half-integer values of $\Delta$, in the timelike near-lightcone regime, the endpoint expression \eqref{Mplus_endpoint} therefore describes the dominant behaviour of $\mathcal M^+$. This expression is plotted in green in figure \ref{fig:M+_L_delta_Delta_slices} for the half-integer $\Delta$ cases. We emphasise that this result arises from a distinct near-lightcone asymptotic expansion and should not be interpreted as a subleading correction to the saddle-point contribution \eqref{Mplus99}, which applies only when $\delta^2-L^2\gg T^2$.

Now we turn to $\M^-$ given by eq.~\eqref{Mminus}. The Gaussian factor in the $v$-integral is centred at $v=-\delta$. In the far-timelike regime
$\delta^2-L^2\gg T^2$, this centre lies deep within the region $v\le -L$, so the contribution from the $v\ge L$ portion of the $|v|>L$ integral is
exponentially suppressed. The leading contribution therefore comes from the
region $v\le -L$, and we get
\begin{equation}
    \frac{\mathcal{M}^-}{\bar{\lambda}^2}\approx
    \frac{\mathcal{M}^-_\textsc{tl}}{\bar{\lambda}^2}
    \ =\ 
    \ii\, T^{2\Delta - 1}\sqrt{\frac{\pi}{2}}\,
    \sin(\pi\Delta)\, e^{ -\frac12 T^2 \Omega^2+\ii \Omega\delta}
    \int_{-\infty}^{-L}\!\dd v\;
    \frac{e^{-\frac{(v+\delta)^2}{2T^2}}}{(v^{2}-L^2)^\Delta}\,,
    \label{Mminus2}
\end{equation}
up to exponentially suppressed corrections.

\begin{figure}
    \centering
    \includegraphics[width=\textwidth]{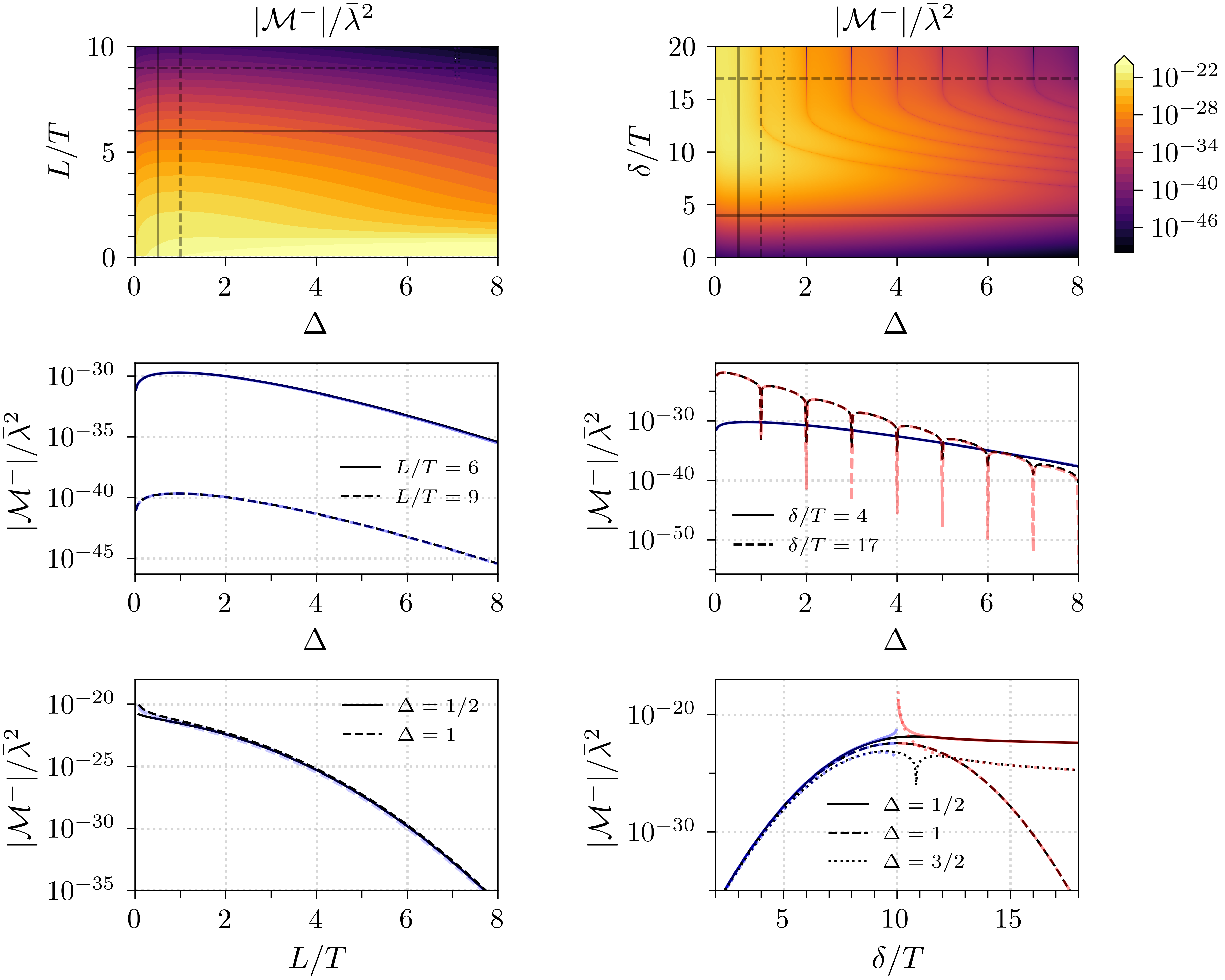}
    \caption{The plots above show $\mathcal M^-$ as a function of $\Delta$, $L/T$, $\delta/T$. The top row of plots appear in figures \ref{fig:M_plus_minus_and_N_plus_minus_Delta-L_heatmaps} and \ref{fig:M_plus_minus_and_N_plus_minus_Delta-delta_heatmaps} in the main text. Below them, in black, we show cross-sections of the top plots at various fixed values. In the left column, we set $\delta/T = 0$ and vary $L/T$ and $\Delta$. In the right column, we set $L/T = 10$ and vary $\delta/T$ and $\Delta$. In all plots, we set $\Omega T = 10$. In blue, we show $\mathcal{M}^{-}_\textsc{sl}$, the approximation for $\mathcal M^-$ in the far-spacelike limit, as given in eq.~\eqref{eq:M-_SL_leading}. In red, we show $\mathcal{M}^{- \, (1)}_\textsc{tl}$, the first-order approximation for $\mathcal M^-$ in the far-timelike limit, as given in eq.~\eqref{eq:M-_TL_leading}.}
    \label{fig:M-_L_delta_Delta_slices}
\end{figure}

Comparing this equation to eq.~\eqref{Mplus2}, we see that in this regime, we have
\begin{equation}
\mathcal{M}^-_\textsc{tl}\ \approx\ -\ii \tan(\pi\Delta)\, \mathcal{M}^+_\textsc{tl}.
   \label{Mminus33} 
\end{equation}
Hence, comparing to eq.~\eqref{Mplus99}, we find
\begin{gather}
    \frac{\mathcal{M}^-_\textsc{tl}}{\bar{\lambda}^2}\ =\  \mathcal{M}^{-\,(1)}_\textsc{tl}
    \left[ 1-\Delta\,\frac{T^2}{\delta^2-L^2}+2\Delta(\Delta+1)\,\frac{\delta^2\,T^2}{(\delta^2-L^2)^2} +\mathcal O\!\left(\frac{T^4}{(\delta^2-L^2)^2}\right)\right], \label{Mminus99}\\
    \mathcal{M}^{-\,(1)}_\textsc{tl}\ =\ \ii \, \pi\,\sin(\pi\Delta)\,e^{-\frac12 T^2\Omega^2+\ii\Omega\delta} \,\frac{T^{2\Delta}}{(\delta^2-L^2)^{\Delta}}. \label{eq:M-_TL_leading}
\end{gather}
We plot this expression in red alongside the numerical evaluation of the exact expression  in figure \ref{fig:M-_L_delta_Delta_slices}. The zeros from the factor $\sin(\pi\Delta)$ are evident, for example, in the $\delta = 17$ cross-section of the right panel of the middle row. Further, we note that when the results in eqs.~\eqref{eq:M+_TL_leading} and \eqref{eq:M-_TL_leading} are added together, the trigonometric factors combine as $\cos(\pi\Delta)+\ii \sin(\pi\Delta)=e^{\ii\pi \Delta}$. Hence the leading-order terms in the timelike regime satisfy   $\mathcal{M}^{+\,(1)}_\textsc{tl} + \mathcal{M}^{-\,(1)}_\textsc{tl} = \mathcal{M}^{(1)}_{\tilde{D}\scriptscriptstyle\gg 1}$, same as the exact expressions. We note again that this is very particular to our choice of setup and Gaussian switching. The extent to which this is true in general is studied in detail in \cite{Zambianco_2024}.

We now consider $\mathcal M^-$ in the far-spacelike regime $L^2-\delta^2\gg T^2$. In this regime the commutator part of the two-point function has no support at spacelike separation, so the integral in \eqref{Mminus} receives contributions only from the exponentially small tails of the Gaussian in the timelike regions $|v|>L$. These tails are dominated by neighbourhoods of the lightcone endpoints $v=\pm L$. Hence, as already noted, $\mathcal M^-$ is negligible compared to $\mathcal M^+$ in this regime. Nevertheless, it is useful to develop an asymptotic expansion in the far-spacelike limit, both to quantify the leading suppression of $\mathcal M^-$ and to compare directly with the numerical evaluation of the exact expression.

For $\delta\ge 0$ (taken for definiteness), we extract the endpoint contribution by expanding the integrand in a neighbourhood of a generic lightcone endpoint. Writing $v=\pm L \mp s$ with $s\ge 0$ and expanding for small $s$, we find
\begin{equation}
    v^2-L^2 \;\approx\; 2Ls,
    \qquad
    e^{-\frac{(v+\delta)^2}{2T^2}}
    \;\approx\;
    e^{-\frac{(L\mp\delta)^2}{2T^2}}
    e^{-\frac{(L\mp\delta)}{T^2}s},
\end{equation}
where we have neglected the subleading factor $e^{-s^2/(2T^2)}$ in the dominant region $s\sim T^2/(L\mp\delta)\ll T$.
where we have neglected the subleading factor $e^{-s^2/(2T^2)}$ in the dominant region $s\sim T^2/(L-\delta)\ll T$. Adding the contributions from the two lightcone endpoints, we obtain
\begin{align}
    \frac{\mathcal{M}^-_\textsc{sl}}{\bar{\lambda}^2}
    \ &\simeq\
    \ii \frac{\pi^{3/2} e^{\ii \Omega\delta}}{2^{\Delta+\frac12}\,\Gamma(\Delta)}\,
    e^{ -\frac12 T^2 \Omega^2}\,
    \frac{T}{L}\left[
      e^{-\frac{(L-\delta)^2}{2T^2}}
      \Big(\frac{L-\delta}{L}\Big)^{\Delta-1}
      +e^{-\frac{(L+\delta)^2}{2T^2}}
      \Big(\frac{L+\delta}{L}\Big)^{\Delta-1}
    \right],
    \label{eq:M-_SL_leading}
\end{align}
where we used $\sin(\pi\Delta)\Gamma(1-\Delta)=\pi/\Gamma(\Delta)$. We plot eq.~\eqref{eq:M-_SL_leading} in blue in figure \ref{fig:M-_L_delta_Delta_slices} in the spacelike regimes. Comparing with the numerical results, we see \eqref{eq:M-_SL_leading} continues to provide an accurate approximation to $\mathcal M^-$ even outside its nominal regime of validity, suggesting that endpoint contributions dominate the integral in eq.~\eqref{Mminus} down to $L/T\approx 1$.

\subsection*{Decomposition of negativity}

Recall our definition of $\mathcal{N}^\pm$ from eq.~\eqref{taxihome}:
\begin{equation}
    \mathcal{N}^{\pm} = \max\left( 0, |\mathcal{M}^{\pm}| - \mathcal{L}_{\A\A} \right) + \mathcal{O}(\lambda^4)\,.
    \label{taxihome2}
\end{equation}
Following the discussion of section \ref{sec:separation}, we can interpret $\mathcal{N}^-/\mathcal{N}$ as quantifying how much of the entanglement is acquired through communication. 

\begin{figure}
    \centering
    \includegraphics[width=\textwidth]{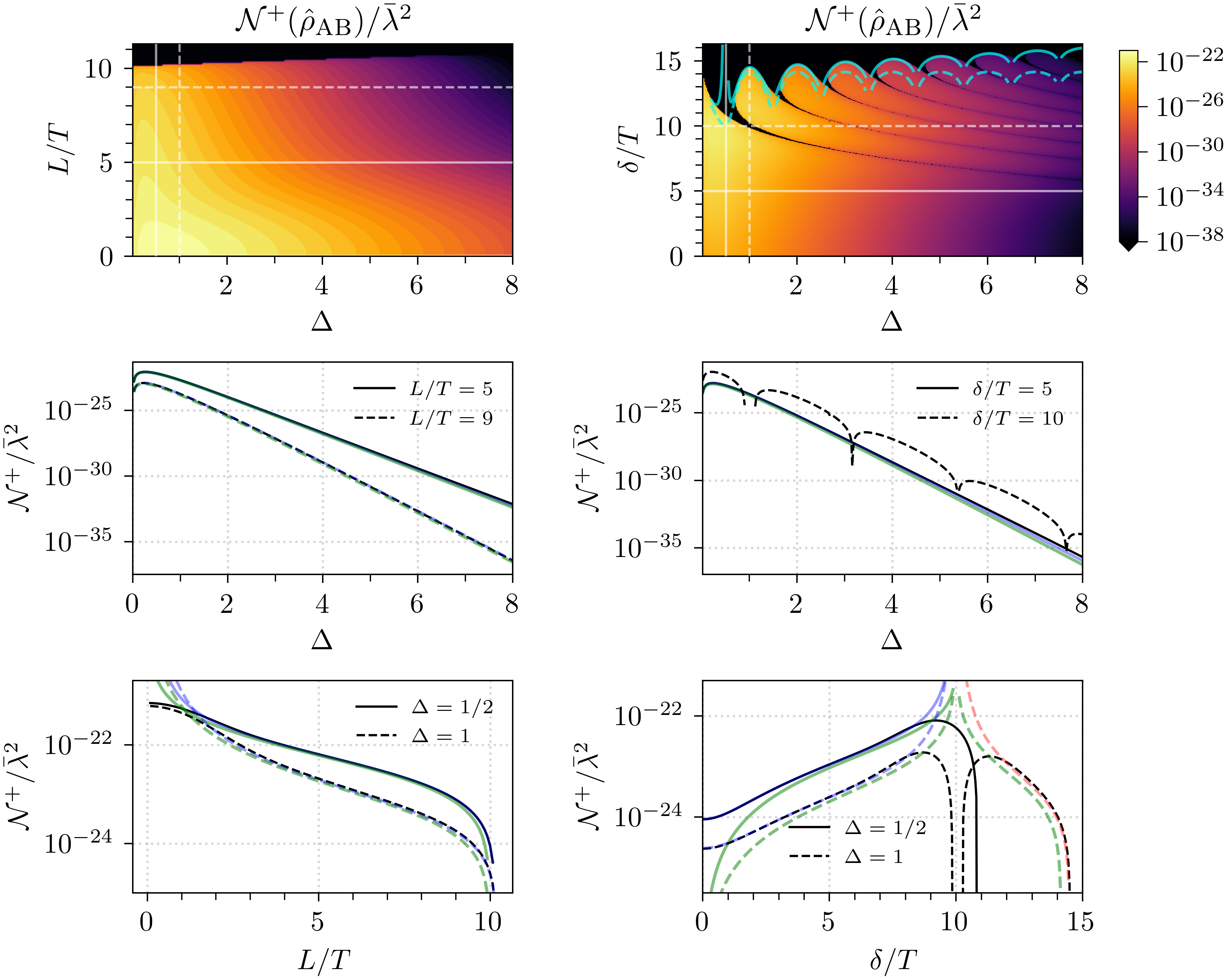}
    \caption{As argued in section \ref{sec:separation}, $\mathcal N^+$ may be interpreted as the negativity arising from genuine entanglement harvesting. The colour plots above appear in figures \ref{fig:M_plus_minus_and_N_plus_minus_Delta-L_heatmaps} and \ref{fig:M_plus_minus_and_N_plus_minus_Delta-delta_heatmaps} in the main text. The cyan lines atop the top right plot indicate the first and second-order approximations to the $\mathcal N = 0$ boundary, provided in eqs.~\eqref{eeq:N+_bdy_1stO} and \eqref{eq:N+_bdy_2ndO} respectively (we have omitted regions around half-integer values of $\Delta$ as these expressions both go to zero).
    Below these plots, in black, we show cross-sections of the top plots at various fixed values. In the left column, we set $\delta/T = 0$ and vary $L/T$ and $\Delta$. In the right column, we set $L/T = 10$ and vary $\delta/T$ and $\Delta$. In all plots, we set $\Omega T = 10$. In blue, we show the spacelike approximation for $\mathcal N^+$ provided in eq.~\eqref{Nplus55}, and in red, the timelike approximation provided by eq.~\eqref{Nplus44}. In green, we show the leading-order approximations to $\mathcal N^+$ given in eqs.~\eqref{eq:N+_SL_0} and \eqref{eq:N+_TL_0} of the main text.}
    \label{fig:N+_L_delta_Delta_slices}
\end{figure}

To arrive at approximations of $\mathcal{N}^{\pm}$ in the far-spacelike regime $L^2 - \delta^2 \gg T^2$, we may use $\mathcal L_\textsc{aa}$ from eq.~\eqref{eq:largeOmega} and $\mathcal{M}^{+}_\textsc{sl}$ from eq.~\eqref{ohare22}. This yields
\begin{align}
  \frac{\mathcal N^{+}}{\bar{\lambda}^2}&\approx  \frac{\mathcal N^{+}_\textsc{sl}}{\bar{\lambda}^2}\ = 
    \pi\,e^{-\frac12 T^2\Omega^2} \Bigg[
    \frac{T^{2\Delta}}{(L^2-\delta^2)^\Delta}
    \Bigg( 1+\Delta\,\frac{T^2}{L^2-\delta^2}
    +2\Delta(\Delta+1)\,\frac{\delta^2\,T^2}{(L^2-\delta^2)^2}\Bigg)
    \nonumber\\
    &\qquad\qquad\qquad\qquad\qquad\qquad\qquad-\frac{1}{(T\Omega)^{2\Delta}} \Bigg(1 - \frac{2\Delta(\Delta+\tfrac12)}{(T\Omega)^2}\Bigg)\Bigg]_+.
    \label{Nplus55}
\end{align}
In figure \ref{fig:N+_L_delta_Delta_slices} we plot the above expression in blue against the results of numerical evaluation. In the left column, $\delta/T = 0$. If we impose this in \eqref{Nplus55}, we arrive at the same boundary for $\mathcal N^{+\,(1)}_\textsc{sl} = 0$ as in eq.~\eqref{eq:nomore} for $\N(\hat{\rho}_{\A\B})^{(1)}_{\tilde{D} {\scriptscriptstyle\gg} 1}$. This makes sense because $\N^-$ is suppressed by an extra exponential in this regime and so we have $\N^+\simeq \N$. 

Let us now consider the far-timelike regime $\delta^2 - L^2 \gg T^2$. Using $\mathcal L_\textsc{aa}$ from \eqref{eq:largeOmega} and $\mathcal{M}^+_\textsc{tl}$ from \eqref{Mplus99}, $\mathcal{N}^+$ becomes
\begin{align}
    \frac{\mathcal{N}^{+}}{\bar{\lambda}^2}\ &\approx\frac{\mathcal{N}^{+}_\textsc{tl}}{\bar{\lambda}^2}\ =  \pi\,e^{-\frac12 T^2\Omega^2} \Bigg[ 
    \frac{|\cos(\pi\Delta)|\,T^{2\Delta}}{(\delta^2-L^2)^{\Delta}}\Bigg( 1-\Delta\,\frac{T^2}{\delta^2-L^2}
    +2\Delta(\Delta+1)\,\frac{\delta^2\,T^2}{(\delta^2-L^2)^2}\Bigg)\nonumber\\
    &\qquad\qquad\qquad\qquad\qquad\qquad\qquad-\frac{1}{(T\Omega)^{2\Delta}} \Bigg(1 - \frac{2\Delta(\Delta+\tfrac12)}{(T\Omega)^2}\Bigg)\Bigg]_+.
    \label{Nplus44}
\end{align}
In figure \ref{fig:N+_L_delta_Delta_slices} we plot the above expression in red against the results of numerical evaluation. Because of the $\cos(\pi\Delta)$ factor, the boundary where $\mathcal{N}^+=0$ oscillates. If we find an approximated expression for the boundary by equating to zero only the leading order terms (those given in eq.~\eqref{eq:N+_TL_0}), we find 
\begin{equation} \label{eeq:N+_bdy_1stO}
    \frac{\delta}{T}\le \sqrt{\frac{L^2}{T^2}+ |\cos(\pi\Delta)|^{1/\Delta}\, T^2\Omega^2}\ \equiv\ \alpha_\Delta\,.
\end{equation}
The boundary where this inequality is saturated is marked by a dashed cyan line in the right colour plot of figure \ref{fig:N+_L_delta_Delta_slices}. If we keep the subleading terms, we find 
\begin{equation}
    \frac{\delta}{T}\ \le\ \alpha_\Delta\,\left[
    1 + \frac{(\Delta+\frac12)\,|\cos(\pi\Delta)|^{1/\Delta}-\frac12}{\alpha_\Delta^2} +\frac{\Delta+1}{|\cos(\pi\Delta)|^{1/\Delta}\,T^2\Omega^2}
    \right]\,. \label{eq:N+_bdy_2ndO}
\end{equation}
The boundary where this is saturated is marked by a solid cyan line in the same figure.

\begin{figure}
    \centering
    \includegraphics[width=\textwidth]{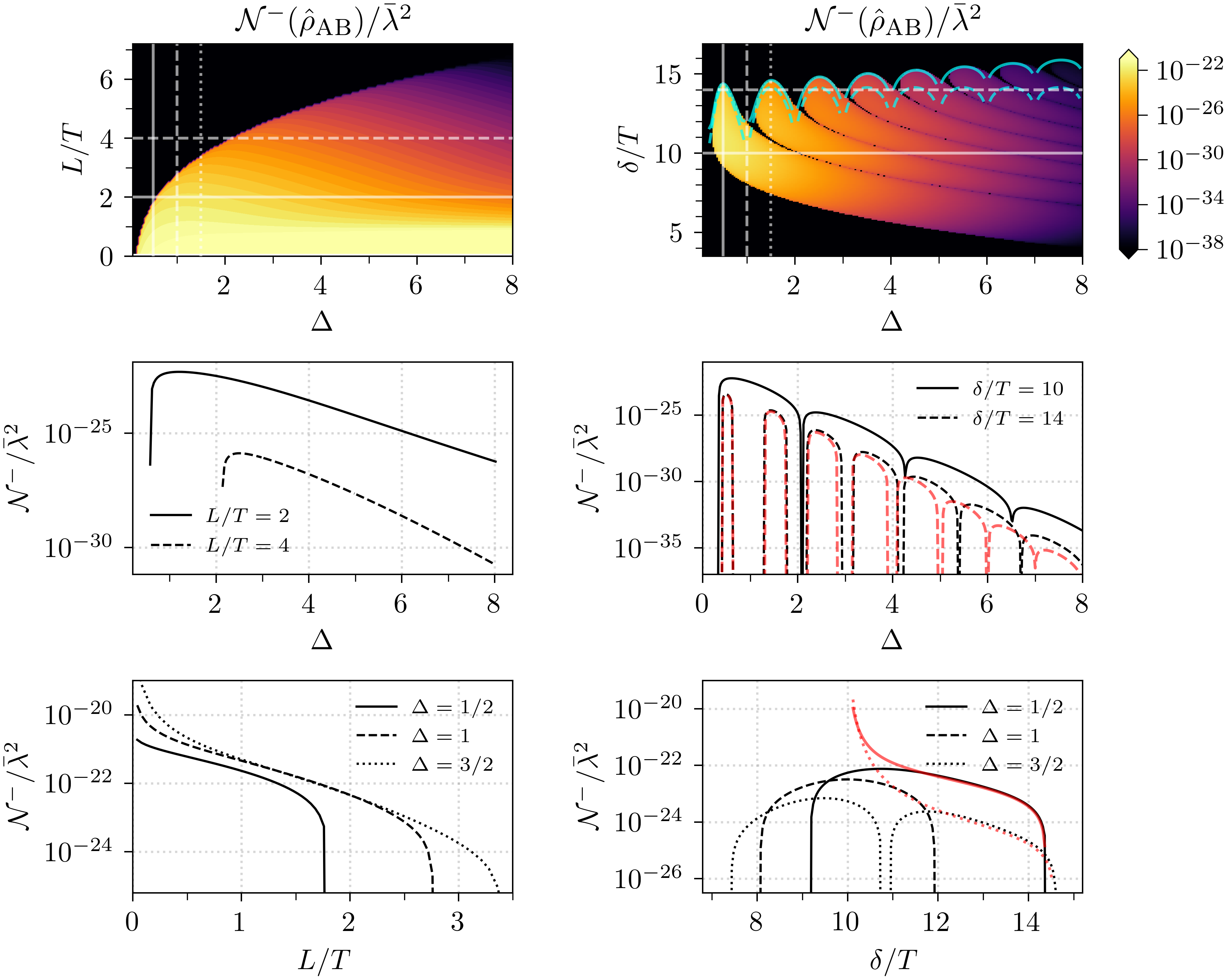}
    \caption{As argued in section \ref{sec:separation}, $\mathcal N^-$ may be interpreted as the negativity arising from communication between the detectors. The colour plots above appear in figures \ref{fig:M_plus_minus_and_N_plus_minus_Delta-L_heatmaps} and \ref{fig:M_plus_minus_and_N_plus_minus_Delta-delta_heatmaps} in the main text. The cyan lines atop the top right plot indicate the first and second-order approximations to the $\mathcal N = 0$ boundary, provided in eqs.~\eqref{eq:N-_bdy_1stO} and \eqref{eq:N-_bdy_2ndO} respectively (we have omitted regions around the integer values of $\Delta$ as these expressions both go to zero). Below them, in black, we show cross-sections of the top plots at various fixed values. In the left column, we set $\delta/T = 0$ and vary $L/T$ and $\Delta$. In the right column, we set $L/T = 10$ and vary $\delta/T$ and $\Delta$. In all plots, we set $\Omega T = 10$. In the far-spacelike regime, we get $\mathcal N_\textsc{sl}^- \simeq 0$, thus asymptotic analysis does not lend itself naturally to parameter regime in the left plots. On the right, in red, we show the far-timelike approximation provided in eq.~\eqref{Nminus44}. }
    \label{fig:N-_L_delta_Delta_slices}
\end{figure}

Finally, we will analyse the contribution of negativity coming from communication, $\mathcal N^-$. In the far-spacelike regime $L^2 - \delta^2 \gg T^2$, we use $\mathcal L_\textsc{aa}$ from eq.~\eqref{eq:largeOmega} and $\mathcal M^-_\textsc{sl}$ from eq.~\eqref{eq:M-_SL_leading}. These combine to yield
\begin{align}
    \frac{\mathcal{N^-_\textsc{sl}}}{\bar{\lambda}^2}\ &\approx\ 0
    \label{Nminus55}
\end{align}
as expected since the commutator has support only in the lightcone and the Gaussian tails exponentially suppress $\mathcal{M}^-$, which cannot compete with the local noise.

For the timelike case, we use $\mathcal L_\textsc{aa}$ from eq.~\eqref{eq:largeOmega} and $\mathcal M^-_\textsc{tl}$ from eq.~\eqref{Mminus99}. These combine to yield
\begin{align}
    \frac{\mathcal{N^-_\textsc{tl}}}{\bar{\lambda}^2}\ &\simeq\ 
    \pi\,e^{-\frac12 T^2\Omega^2} \Bigg[ 
    \frac{|\sin(\pi\Delta)|\,T^{2\Delta}}{(\delta^2-L^2)^{\Delta}}\Bigg(
    1-\Delta\,\frac{T^2}{\delta^2-L^2}
    +2\Delta(\Delta+1)\,\frac{\delta^2\,T^2}{(\delta^2-L^2)^2}\Bigg)
    \nonumber\\
    &\qquad\qquad\qquad\qquad\qquad-\frac{1}{(T\Omega)^{2\Delta}} \Bigg(1 - \frac{2\Delta(\Delta+\tfrac12)}{(T\Omega)^2}\Bigg)\Bigg]_+\,.
    \label{Nminus44}
\end{align}
In figure \ref{fig:N-_L_delta_Delta_slices} we plot the above expression in red alongside the results of numerical integration. Because of the $\sin(\pi\Delta)$ factor, the boundary where $\mathcal{N}^-=0$ oscillates. In the far-timelike regime (\ie $\delta\gg L\gg T$), eq.~\eqref{Mminus33} holds and so the only real difference between eqs.~\eqref{Nplus44} and \eqref{Nminus44} is the replacement $\cos(\pi\Delta)\to\sin(\pi\Delta)$. Hence, if we consider only leading-order terms in \eqref{Nminus44}, the boundary is given by 
\begin{equation} \label{eq:N-_bdy_1stO}
    \frac{\delta}{T}\le \sqrt{\frac{L^2}{T^2}+ |\sin(\pi\Delta)|^{1/\Delta}\, T^2\Omega^2}\equiv \tilde\alpha_\Delta\,.
\end{equation}
Again, the boundary where this inequality is saturated is marked by a dashed cyan line in the right colour plot of figure \ref{fig:N+_L_delta_Delta_slices}. Including the subleading terms of \eqref{Nminus44}, we get 
\begin{equation}
    \frac{\delta}{T}\le \tilde\alpha_\Delta\left[
    1 + \frac{(\Delta+\frac12)\,|\sin(\pi\Delta)|^{1/\Delta}-\frac12}{\tilde\alpha_\Delta^2} +\frac{\Delta+1}{|\sin(\pi\Delta)|^{1/\Delta}\,T^2\Omega^2}
    \right]_+\,. \label{eq:N-_bdy_2ndO}
\end{equation}
The boundary where this is saturated is marked by a solid cyan line in the same figure. Because the spacelike approximation $\mathcal N^-_\textsc{sl}$ is zero and the timelike approximation $\mathcal N^-_\textsc{tl}$ applies technically only for far-timelike regions, where the negativity is zero, we find that our asymptotic analysis does not cover much of the range where $\mathcal N^-_\textsc{tl} \neq 0$.

\bibliographystyle{JHEP}
\bibliography{biblio.bib}
\end{document}